\documentclass[11pt,a4paper]{article}
\pdfoutput=1

\usepackage{jheppub} 

\usepackage{amsmath,amssymb,epsfig,amsfonts, mathtools}
\usepackage{graphicx}
\usepackage{epsf}
\usepackage{makeidx}
\usepackage{multirow}
\usepackage[all]{xy}
\usepackage{hyperref}
\usepackage{verbatim} 
\usepackage{rotating}
\usepackage{xcolor}
\usepackage{bm}
\usepackage{subcaption}
\usepackage{cleveref}
\usepackage{slashed}
\usepackage[normalem]{ulem}

\usepackage{graphicx}
\usepackage{latexsym,amsmath,amsfonts,amssymb}
\usepackage{mathrsfs}
\usepackage{bbm}
\usepackage{bm}
\usepackage{paralist}
\usepackage{youngtab}
\usepackage{bclogo}
\usepackage{array}

\usepackage{tikz}
\usepackage{tikz-cd}
\usepackage{diagbox}
\usetikzlibrary{positioning}
\usetikzlibrary{calc}
\usetikzlibrary{decorations.pathreplacing,calligraphy}

\usepackage{xstring}
\usetikzlibrary{decorations.pathmorphing} 
\usetikzlibrary{decorations.markings} 
\usetikzlibrary{snakes}
\usetikzlibrary{arrows, arrows.meta, patterns, patterns.meta} 
\usetikzlibrary{shapes} 
\usetikzlibrary{matrix} 
\usetikzlibrary{positioning} 
\usepackage[english]{babel} 
\usepackage[autostyle]{csquotes}

\usepackage{tcolorbox}

\usepackage[font=small,labelfont=bf]{caption}

\tikzstyle{brane}=[draw]
\tikzset{D7/.style={circle, draw=black, inner sep=0pt, fill=white, minimum size=3mm}}
\tikzset{hasse/.style={circle, fill,inner sep=2pt}}
\tikzset{flavour/.style={regular polygon,fill=white,regular polygon sides=4,inner sep=2.5pt, draw}}
\tikzset{gauge/.style={circle, draw,inner sep=2.5pt}}
\tikzset{gaugeb/.style={circle, draw,fill=black,inner sep=2.5pt}}
\tikzset{gauger/.style={circle, draw,fill=cyan,inner sep=2.5pt}}
\tikzset{gaugeg/.style={circle, draw,fill=red,inner sep=2.5pt}}
\tikzset{bd/.style={circle, draw=black, inner sep=0pt, fill=black, minimum size=2mm}}
\tikzset{wd/.style={circle, draw=black, inner sep=0pt, fill=white, minimum size=2mm}}
\tikzset{SUd/.style={circle, draw=black, inner sep=0pt, fill=yellow, minimum size=2mm}}
\tikzset{Dynkin/.style={circle, draw=black, inner sep=0pt, fill=white, minimum size=2mm}}
\tikzstyle{ligne}=[draw, thick] 
\tikzset{doublearrow/.style={ draw=black!75, color=black!75, thick, double distance=3pt, }}

\tikzset{->-/.style={decoration={
  markings,
  mark=at position #1 with {\arrow{>}}},postaction={decorate}}}
\tikzset{->>-/.style={decoration={
  markings,
  mark= between positions #1-0.05 and #1+0.05 step 0.1 with {\arrow{>}}
  },postaction={decorate}}}
\tikzset{->>>-/.style={decoration={
  markings,
  mark= between positions #1-0.1 and #1+0.1 step 0.1 with {\arrow{>}}
  },postaction={decorate}}}
  \tikzset{->>>>-/.style={decoration={
  markings,
  mark= between positions #1-0.2 and #1+0.2 step 0.1 with {\arrow{>}}
  },postaction={decorate}}}
    \tikzset{->>>>>-/.style={decoration={
  markings,
  mark= between positions #1-0.3 and #1+0.2 step 0.1 with {\arrow{>}}
  },postaction={decorate}}}
  \tikzset{->>>>>>-/.style={decoration={
  markings,
  mark= between positions #1-0.1 and #1+0.5 step 0.1 with {\arrow{>}}
  },postaction={decorate}}}

\numberwithin{equation}{section}

\newcommand{\mat}[1]{\begin{pmatrix} #1 \end{pmatrix}}

\newcommand{\be}{\begin{equation}} 
\newcommand{\ee}{\end{equation}}
\newcommand{\bea}{\begin{equation} \begin{aligned}} \newcommand{\eea}{\end{aligned} \end{equation}}

\newcommand{\bit}{\begin{itemize}} 
\newcommand{\eit}{\end{itemize}}

\newcommand{\cF}{\mathcal{F}}

\newcommand{\cH}{\mathcal{H}}
\newcommand{\cI}{\mathcal{I}}

\newcommand{\cN}{\mathcal{N}}
\newcommand{\cO}{\mathcal{O}}

\newcommand{\cS}{\mathcal{S}}
\newcommand{\cT}{\mathcal{T}}
\newcommand{\cU}{\mathcal{U}}

\newcommand{\cZ}{\mathcal{Z}}

\renewcommand{\t}{\widetilde }

\newcommand{\CE}{\mathcal{E}}
\newcommand{\CF}{\mathcal{F}}

\newcommand{\CN}{\mathcal{N}}

\newcommand{\CT}{\mathcal{T}}

\newcommand{\Fg}{\mathfrak{g}}

\newcommand{\ov}{\over}

\newcommand{\KK}{D_{S^1}}

\makeatletter
\newcommand{\thickhline}{%
    \noalign {\ifnum 0=`}\fi \hrule height 1pt
    \futurelet \reserved@a \@xhline
}
\newcolumntype{"}{@{\hskip\tabcolsep\vrule width 1pt\hskip\tabcolsep}}
\makeatother

\title{Seiberg-Witten geometry, modular rational elliptic surfaces and BPS quivers}

\abstract{We study the Coulomb branches of four-dimensional supersymmetric quantum field theories with $\cN = 2$ supersymmetry, including the KK theories obtained from the circle compactification of the 5d $\cN = 1$ $E_n$ Seiberg theories, with particular focus on the relation between their Seiberg-Witten geometries and rational elliptic surfaces. More attention is given to the modular surfaces, which we completely classify using the classification of subgroups of the modular group ${\rm PSL}(2,\mathbb{Z})$, deriving closed-form expressions for the modular functions for all congruence and some of the non-congruence subgroups. Moreover, in such cases, we give a prescription for determining the BPS quivers from the fundamental domains of the monodromy groups and study how changes of these domains can be interpreted as quiver mutations. This prescription can be also generalized to theories whose Coulomb branches contain `undeformable' singularities, leading to known quivers of such theories.

}

\author{Horia Magureanu

{\it Mathematical Institute, University of Oxford, \\
Andrew-Wiles Building,  Woodstock Road, Oxford, OX2 6GG, UK}\\

}

\makeatletter
\gdef\@fpheader{}
\makeatother

\begin{document}
\maketitle

\numberwithin{equation}{section}  
\allowdisplaybreaks  



\section{Introduction}

Supersymmetry often allows for a better grasp of the strongly coupled regime of quantum field theories. Our understanding of four-dimensional quantum field theories with $\mathcal{N}=2$ supersymmetry, in particular, was improved by the introduction of the Seiberg-Witten (SW) geometry in \cite{Seiberg:1994aj, Seiberg:1994rs}. In that approach, the Coulomb branch (CB) physics was encoded in a non-trivial fibration of a torus over the CB, with the complex structure parameter of the fiber being identified with the low energy $U(1)$ effective gauge coupling, in the specific example of one (complex) dimensional Coulomb branches. 

These geometries are rational elliptic surfaces (RES), for which a classification of the allowed configurations of singular fibers exists in the mathematical literature \cite{Persson:1990, Miranda:1990}. This mathematical formalism was recently discussed in this context in \cite{Caorsi:2018ahl, Closset:2021lhd}, where it was proposed to identify a four-dimensional $\cN = 2$ theory by specifying the singular fiber at infinity. In this approach, RG flows to Argyres-Douglas \cite{Argyres:1995jj, Argyres:1995xn} or Minahan Nemeschansky \cite{Minahan:1996cj, Minahan:1996fg} theories are easily identifiable. Additionally, it was shown in \cite{Closset:2021lhd}, based on F-theory methods (see \textit{e.g.} \cite{Banks:1996nj, Mikhailov:1998bx, Aspinwall:1998xj, Mayrhofer:2014opa, Cvetic:2017epq}), that this formalism also encodes information about global aspects of the flavour symmetry and, conjecturally, about the one-form symmetries of these theories \cite{Gaiotto:2014kfa}. This leads to results that are consistent with other previous works on the subject \cite{Morrison:2020ool, Albertini:2020mdx, Apruzzi:2021vcu, Bhardwaj:2021ojs, Closset:2020scj, DelZotto:2020esg, Buican:2021xhs, Bhardwaj:2020phs, Bhardwaj:2021pfz, Bhardwaj:2021mzl}.

A particularly interesting and constrained subset of rational elliptic surfaces consists of the modular RES, first constructed by Shioda \cite{Shioda:1972}. This construction assigns a finite index subgroup of $\Gamma \subset {\rm PSL}(2,\mathbb{Z})$ to an elliptic surface, with the singular fibers being mapped to cusps and elliptic points of $\Gamma$. An important use of modularity is the evaluation of the so-called $u$-plane integral \cite{Moore:1997pc, Labastida:2005zz, Malmendier:2008db, Manschot:2021qqe, Korpas:2019cwg, Aspman:2021kfp}, which makes use of the map from the Coulomb branch to a subspace of the upper half-plane, called the fundamental domain of $\Gamma$. This map was more recently studied to great depths in  \cite{Aspman:2020lmf, Aspman:2021evt, Aspman:2021vhs} for 4d SQCD theories. See also \cite{Matone:1995jr, Matone:1995rx} for relevant aspects of this map.

The present work has two main objectives. First, we completely classify all 4d $\cN = 2$ modular Coulomb branches of rank-one from the perspective of subgroups $\Gamma$ of the modular group. Additionally, using the major simplification of the low energy dynamics provided by modularity, we derive a simple way of obtaining BPS quivers \cite{Fiol:2000wx, Denef:2002ru, Alim:2011kw, Alim:2011ae, Closset:2019juk} from fundamental domains of 4d $\cN = 2$ theories, generalising the examples presented in \cite{Closset:2021lhd}. In particular, the simplest type of singularity on the CB -- which corresponds to a single light BPS state becoming massless, or, equivalently, to an $I_1$ singular fiber in Kodaira's classification -- can be mapped to a rational number on the boundary of the upper half-plane, from which we obtain the corresponding BPS charges as follows:
\be \label{Identification}
    \tau = {q\ov m} \in \mathbb{Q} \quad \longleftrightarrow \quad \pm (m,-q)~,
\ee
where our conventions for the charges are (magnetic, electric). This identification is based on the fact that fundamental domains encode information about the generators of the monodromy group, and, thus, about the monodromies around the singular points on the Coulomb branch. We generalise this identification for other Kodaira-types of singular fibers, which ultimately leads to a construction of BPS quivers from the singular fibers of the SW geometry. Moreover, we give an interpretation for how quiver mutations are related to changes of the fundamental domains. However, let us mention that we do not have a prescription for obtaining the quiver superpotential from the modular properties at the moment. In certain cases, this can be found using exceptional collections \cite{Herzog:2003zc} or brane-tiling methods \cite{Franco:2005rj, Hanany:2005ss}. For more recent work on these subjects we refer to \cite{Closset:2018bjz, Closset:2019juk, Banerjee:2018syt, Banerjee:2020moh, Bonelli:2020dcp, Mozgovoy:2020has, Beaujard:2020sgs, Longhi:2021qvz, DelMonte:2021ytz}.

In this set-up, the `natural' rank-one theories are the circle compactification of the five-dimensional Seiberg $E_n$ SCFTs \cite{Seiberg:1996bd, Morrison:1996xf, Nekrasov:1996cz}, which we will often denote by $\KK E_n$. In five-dimensions, these SCFTs are the UV completion of $SU(2)$ gauge theories with $N_f = n-1$ fundamental flavours. These theories are the simplest examples of 5d SCFTs, which have received much attention in the recent years -- see \textit{e.g.} \cite{Jefferson:2017ahm, Jefferson:2018irk, Apruzzi:2018nre, Apruzzi:2019enx, Bhardwaj:2018yhy, Bhardwaj:2018vuu, Apruzzi:2019opn, Bhardwaj:2019jtr, Bhardwaj:2019ngx, Closset:2020afy, Bhardwaj:2020gyu, Bhardwaj:2020kim, vanBeest:2020kou, Hubner:2020uvb, Apruzzi:2019kgb, Bhardwaj:2020ruf, vanBeest:2020civ, vanBeest:2021xyt,  Closset:2021lwy, Tian:2021cif} and references therein. 
Their SW geometries have been determined using local mirror symmetry arguments in \cite{Hori:2000ck, Hori:2000kt}, and from the SW curve of the six-dimensional E-string theory \cite{Ganor:1996pc, Eguchi:2002fc, Eguchi:2002nx}. The usual 4d SQCD theories can then be obtained by taking the geometric-engineering limit \cite{Katz:1996fh, Klemm:1996bj, Douglas:1996xp, Katz:1997eq}.

As mentioned before, modular configurations are a particularly useful subset of rational elliptic surfaces. 
These configurations have been classified in \cite{Doran:1998hm}, leading to 33 modular RES, which we extend to 47 different surfaces by considering `quadratic twists', which, however, do not change the modular properties of the elliptic surface. In this work, we review this classification from a different perspective, namely by using the classification of subgroups of ${\rm PSL}(2,\mathbb{Z})$ \cite{Sebbar:2001, Cummins2003, STROMBERG2019436}. These are of two types, called congruence or non-congruence, with the former having the principal congruence subgroup $\Gamma(N)$ as a subgroup. Note that the subgroups of interest in our scenario have index less or equal to 12, due to the rationality condition of the elliptic surfaces. Out of these 33 possibilities, 22 correspond to congruence subgroups, for which we derive closed-form expressions for their modular functions, with particular focus on the `non-standard' cases, that is the groups that are not of $\Gamma^0(N)$ or $\Gamma^1(N)$ type. Many of these have already been known, see \textit{e.g.} \cite{Conway:1979qga, Harnad:1998hh, Maier2006, 2016arXiv160503988S}, with the results being summarized in table~\ref{tab: Modular Functions Summary}, where the groups are ordered by their index. Note additionally that certain subgroups are conjugate in ${\rm PSL}(2,\mathbb{R})$ as indicated, with their Hauptmoduln related by a rescaling of $\tau$.%
\renewcommand{\arraystretch}{1.7}
\begin{table}[h]
\small
\centering
\begin{tabular}{ |c||c|c|c|c|c|} 
 \hline
$\Gamma$ & $n_{\Gamma}$& $(w_i)$ & $(e_2,e_3)$  & $f(\tau)$ & \textit{Monster}
\\ \hline \hline
$\Gamma(1)$ & $1$ & $(1)$ & $(1,1)$ & $j(\tau)$ & $1A$ \\ \hline
$\Gamma^2$ & $2$ & $(2)$ & $(0,2)$ & ${E_6(\tau) \ov \eta(\tau)^{12}}$ & $2a$\\ \hline
$\Gamma^3$ & $3$ & $(3)$ & $(0,2)$ & ${\vartheta_2(\tau)^8 + \vartheta_3(\tau)^8 + \vartheta_4(\tau)^8 \ov 2\eta(\tau)^8}$  & $3C$ \\ \hline
$\Gamma_0(2)$ & $3$ & $(2,1)$ & $(1,0)$ & $ \left( {\eta(\tau) \ov \eta(2\tau)}\right)^{24}$  & $2B$ \\ \hline
$\Gamma_0(3)$ & $4$ & $(3,1)$ & $(0,1)$ & $ \left( {\eta(\tau) \ov \eta(3\tau)}\right)^{12}$  & $3B$ \\ \hline
$4A^0$ & $4$ & $(4)$ & $(2,1)$ & ${\vartheta_2(\tau)^6 + i \vartheta_3(\tau)^6 + \vartheta_4(\tau)^6 \ov \vartheta_2(\tau)^2 \vartheta_3(\tau)^2 \vartheta_4(\tau)^2}$  & $-$ \\ \hline
$5A^0$ & $5$ & $(5)$ & $(1,2)$ & \eqref{5A0}  & $5a$ \\ \hline
$6A^0$ & $6$ & $(6)$ & $(0,3)$ & ${ \vartheta_3(\tau)^4 + e^{2\pi i \ov 3} \vartheta_2(\tau)^4 \ov \eta(\tau)^4}$  &  $-$\\ \hline \hline
$\Gamma_0(4)$ & $6$ & $(4,1,1)$ & $(0,0)$ & $ \left( {\eta(\tau) \ov \eta(4\tau)}\right)^{8}$  & $4B$\\ \hline
$\Gamma(2)$ & $6$ & $(2,2,2)$ & $(0,0)$ & $ \left( {\eta\left({\tau\ov 2}\right) \ov \eta(2\tau)}\right)^{8}$  & $4C$ \\ \hline \hline
$\Gamma_0(5)$ & $6$ & $(5,1)$ & $(2,0)$ & $ \left( {\eta(\tau) \ov \eta(5\tau)}\right)^{6}$  & $5B$\\ \hline
$3C^0$ & $6$ & $(3,3)$ & $(2,0)$ & $ \left( {\eta(\tau)^2 \ov \eta\left({\tau\ov 3}\right) \eta(3\tau)}\right)^{6}$  & $9A$\\ \hline
$4C^0$ & $6$ & $(4,2)$ & $(2,0)$ & $\left( {\eta(\tau)^2 \ov \eta\left({\tau\ov 2}\right) \eta(2\tau)}\right)^{12}$  &  $4D$\\ \hline
$\Gamma_0(7)$ & $8$ & $(7,1)$ & $(0,2)$ & $ \left( {\eta(\tau) \ov \eta(7\tau)}\right)^{4}$  & $7B$\\ \hline
$6C^0$ & $8$ & $(6,2)$ & $(0,2)$ & $\left( {\eta(2\tau) \ov  \eta(6\tau)}\right)^{6}$  & $6c$\\ \hline
$4D^0$ & $8$ & $(4,4)$ & $(0,2)$ & \eqref{4D0}  & $-$ \\ \hline
$\Gamma_1(5)$ & $12$ & $(5,5,1,1)$ & $(0,0)$ & $ {1\ov q}\prod_{n=1}^{\infty} \left(1-q^n\right)^{-5\left( {n\ov 5}\right)}$  & $-$ \\ \hline
$\Gamma_0(6)$ & $12$ & $(6,3,2,1)$ & $(0,0)$ & $ \left({\eta(\tau)^5 \eta(3\tau) \ov \eta(6\tau)^5 \eta(2\tau)}\right)$  & $6E$ \\ \hline \hline
$\Gamma_0(8)$ & $12$ & $(8,2,1,1)$ & $(0,0)$ & $ \left( {\eta(4\tau)^3 \ov \eta(2\tau) \eta(8\tau)^2}\right)^{4}$  & $8E$ \\ \hline
$\Gamma_0(4) \cap \Gamma(2)$ & $12$ & $(4,4,2,2)$ & $(0,0)$ & $ \left( {\eta\left(2\tau \right)^3 \ov \eta(\tau) \eta(4\tau)^2}\right)^{4}$  & $8D$ \\ \hline \hline
$\Gamma_0(9)$ & $12$ & $(9,1,1,1)$ & $(0,0)$ & $ \left( {\eta(\tau) \ov \eta(9\tau)}\right)^{3}$  & $9B$\\ \hline
$\Gamma(3)$ & $12$ & $(3,3,3,3)$ & $(0,0)$ & $ \left( {\eta\left({\tau\ov 3}\right) \ov \eta(3\tau)}\right)^{3}$  & $9B$\\ \hline
\end{tabular}
    \caption{Modular functions of congruence subgroups of ${\rm PSL}(2,\mathbb{Z})$ that correspond to rational elliptic surfaces; $n_{\Gamma}$ is the index, $(w_i)$ are the widths of the cusps and $(e_2, e_3)$ are the number of elliptic elements. $\left(n\ov p\right)$ is the Legendre symbol. For more details we refer the reader to section \ref{section: modular RES classification}.}
    \label{tab: Modular Functions Summary}
\end{table}\renewcommand{\arraystretch}{1}%
%
Fundamental domains can be drawn for these groups using the isomorphism with pairs of permutations derived by Millington \cite{10.1112/jlms/s2-1.1.351}. This is summarized in appendix \ref{Appendix Permutations}. 

Non-congruence subgroups are still an active field of research, being much less studied compared to congruence subgroups. For a nice review of the existing literature, see \cite{long2007arithmetic}. Their systematic study was initiated by Atkin and Swinnerton-Dyer in \cite{atkin1971modular}, when it was observed that the Fourier coefficients of the associated modular forms have \emph{unbounded denominators}, a conjecture only recently proved \cite{2021arXiv210909040C}. Closed-form expressions for these modular forms can still be found in certain cases, in particular when the non-congruence subgroup of interest is also a subgroup of a proper congruence subgroup \cite{Scholl1985, Scholl1988TheLR, Scholl2010OnLR}, but, as we will see, they involve fractional powers of the usual modular functions. This procedure is based on a connection to Galois theory explained in section \ref{section: modular RES classification}, and will only work for 2 out of the 11 non-congruence subgroups. 



The rest of the paper is organised as follows. In section \ref{section: intro} we review the necessary tools of (modular) rational elliptic surfaces, while in section \ref{section: 3} we derive the identification between cusps of fundamental domains and BPS states, and subsequently apply this correspondence to the four-dimensional SQCD theories, analysing how changes of the fundamental domains are reflected on the BPS quivers. Section \ref{section: modular RES classification} reviews the classification of modular rational elliptic surfaces for both congruence and non-congruence monodromy groups, while in section \ref{Rank-one SCFTs} we discuss some aspects of four-dimensional $\cN = 2$ SCFTs, including theories with undeformable singularities.


\section{The physics of Rational Elliptic Surfaces}\label{section: intro}

In this section, we discuss the link between fundamental domains and BPS quivers. We start by reviewing the link between Coulomb branches of 4d $\cN = 2$ theories and rational elliptic surfaces \cite{Caorsi:2018ahl, Closset:2021lhd}, focusing in particular on the modular rational elliptic surfaces. In these cases, the $U$-plane, \textit{i.e.} the Coulomb branch, can be mapped to a finite region of the upper half-plane by an isomorphism, where the region is the fundamental domain of the associated monodromy group.

Different choices of fundamental domains can be understood on the $U$-plane as changes in the monodromies around the singularities induced by non-trivial paths around singularities. We will try to clarify how such changes of the fundamental domain can be thought of as quiver mutations.


\subsection{The $U$-plane of $\cN = 2$ theories}

The main focus of this section will be on rank-one 4d $\cN = 2$ theories, namely the theories having a one (complex) dimensional Coulomb branch. The simplest representation of the $\cN = 2$ superalgebra is the vector multiplet, which contains a scalar field $\phi$, the gauge connection, as well as the fermionic superpartners. Generic vacua of such theories are described by the vanishing of the scalar potential, in which the gauge group is broken to its maximal torus $SU(2) \rightarrow U(1)$, while the scalar field $\phi$ receives a VEV via the Higgs mechanism. For pure 4d theories, one has:
\be
    \langle \phi\rangle = \left( \begin{matrix} a & 0 \\ 0 & - a \end{matrix} \right)~, \qquad a \in \mathbb{C}~,
\ee
semiclassically. The Coulomb branch is then parametrised by the gauge invariant operator:
\be
    u = \langle Tr \phi^2 \rangle \approx a^2 + \ldots~,
\ee
where the dots signal the presence of quantum corrections. For the KK theories, the real scalar $\sigma$ in the 5d $\cN = 1$ vector multiplet is paired with the $U(1)$ holonomy along the circle direction to form a complex scalar which, by abuse of notation, we will also denote by $a$. Due to the five-dimensional large gauge transformations along the circle direction, the gauge invariant parameter becomes instead \cite{Nekrasov:1996cz, Closset:2021lhd}:
\be \label{U as expectation value}
    U = e^{2\pi i a} + e^{-2\pi i a} + \ldots ~,
\ee
which corresponds to the expectation value of a Wilson line in the fundamental representation of $SU(2)$ in five dimensions.\footnote{We will often use $U$ for both the 5d and 4d Coulomb branch parameter.} At this stage, it is useful to view the Coulomb branch as a $\mathbb{P}^1$, by compactifying the point at infinity, and, thus, the Seiberg-Witten geometry corresponds to the elliptic fibration:
\be
    E \longrightarrow \cS \longrightarrow \mathbb{P}^1 \cong \{ U\}~.
\ee
This is also the definition\footnote{An elliptic surface $\cS$ is rational if it is birationally equivalent to $\mathbb{P}^2$.} of a rational elliptic surface $\cS$ (RES), which can be described in the Weierstrass model by:
\be \label{Weierstrass form}
    y^2 = 4x^3 - g_2(U,\boldsymbol{M}) x - g_3(U,\boldsymbol{M})~.
\ee
We will follow the conventions of \cite{Closset:2021lhd} for all $\KK E_n$ and 4d SQCD Seiberg-Witten curves. In \eqref{Weierstrass form} we also indicate the dependence on the complexified mass parameters of the theory, which, similarly to \eqref{U as expectation value}, are flavour Wilson lines along the circle direction. Let us also define the discriminant and $J$-invariant as:
\be
    \Delta(U,\boldsymbol{M}) = g_2(U,\boldsymbol{M})^3 - 27 g_3(U,\boldsymbol{M})^2~, \qquad \quad J(U,\boldsymbol{M}) = {j(U,\boldsymbol{M})\ov 1728} = {g_2(U,\boldsymbol{M})^3 \ov \Delta(U,\boldsymbol{M})}~.
\ee
The low energy effective action of the $\cN = 2$ theory is obtained by integrating out all massive degrees of freedom, being fully determined by a holomorphic function called the prepotential $\cF$(a). Then, the effective gauge coupling reads:
\be
    \tau = {\partial^2 \cF(a) \ov \partial a^2}~,
\ee
which gets identified with the complex structure parameter of the elliptic fiber. This description breaks down at the loci where certain BPS states become massless. The elliptic fibers above these loci are thus singular and can be found as the loci where the discriminant vanishes. The possible types of singular fibers are given by the Kodaira classification, as shown in table~\ref{tab:kodaira}, while the effective gauge coupling $\tau$ transforms by some elements of ${\rm SL}(2,\mathbb{Z})$. Introducing the dual field to $a$:
\be
    a_D = {\partial \cF \ov \partial a}~, \qquad \tau = {\partial a_D \ov \partial a}~,
\ee
the low energy effective theory is alternatively fully determined by the data $(a_D, a)$, which is a section of a rank-two holomorphic affine bundle over the $U$-plane. For the five-dimensional theories, these periods correspond in the large volume limit to $D4$ and $D2$ branes wrapping the (vanishing) $4$-cycle $[dP_n]$ and a $2$-cycle $[C_f]$ with vanishing self-intersection, 
respectively \cite{Closset:2018bjz, Closset:2021lhd}. Note that in a conveniently chosen basis, the remaining $D2$-brane periods do not receive quantum corrections, which, in practice, boils down to the reduction of the associated Picard-Fuchs equations to a single second order differential equation. The ${\rm SL}(2,\mathbb{Z})$ transformations act on these periods by:
\be
    \left(\begin{matrix} a_D \\ a \end{matrix}\right) \longrightarrow \mathbb{M} \left(\begin{matrix} a_D \\ a \end{matrix}\right)~,
\ee
with the monodromy matrices satisfying the constraint:
\be
    \prod_v \mathbb{M}_v = \mathbb{I}~.
\ee
In the Weierstrass model, the singular fibers are either nodal or cuspidal curves; these can be resolved, with the exceptional divisors intersecting according to the associated extended ADE Dynkin diagram. Each exceptional fiber $F_v$ has an Euler number $e_v$ that is equal to the number of components $m_v$ if the fiber is of type $I_{k>0}$, or equal to $m_v + 1$ otherwise. In particular, $e(F_v) = \text{ord}(\Delta)$, with the latter shown explicitly in table~\ref{tab:kodaira}. Then, one important constraint of a rational elliptic surface is \cite{schuttshioda}:
\be \label{euler number sum}
    \sum_v e(F_v) = 12~,
\ee
where the sum is taken over all singular fibers. A further constraint comes from the flavour algebra associated to each singular fibre, namely:
\be
    {\rm rank}(F_v) = {\rm rank}(\frak{g}_v) = m_v - 1~,
\ee
with:
\be
    \sum_v {\rm rank}(F_v) \leq 8~.
\ee
These constraints severely restrict the possible rational elliptic surfaces, which were classified by Persson and Miranda by their singular fibers and their Mordell-Weil group \cite{Persson:1990, Miranda:1990}. For an elliptic curve, this group law is obtained by declaring that three points add to the marked point at infinity if and only if they are collinear. In the context of elliptic surfaces, the global sections are in one-to-one correspondence with the rational points of the elliptic fibers \cite{schuttshioda}, and, by the Mordell-Weil theorem one has:
\be
    \Phi = \mathbb{Z}^{{\rm rk}(\Phi)} \oplus \mathbb{Z}_{k_1} \oplus \ldots \oplus \mathbb{Z}_{k_l}~.
\ee
Focusing for now on 4d $SU(2)$ SQCD theories with $N_f$ fundamental flavours, in the weak coupling regime $U \approx \infty$, where the semi-classical picture holds, the one-loop correction to the effective gauge coupling can be interpreted as a monodromy matrix acting on the $(a_D, a)$ periods, which leads to the singularity:
\be
    4d~ SU(2)\,+\, N_f~: \qquad F_{\infty} = I^*_{4-N_f}~.
\ee
A similar argument was presented in \cite{Closset:2021lhd} for the $\KK E_n$ theories, with:
\be
    \KK E_n~: \qquad F_{\infty} = I_{9-n}~.
\ee
This led to the proposal of identifying a theory by its fiber at infinity, where we have the following identifications:
\bea    \label{Fiber at infinity}
    II\, , \, III\, \text{ and } \, IV ~ \text{ SCFTs} ~: & \qquad F_{\infty} = II^*\, , \, III^*\, \text{ and } \, IV^* \, \text{ respectively}~, \\
    II^*\, , \, III^*\, \text{ and } \, IV^* ~ \text{ SCFTs} ~:  & \qquad F_{\infty} = II\, , \, III\, \text{ and } \, IV \, \text{ respectively}~,
\eea
while the $I_0^*$ SCFTs have $F_{\infty} = I_0^*$. These identifications are based on the four rational elliptic surfaces with only two singular fibers, namely:
\be
    (I_0^*, I_0^*)~, \qquad (II^*, II)~, \qquad (III^*, III)~, \qquad (IV^*, IV)~,
\ee
which are thus interpreted as describing simultaneously two possibly different SCFTs. Let us stress that the identification \eqref{Fiber at infinity} only partially fixes the theory, as one further needs to specify the deformation pattern, according to \cite{Argyres:2015ffa, Argyres:2015gha, Argyres:2016xmc, Argyres:2016xua}, as well as the Mordell-Weil group of the maximally deformed Coulomb branch. The latter is due to the conjecture of \cite{Closset:2021lhd}, in which the MW group encodes information about the one-form symmetries \cite{Gaiotto:2014kfa}. More precisely, it is the group generated by the sections of the rational elliptic surfaces intersecting `trivially'\footnote{Recall that in the Weierstrass model the singular curves are either nodal or cuspidal curves. Here, by trivial intersection we mean that the sections do not intersect the singular points of these fibers.} all bulk singular fibers that corresponds to the 1-form symmetry. 

Such an example where the MW needs to be specified is $F_{\infty} = I_8$, which corresponds to the $\KK E_1$ and $\KK \t E_1$ theories, namely the theories obtained by the circle compactification of the UV completions of 5d $\cN=1$ $SU(2)$ and $SU(2)_{\pi}$ theories. The $\KK E_1$ theory has a $\mathbb{Z}_2^{[1]}$ one-form symmetry \cite{Morrison:2020ool, Albertini:2020mdx}, while $\KK \t E_1$ does not, and thus the $(I_8, 4I_1)$ configuration with $\Phi = \mathbb{Z}_2$ corresponds to the $\KK E_1$ theory. 

It was also shown in \cite{Closset:2021lhd}, based on F-theory methods, that the free generators of the MW group correspond to abelian factors of the flavour symmetry. Moreover, the torsion part of the Mordell-Weil group encodes information about the global form of the flavour symmetry. However, at this stage, this identification only holds for the theories whose Coulomb branches do not contain \emph{undeformable singularities}. These theories will be the main focus of this work, and we will refer to them as theories with `trivial' or `pure' Coulomb branches, as their maximally deformed CB only involve $I_1$ singular fibers. We will comment further on theories with undeformable singularites in the next section.

One of the starting motivations of \cite{Closset:2021lhd} involved non-trivial RG flows from the 5d $\cN = 1$ $E_n$ SCFTs to four-dimensional theories. The existence of these flows was suggested by the relation between their BPS quivers and q-deformed Painlev\'e equations \cite{Bonelli:2020dcp, Bonelli:2016idi, Bonelli:2016qwg, Bonelli:2017gdk}. Using the formalism of rational elliptic surfaces, the allowed configurations of singular fibers of a theory with trivial Coulomb branch described by the fiber $F_{\infty}$ is given by the subset of rational elliptic surfaces in Persson's classification \cite{Persson:1990, Miranda:1990} that contain the fiber $F_{\infty}$. The only exception to this rule is the aforementioned case $F_{\infty} = I_8$, where the existence of a one-form symmetry implies that the MW group of an allowed configuration must contain a $\mathbb{Z}_2$ torsion subgroup.

This argument can be generalised to theories containing undeformable singularities as follows. One first associates to the maximally deformed Coulomb branch a `naive' flavour lattice $T_{def}$, by using the data in table~\ref{tab:kodaira}. Note that this is not the correct flavour symmetry of such theories, which was determined using different methods in \cite{Caorsi:2018ahl, Argyres:2018taw}. Then, we claim that the allowed configurations for such theories are those that not only contain the singular fiber $F_{\infty}$, but, additionally, their associated naive flavour lattice contains $T_{def}$ as a sublattice. As stated before, this argument might be subject to slight changes if the theory has a non-trivial one-form symmetry.


\subsection{Modular elliptic surfaces}

An important subset of RES consists of the modular rational elliptic surfaces, in which case the $U$-plane is isomorphic to a region of the upper half-plane, $U(\tau)$ being a meromorphic function in $\tau$. In the original work of Seiberg and Witten \cite{Seiberg:1994aj, Seiberg:1994rs}, it was pointed out that the curve of the pure $SU(2)$ theory is the modular curve $\mathbb{H} / \Gamma^0(4)$, where $\Gamma^0(4)$ is a congruence subgroup of ${\rm PSL}(2,\mathbb{Z})$. The modular properties of the other SQCD theories were found in \cite{Nahm:1996di}, and more recently discussed in great detail in \cite{Aspman:2020lmf, Aspman:2021evt, Aspman:2021vhs}. Furthermore, this latter work also includes fundamental domains of configurations that are not modular, in which case the map $U \mapsto \tau$ is not one-to-one anymore. In such cases, the fundamental domains cannot be constructed as quotients $\mathbb{H}/ \Gamma$ due to the existence of branch points and branch cuts on the upper half-plane. 

For our purposes, it will be useful to introduce certain subgroups of the modular group ${\rm PSL}(2,\mathbb{Z})$. We first introduce the principal congruence subgroups of level $N$, described as:
\bea
    \Gamma(N) & = \Bigg\{ \left(\begin{matrix} a & b \\ c & d\end{matrix}\right) \in {\rm PSL}(2,\mathbb{Z}): \left(\begin{matrix} a & b \\ c & d\end{matrix}\right) = \left(\begin{matrix} 1 & 0 \\ 0 & 1\end{matrix}\right)~ {\rm mod} ~ N \Bigg\}~,
\eea
which can be viewed as the kernel of the group homomorphism ${\rm PSL}(2,\mathbb{Z}) \rightarrow {\rm PSL}(2,\mathbb{Z}_N)$. The subgroups $\Gamma$ of ${\rm PSL}(2,\mathbb{Z})$ containing the principal congruence subgroup $\Gamma(N)$ are called \textit{congruence subgroups}, with the \textit{level} being the smallest such positive integer $N$. The most common level-$N$ congruence subgroups are:
\bea
    \Gamma_0(N) & = \Bigg\{ \left(\begin{matrix} a & b \\ c & d\end{matrix}\right) \in {\rm PSL}(2,\mathbb{Z}): c = 0~ {\rm mod} ~ N \Bigg\}~.\\
    \Gamma_1(N) & = \Bigg\{ \left(\begin{matrix} a & b \\ c & d\end{matrix}\right) \in {\rm PSL}(2,\mathbb{Z}): \left(\begin{matrix} a & b \\ c & d\end{matrix}\right) = \left(\begin{matrix} 1 & b \\ 0 & 1\end{matrix}\right)~ {\rm mod} ~ N \Bigg\}~.
\eea
and similarly for the groups $\Gamma^0(N)$ and $\Gamma^1(N)$, which are instead defined by imposing $b = 0~{\rm mod}~N$. Note that these are related to the $\Gamma_0(N)$ and $\Gamma_1(N)$ groups by conjugation. These subgroups satisfy the inclusion:
\be
    \Gamma(N) \subset \Gamma_1(N) \subset \Gamma_0(N)~,
\ee
with $\Gamma_1(N) \cong \Gamma_0(N)$ for $N = 2, 3, 4$ and $6$. Note that these congruences are no longer satisfied in ${\rm SL}(2,\mathbb{Z})$, unless $N = 2$.

Non-congruence subgroups are those that do not contain $\Gamma(N)$ as a subgroup, and are much less studied in the mathematical literature. In most cases their modular forms do not have closed-form expressions; moreover, it was conjectured that the Fourier coefficients of modular forms of non-congruence subgroups have unbounded denominators, a conjecture that has been recently proved in \cite{2021arXiv210909040C}.

Given a subgroup $\Gamma \in {\rm PSL}(2,\mathbb{Z})$, its index $n_{\Gamma} = [\Gamma(1):\Gamma] $ in $\Gamma(1) \cong {\rm PSL}(2,\mathbb{Z})$ is the number of right-cosets of $\Gamma$ in the modular group. As a result, we have:
\be
    {\rm PSL}(2,\mathbb{Z}) = \bigsqcup_{i=1}^{n_{\Gamma}} \Gamma~ \alpha_i~, \qquad \alpha_i \in {\rm PSL}(2,\mathbb{Z})~,
\ee
for a list of coset representatives $\{\alpha_i\}$. The elements of the modular group act on the upper half-plane $\mathbb{H}$ as:
\be
    \tau \mapsto {a\tau + b \ov c\tau + d}~, \qquad \gamma = \left( \begin{matrix} a & b \\ c & d \end{matrix}\right)\in {\rm PSL}(2,\mathbb{Z})~, \qquad \forall \tau \in \mathbb{H}~.
\ee
It then follows that a fundamental domain for the subgroup $\Gamma$ is defined as an open subset $\cF_{\Gamma} \subset \mathbb{H}$ of the upper half-plane, such that no two distinct points are equivalent under the action of $\Gamma$, unless they are on the boundary of $\mathcal{F}_{\Gamma}$; furthermore, under the action of $\Gamma$, any point of $\mathbb{H}$ is mapped to the closure of $\cF_{\Gamma}$. Let us denote the fundamental domain of ${\rm PSL}(2,\mathbb{Z})$ by $\cF_0$. The upper half-plane $\mathbb{H}$ is then obtained by the action of the modular group as:
\be
    \mathbb{H} = {\rm PSL}(2,\mathbb{Z})\cF_0~.
\ee
The fundamental domain of $\Gamma\subset {\rm PSL}(2,\mathbb{Z})$ can be obtained from a list of coset representatives $\{\alpha_i\}$, since:
\be
    \mathbb{H} = \left(\bigsqcup_{i=1}^{n_{\Gamma}} \Gamma~ \alpha_i \right)\cF_0 = \Gamma \left(\bigsqcup_{i=1}^{n_{\Gamma}}  \alpha_i \cF_0 \right)~.
\ee
Thus, the fundamental domain of $\Gamma$ is the disjoint union $\cF_{\Gamma} = \bigsqcup \alpha_i\cF_0$, with the coset representatives chosen such that $\cF_{\Gamma}$ has a connected interior.\footnote{For the standard groups introduced above, fundamental domains can be drawn with Helena Verrill's Java applet \cite{Verrill2000}.} 

A \textit{cusp} of $\Gamma$ is defined as an equivalence class in $\mathbb{Q} \cup \{\infty\}$ under the action of $\Gamma$. The ${\rm PSL}(2,\mathbb{Z})$ group has only one cusp, with the representative usually chosen as $\tau_{\infty} = i\infty$. The \textit{width} of the cusp $\tau_{\infty}$ in $\Gamma$ is the smallest integer $w$ such that $T^w \in \Gamma$. More generally, for a cusp $\widetilde{\tau} = \gamma\tau_{\infty}$, the width is defined as the width of $\tau_{\infty}$ for the group $\gamma^{-1}\Gamma\gamma$. The cusps other than $\tau_{\infty}$ are typically chosen as the points of intersection of the fundamental domain with the real axis.

The other special points in the fundamental domain are the \textit{elliptic} points, which are those points with non-trivial stabilizer, \textit{i.e.} $\gamma_e \tau_* = \tau_*$ for some non-trivial element $\gamma_e \in \Gamma$. The elements $\gamma_e$ are called the elliptic elements of $\Gamma$. It can be shown that the elliptic points always lie on the boundary of the fundamental domain. Finally, the order of an elliptic point $\tau_*$ is the order of the stabilizing subgroup of $\tau_*$ in $\Gamma$. For ${\rm PSL}(2,\mathbb{Z})$ the only elliptic points are $\tau_0\in\{i, e^{2\pi i\ov 3}\}$, with stabilizers $\langle S \rangle$ and $\langle ST \rangle$, of order $2$ and $3$, respectively. One can prove that,  for a given finite index subgroup $\Gamma$ with fundamental domain $\mathcal{F}$, the elliptic points $\tau_* \in \cF$ are always in the ${\rm SL}(2,\mathbb{Z})$ orbit of the above elliptic points, \textit{i.e.} $\tau_* = \gamma\tau_0$,  and thus must have orders $2$ or $3$.

An important subclass of elliptic surfaces is composed of the modular elliptic surfaces, with their construction based on subgroups of $\Gamma \subset {\rm PSL}(2,\mathbb{Z})$ as first discussed by Shioda \cite{Shioda:1972}.\footnote{Shioda's construction is slightly more general, being based on subgroups of ${\rm  SL}(2,\mathbb{Z})$.} Let $n_{\Gamma}$, $\epsilon_2$, $\epsilon_3$ and $\epsilon_{\infty}$ be the index of $\Gamma$ in ${\rm PSL}(2,\mathbb{Z})$, the number of elliptic elements of order two and three, and the number of cusps of $\Gamma$, respectively. The quotient $\mathbb{H}/\Gamma$ together with a finite number of cusps $\epsilon_{\infty}$ forms a compact Riemann surface $\Delta_{\Gamma}$. There then exists a holomorphic map onto the projective line:
\be
    J_{\Gamma} : \Delta_{\Gamma} \rightarrow \mathbb{P}^1~,
\ee
as follows. Let $\Gamma \subset \Gamma_1$ with a canonical map between the Riemann surfaces $\mathbb{H}/\Gamma \rightarrow \mathbb{H}/\Gamma_1$, \textit{i.e.} the map arising naturally from the surjection $\Gamma \rightarrow \Gamma_1$. Taking $\Gamma_1 = {\rm PSL}(2,\mathbb{Z})$, one has $\Delta_{\Gamma_1} \cong \mathbb{P}^1 \cong S^2$, where the $J_{\Gamma_1}$-map is the usual $J$-invariant defined on $\mathbb{H}$, which due to ${\rm PSL}(2,\mathbb{Z})$ invariance descends to a holomorphic map $\Delta_{\Gamma_1} \rightarrow \mathbb{P}^1$. The map $J_{\Gamma}$ is then simply the canonical map between the Riemann surfaces.

Given such a Riemann surface, it can be shown from the ramification of this canonical map and the Riemann-Hurwitz theorem that its genus is given by:
\be \label{genus}
    g_{\Gamma} = 1 + {n_{\Gamma} \ov 12} - {\epsilon_2 \ov 4} - {\epsilon_3 \ov 3} - {\epsilon_{\infty} \ov 2}~.
\ee
Following \cite{Shioda:1972}, let $\Sigma$ be the set of cusps and elliptic points of $\Gamma$, and let:
\be
    \Delta_{\Gamma}' = \Delta_{\Gamma} \setminus \Sigma \subset \mathbb{H}/\Gamma~.
\ee
From a universal covering $\cU$ of $\Delta_{\Gamma}'$, with the projection $\pi: \cU \rightarrow \Delta_{\Gamma}'$, 
there is a holomorphic map $\omega : \cU \rightarrow \mathbb{H} $ such that:
\be
    J_{\Gamma}(\pi(u)) = j(\omega(u))~, \qquad u \in \cU~,
\ee
with $j$ the elliptic modular function on $\mathbb{H}$. That is, the following diagram commutes:
\be
\xymatrix{
   \cU \ar[r]^{\pi} \ar@<-2pt>[d]_{\omega} & \Delta_{\Gamma}' \ar[r] \ar@<-2pt>[d]_{J_{\Gamma}} & \Delta_{\Gamma} \\
   \mathbb{H} \ar[r]^{j}  & \mathbb{P}^1 &
}
\ee
Using Kodaira's construction, one finds an elliptic surface $B_{\Gamma}$ over $\Delta_{\Gamma}$, with a global section having $J_{\Gamma}$ as a functional invariant. Additionally, the map introduced before, $\omega(u)$, which is a single valued function on $\cU$, becomes the `period' map of the elliptic fibre at $u \in \cU$, being a multi-valued function on $\Delta_{\Gamma}$.

The singular fibers lie above the set $\Sigma$ of elliptic points and cusps. The type of singular elliptic fibre (as shown in table~\ref{tab:kodaira}) is determined by some additional data. That is, there is a unique representation $\varphi$ of the fundamental group $\pi_1(\Delta_{\Gamma}')$ of $\Delta_{\Gamma}'$:
\be
    \varphi: \pi_1(\Delta_{\Gamma}') \rightarrow \Gamma \subset {\rm PSL}(2,\mathbb{Z})~,
\ee
satisfying:
\be
    \omega( \gamma \cdot u) = \varphi(\gamma) \cdot \omega(u)~, \qquad \gamma \in \pi_1(\Delta_{\Gamma}')~.
\ee
Note that on the right-hand side of the above equation we have the usual action of an element of $\Gamma$ on a point on the upper half-plane, as $\varphi(\gamma)\in \Gamma$. Thus, for $\gamma_u$ a positively oriented loop around $u\in\Delta_{\Gamma}$, the matrices $\varphi(\gamma_u)$ determine the type of singular fibers. %
\begin{table}%
\centering 
 \begin{tabular}{|c | c||c|c|c|| c|c|c|  } %
\hline
$F_v$ & $\tau$  &ord$(g_2)$  &ord$(g_3)$ & ord$(\Delta)$        &  $\mathbb{M}_\ast$ & $\mathfrak{g}$ & $m_v$ \\  \hline\hline
$I_k$  & $i \infty $&  $0$ &$0$ & $k$        & $T^k$  & $\mathfrak{su}(k)$ & $k$\\  \hline
$I_{k}^\ast$  &$i \infty$&   $2$ &$3$ & $k+6$         & $PT^{k}$ & $\mathfrak{so}(2k+8)$ & $k+5$ \\  \hline
$I_{0}^\ast$  &$\tau_0$&  $\geq 2$ &$\geq 3$ & $6$        & $P$  & $\mathfrak{so}(8)$ & $5$\\  \hline
 $II$ &$e^{{2\pi i \ov 3}}$& $\geq 1$ &$1$ & $2$         & $(ST)^{-1}$  & - & $1$ \\  \hline
 $II^\ast$ &$e^{{2\pi i \ov 3}}$& $\geq 4$ &$5$ & $10$         & $(ST)$ &$ \mathfrak{e}_8$ & $9$ \\  \hline
  $III $&$i$  & $1$ &$\geq 2$ & $3$         & $S^{-1}$  & $\mathfrak{su}(2)$ & $2$\\  \hline
$  III^\ast $&$i$& $3$ &$\geq 5$ & $9$         & $S$ &$ \mathfrak{e}_7$ & $8$\\  \hline
   $IV$ &$e^{{2\pi i \ov 3}}$& $\geq 2$ &$2$ & $4$         & $(ST)^{-2}$  & $\mathfrak{su}(3)$ & $3$\\  \hline
$ IV^\ast$ &$e^{{2\pi i \ov 3}}$ & $\geq 3$ &$4$ & $8$         &  $(ST)^{2}$ &$ \mathfrak{e}_6 $ & $7$ \\  \hline
\end{tabular}
\caption{Kodaira classification of singular fibers based on the orders of vanishing of $g_2$, $g_3$ and $\Delta$ in the Weierstrass model. We indicate the value of the complex structure parameter $\tau$, the monodromy $\mathbb{M}$, the associated flavour algebra $\frak{g}$, as well as the number of components $m_v$ of the singular fibers. Note also that $P = -\mathbb{I}$.}
\label{tab:kodaira}
\end{table} 

The modular group has two elements of order $2$ and $3$, namely the generators $S$ and $ST$, respectively, which satisfy:
\be
    S^2 = 1~, \qquad \quad (ST)^3 = 1~,
\ee
in ${\rm PSL}(2,\mathbb{Z})$. Thus, if $u$ is an elliptic point of order 2, then $\varphi(\gamma_u)$ must be conjugate to $S$, and similarly for elliptic points of order 3. Moreover, from Kodaira's classification of singular fibers in table~\ref{tab:kodaira}, we see that the $J_{\Gamma}$ map `forgets' about starred fibers when we restrict to subgroups of ${\rm PSL}(2,\mathbb{Z})$. This is, in fact, related to the notion of `quadratic twisting', in which the $J$-map remains the same, but the starred singular fibers change according to:
\bea    \label{quadratic twist}
    I_n \longleftrightarrow I_n^*~, \qquad II \longleftrightarrow IV^*~, \qquad III \longleftrightarrow III^*~, \qquad IV \longleftrightarrow II^*~.
\eea
The same argument holds for elliptic points of order 3, as well as for the cusps of $\Gamma$. To summarise, we have the following map between the `special' points of $\Gamma$ and the singular elliptic fibers:
\bea    \label{summary Gamma - ES relation}
    \textit{elliptic point of order 2} \quad (J = 1)~:&\qquad  III~,~ III^*~, \\
    \textit{elliptic point of order 3} \quad (J = 0)~:&\qquad  II~,~ IV~,~ IV^*~, ~ II^*~, \\
    \textit{cusp of width $n$} \quad (J \rightarrow \infty)~:&\qquad  I_n~, ~ I_n^* ~.
\eea
It is worth pointing out that to a subgroup $\Gamma \subset {\rm PSL}(2,\mathbb{Z})$, we can thus associate more than one elliptic surface. As we are interested in rational elliptic surfaces, a necessary condition is that the sum of the Euler numbers associated to all fibers is 12, as indicated before in \eqref{euler number sum}. This constraint can be satisfied in more than one way. First, we have the (rational) elliptic surfaces that differ by a quadratic twist, with the singular fibers related as in \eqref{quadratic twist}. Additionally, if $\Gamma$ has elliptic points of order $3$, there is an additional choice between $II$ and $IV$ singular fibers. For instance, consider the subgroup $\Gamma^0(3)$, which has two cusps of widths 3 and 1, respectively, and one elliptic point of order 3. Then, we can form the following rational elliptic surfaces:
\be
     (I_3, I_1, IV^*)~, \qquad (I_3^*, I_1, II)~, \qquad (I_3, I_1^*, II)~, \qquad (I_3, I_1, II, I_0^*)~,
\ee
which are all related by quadratic twists. However, we can form other non-rational surfaces, such as $(I_3, I_1, II)$, for instance. Note, however, that if one restricts attention to subgroups of ${\rm SL}(2,\mathbb{Z})$ instead, then there is no such ambiguity. 

Let us finish this section with a corollary from \cite{Doran:1998hm}. If a rational elliptic surface is modular, then it does not contain singular fibres of types $IV$ or $II^*$. This can be seen from Persson's list of allowed configurations of singular fibers \cite{Persson:1990, Miranda:1990}. In particular, the only configurations with a $II^*$ singular fiber are $(II^*, II)$ and $(II^*, 2I_1)$, which cannot correspond to any subgroup $\Gamma \subset {\rm PSL}(2,\mathbb{Z})$. Thus, since $II^*$ is not allowed, then we cannot have modular rational elliptic surfaces with both a type $IV$ fibre and a `starred' fibre, due to the quadratic twisting procedure. This only leaves the configurations without `starred' fibers that involve $IV$ singular fibers. For a complete proof of this statement, we refer to \cite{Doran:1998hm}. This corollary simplifies \eqref{summary Gamma - ES relation} to:
\bea    \label{summary Gamma - RES relation}
    \textit{elliptic point of order 2} \quad (J = 1)~:&\qquad  III~,~ III^*~, \\
    \textit{elliptic point of order 3} \quad (J = 0)~:&\qquad  II~,~  IV^*~, \\
    \textit{cusp of width $n$} \quad (J \rightarrow \infty)~:&\qquad  I_n~, ~ I_n^*~.
\eea
Doran showed that there are 33 modular RES \cite{Doran:1998hm}, which can be extended to 47 distinct RES, as we will see in the next subsection. Our approach is different from that of \cite{Doran:1998hm} and closer to the construction of Shioda \cite{Shioda:1972}; we will use the classification of subgroups of ${\rm PSL}(2,\mathbb{Z})$ \cite{Sebbar:2001, Cummins2003, STROMBERG2019436} and the map \eqref{summary Gamma - RES relation} in order to find all modular rational elliptic surfaces. This will be done explicitly in section \ref{section: modular RES classification}.


\section{Fundamental Domains and BPS quivers}   \label{section: 3}

The goal of this section is to derive the identification \eqref{Identification} for more general Kodaira-type singularities. Moreover, we discuss the link between fundamental domains and BPS quivers, analysing how quiver mutations are reflected on the upper half-plane. We apply this logic to the 4d SQCD theories.


\subsection{Quiver mutations} 

Consider for now $\CT_{\boldsymbol{X}}$ a 5d theory on $S^1$, partially\footnote{We focus on theories with `trivial' CB, which are fully determined by also specifying the MW group of the maximally deformed CB.} determined by the fiber at infinity $F_{\infty} = I_{9-n}$. Recall that these theories are engineered in M-theory compactifications on non-compact Calabi-Yau threefolds $\boldsymbol{X}$, which are just the canonical line bundle over del Pezzo surfaces $dP_n$ or the Hirzebruch surface $\mathbb{F}_0$. The BPS spectrum $\{ \gamma\}$ of these theories is generally very complicated to determine, being still an open problem. However, at \emph{quiver points}, namely loci where the central charges of the $n+3$ elementary particles almost align, the BPS spectrum is conjecturally given by bound states of these particles (see \textit{e.g.} \cite{Closset:2019juk}).

In the limit of vanishing string coupling where the D-branes wrapped over holomorphic cycles are accurately described as $B$-branes, the latter are objects of the (bounded) derived category of coherent sheaves. In this limit, the category of BPS states on the Coulomb branch is expected to be accurately described by this category:
\be
    \cI_{\CT_{\boldsymbol{X}}} \cong D^b (\hat{\boldsymbol{X}})~.
\ee
Here, the extended Coulomb branch is identified with the extended K\"ahler cone of $\boldsymbol{X}$, while $\hat{\boldsymbol{X}}$ is a resolution of the Calabi-Yau singularity. 

It is also the case that this category is equivalent to the derived category of quiver representations, explaining thus the importance of BPS quivers. Note that the BPS spectrum is a subset of the set of all BPS states in this category which satisfy a set of stability conditions \cite{Douglas:2000ah, Douglas:2000qw, Aspinwall:2004jr}. It is well known that the BPS quivers for the theories associated to the toric geometries can be derived using brane-tiling techniques, see \textit{e.g.} \cite{Franco:2005rj, Hanany:2005ss, Hanany:2012hi}. Alternatively, if a basis of BPS states with electromagnetic charges $\gamma_i = (m_i, q_i)$ is found, the BPS quiver is obtained by assigning a quiver node $(i)\sim \CE_{\gamma_i}$ to each light dyon, and a (effective) number $n_{ij}$ of arrows from node $(i)$ to $(j)$ given by the Dirac pairing: 
\be
n_{ij}= \langle \gamma_i, \gamma_j \rangle =  m_i q_j - q_i m_j~.
\ee
These charges are typically obtained using the quantum periods of the $\CT_{\boldsymbol{X}}$ theory, evaluated at the quiver point, which was done explicitly in \cite{Closset:2021lhd} for some simple examples. These states correspond to the so-called fractional branes associated to the singularity $\boldsymbol{X}$ in type-II string theory, but computing the exact form of the periods is usually a difficult task, which we can bypass by using the fundamental domains.

We will limit our analysis to the modular rational elliptic surfaces only, and denote by $\cF_{\cT}$ the fundamental domain of $\CT$. Here, we relax the definition of $\CT$, allowing it to be any 4d $\cN=2$ theory, or one of the 5d theories on a circle. Fundamental domains can be obtained for non-modular configurations as well, but these involve branch cuts as discussed in \cite{Aspman:2021vhs, Aspman:2021evt}, which we do not discuss here. However, we will show how to relate different modular configurations by following quiver mutations.

Recall that the simplest type of singularity occurring in the interior of the $U$-plane is when a single charged particle becomes massless. In the appropriate duality frame, the low-energy physics at that point is then governed by SQED. Due to the $\beta$-function of SQED, the local monodromy is given by $T$ and  it thus follows that a massless dyon of charges $(m,q)$ at $U_\ast$ induces a monodromy:
\be\label{gen Mmq}
\mathbb{M}_\ast^{(m,q)} = B^{-1}T B = \mat{1+m q&\, q^2\\ -m^2 & 1-m q}~.
\ee
These are the $I_1$ singularities in Kodaira's classification, as listed in table~\ref{tab:kodaira}. In the case of SQED with $n$ electrons (or some hypermultiplets of charges $q_j$ such that $\sum_j q_j^2=n$), the monodromy is conjugate to $T^n$ \cite{Seiberg:1994rs, Seiberg:1994aj}, corresponding thus to an $I_n$ singularity. The $I_n$ fibers are sometimes called \emph{multiplicative} fibers. The remaining types of singularities in table~\ref{tab:kodaira} are all \emph{additive}, and can appear due to mutually non-local light BPS states becoming massless simultaneously. Thus, the monodromy induced by the additive fibers should be viewed as a product of monodromies of the type \eqref{gen Mmq}. Thus, we will typically deform the Coulomb branches containing additive singular fibers, such that the singularities are broken to multiplicative fibers only.

The classification programme of 4d $\CN = 2$ rank-one SCFTs \cite{Argyres:2015ffa, Argyres:2015gha, Argyres:2016xmc, Argyres:2016xua}, recently reviewed in \cite{Caorsi:2019vex, Cecotti:2021ouq}, also involves theories whose Coulomb branches include \emph{undeformable} singularities. In this context, the $I_n$ singularities are to be interpreted as due to a single massless hypermultiplet, but of charge $Q = \sqrt{n}$, in a purely electric duality frame. The flavour algebra of these theories can be understood from the so-called flavour root system of \cite{Caorsi:2018ahl}, yet there is no clear picture of how the MW group restricts the global form of the flavour symmetry. However, as the SW geometries of these theories are already known (see \cite{Argyres:2015gha}, for instance), we will be able to bypass these issues and analyse their modular properties as well. This is done explicitly in section~\ref{Rank-one SCFTs}.

A quiver description depends on the basis of BPS states and, thus, is not unique. Such a basis choice splits the spectrum into particles and anti-particles, which have central charge vectors of equal magnitude but opposite direction. We will denote by $\cH_{\cZ}$ a choice of half-plane on the central-charge $\cZ$-plane. A change of the basis states can be implemented through a quiver mutation, which effectively rotates the central charge half-plane $\cH_{\cZ}$, leading to a relabelling of the particles and anti-particles. Following the conventions of \cite{Alim:2011kw}, a quiver mutation arising from clockwise rotations of the central charge half-plane, \textit{i.e.} a left mutation on the node $\gamma_i$, leads to the basis change: 
\bea
    \gamma_i~ & \longrightarrow && ~-\gamma_i~, \\
    \gamma_j~ & \longrightarrow && ~\begin{cases}
    \gamma_j - \langle \gamma_j, \gamma_i \rangle \gamma_i~, & \text{if }  \langle \gamma_j, \gamma_i \rangle<0~, \\
    \gamma_j ~, & \text{if }  \langle \gamma_j, \gamma_i \rangle \geq 0~. \\
    \end{cases} 
\eea
while for counter-clockwise rotations, \textit{i.e.} right mutations on node $\gamma_i$, we have:
\bea
    \gamma_i~ & \longrightarrow && ~-\gamma_i~, \\
    \gamma_j~ & \longrightarrow && ~\begin{cases}
    \gamma_j + \langle \gamma_j, \gamma_i \rangle \gamma_i~, & \text{if }  \langle \gamma_j, \gamma_i \rangle>0~, \\
    \gamma_j ~, & \text{if }  \langle \gamma_j, \gamma_i \rangle \leq 0~. \\
    \end{cases} 
\eea
In the following, we will denote right mutations by $\hat{\gamma}_i$. Additionally, we will denote by $\gamma_i - \gamma_j$ the mutation sequence that corresponds to mutating first on node $\gamma_i$ and then on the node $\gamma_j$, while $\gamma_{ij}$ will correspond to simultaneous mutations on nodes $\gamma_i$ and $\gamma_j$, and similarly for right mutations. This notation will be used for mutations on blocks, being essentially equivalent to successive mutations on the nodes forming the blocks.


\subsection{Fundamental domains.}  \label{sec: BPS states - Singularities identification}

Having introduced BPS quivers, we will now show how to derive them from the fundamental domains. The latter are based on the so-called Dedekind tessellation of the upper half-plane $\mathbb{H}$, which is obtained by the M\"obius action on the modular group ${\rm PSL}(2,\mathbb{Z})$. Fundamental domains for subgroups $\Gamma$ of the modular group can be constructed using a set of coset representatives $\left(\alpha_i\right)$, for $i = 1, \ldots, n_{\Gamma}$ from the disjoint union:
\be
    \cF_{\Gamma} =  \bigsqcup_{i=1}^{n_{\Gamma}} \alpha_i \cF_0~,
\ee
where $\cF_0$ is the fundamental domain of ${\rm PSL}(2,\mathbb{Z})$. Recall that the CB can be isomorphically mapped to a subregion of the upper half-plane when the associated rational elliptic surface is modular. The cusps of the monodromy group of the theory correspond to $U$-plane singularities and can be mapped to the real axis $\mathbb{R}$ of the upper half-plane (except the cusp $F_{\infty}$ at $\infty$, which is fixed). 

Note that the cusps are in the ${\rm PSL}(2,\mathbb{Z})$ orbit of $\tau_{\infty} = i\infty$, which, by convention, is the natural position of the unique width-one cusp of the modular group. The action of an element of the modular group on the cusp $\tau_{\infty}$ leads to rational numbers of the form ${q\ov m}$, with $q, m \in \mathbb{Z}$, as follows:
\be
    \sigma \tau_{\infty} = {a \tau_{\infty} + b \ov c \tau_{\infty} + d} = {q \ov m}~, \qquad \quad \sigma = \left( \begin{matrix} a & b \\ c & d \end{matrix} \right) \in {\rm PSL}(2,\mathbb{Z})~.
\ee
where $\sigma$ should be viewed as one of the coset representatives $\alpha_i$ and ${q\ov m}$ is an irreducible fraction. For this equality to be satisfied, one requires:
\be 
    {a \ov c} = {q \ov m}~.
\ee
Assuming that this cusp corresponds to an $I_1$ singular fiber, this leads to the monodromy matrix:
\be \label{Matrix from fraction}
    \sigma T \sigma^{-1} = \mathbb{M}_{(c,-a)} = \mathbb{M}_{(k m,-k q)} \in \Gamma \subset {\rm PSL}(2,\mathbb{Z})~, 
\ee
for some non-zero $k \in \mathbb{C}$. For the theories that do not contain undeformable singularities, it is generally enough to restrict attention to $k = 1$, with the states having $k>1$ being unstable. Thus, we have the following correspondence:

\begin{tcolorbox}[colback=blue!3!white]
To an $I_n$ singularity that corresponds to a width $n$ cusp at $\tau = {q\ov m} \in \mathbb{Q}$ on the upper half-plane we assign $n$ light BPS states of charge $\pm (m, -q)$.
\end{tcolorbox}

\noindent Note that since this approach is solely based on the monodromies, there is a sign ambiguity in choosing the charges of the BPS states. A constraint that follows from the central charges is that one needs to make the same sign choice for all $n$ states forming an $I_n$ cusp. In principle, this sign ambiguity can be fixed by solving the associated Picard-Fuchs equation and evaluating the central charges at the quiver point. 

All BPS quivers studied in this work that arise from fundamental domains with no elliptic points satisfy a curious property. This observation leads to the following conjecture that remedies the sign ambiguity of the BPS charges. Let $(\tau_i)$ be the \textit{ordered} vector of the positions of the distinct $(I_{n_i})$ cusps:
\be
    (\tau_1~, \ldots , \tau_k)~, \qquad \quad \tau_1< \ldots < \tau_k~.
\ee
and let $j_{(i)}$ be the position of the cusp $\tau_i$ in this vector. As shown above, we can write each such $\tau_i$ as a rational number:
\be
    \tau_i = {q_i \ov m_i} \in \mathbb{Q}~, \qquad m_i \in \mathbb{Z}_{>0}~,
\ee
where we make a choice for the sign of the denominator, which also fixes the sign of the numerator. Then, we conjecture that for modular configurations containing only multiplicative cusps $(I_{n_i})$, the assignment of $n_i$ light BPS states of charges:
\be
      (-1)^{j_{(i)}} (m_i,-q_i)~.
\ee
leads to the correct BPS quiver description. Note that an overall sign change of all BPS states will still lead to a consistent quiver, as this simply replaces all particles by their anti-particles.

These statements can be further generalized to include undeformable multiplicative singularities as follows. Recall that an undeformable $I_n$ singularity corresponds to a single massless hypermultiplet of charge $Q = \sqrt{n}$ in an electric frame. Thus, the proportionality factor $k \in \mathbb{C}$ precisely amounts for this charge renormalization, leading to:

\begin{tcolorbox}[colback=blue!3!white]
To an $I_n$ singularity that corresponds to a width $n$ cusp at $\tau = {q\ov m} \in \mathbb{Q}$ on the upper half-plane, with deformation pattern $I_n \rightarrow \bigoplus_j I_{k_j}$, where each $I_{k_j}$ is undeformable, we assign light BPS states of charges $\Big(\pm \sqrt{k_j}(m, -q)\Big)_{j}$.
\end{tcolorbox}

\noindent As stated before, the main focus will be on the theories for which the maximally deformed Coulomb branches involve only $I_1$ type singularities, but we will briefly discuss undeformable singularities in section~\ref{Rank-one SCFTs}. 

One of the assumptions behind the above identification is that there exists a BPS chamber that contains the states associated to the $I_n$ cusps. The $U$-plane has generally walls of marginal stability connecting the singularities, which separate these BPS chambers. 
While for massless SQCD this assumption is known to be true as there are only two BPS chambers \cite{Bilal:1996sk, Ferrari:1996sv, Bilal:1997st}, the structure of the $U$-plane for the KK theories is much more intricate and, in general, there might not be such a chamber. However, in order to make sure that the states are part of the spectrum, we will relate the different modular configurations to known BPS quivers of the theories of interest. 

As we have already mentioned, the category of BPS states is equivalent to the derived category of quiver representations. To determine this category, one also needs the quiver superpotential. We do not have a prescription for determining superpotentials from the fundamental domains, but these can be found in some cases using exceptional collections \cite{Herzog:2003zc} or, alternatively, brane-tiling methods - see \textit{e.g.} \cite{Hanany:2012hi, Closset:2019juk}. 

Let us finally comment on the matrices $\sigma \in {\rm PSL}(2,\mathbb{Z})$ appearing in \eqref{Matrix from fraction} that satisfy $\sigma T \sigma^{-1} = \mathbb{M}_{(m,-q)}$. These are some of the coset representatives of $\Gamma \subset {\rm PSL}(2,\mathbb{Z})$, and do play an important role in the identification \eqref{Identification}. Finding such matrices is a non-trivial task, but there exists an algorithmic way of solving this problem using continued fractions \cite{Conrad}. One has:\footnote{For this, we use \texttt{Mathematica}'s inbuilt function \texttt{ContinuedFraction[]}.}
\be
    {q\ov m} = p_1 + { 1 \ov p_2 + {1\ov p_3 + {1 \ov p_4 + \ldots }}}~,
\ee
for some $p_i \in \mathbb{Z}$. Note that since ${q\ov m}$ is rational, the sequence $\{p_1, p_2, \ldots\}$ must terminate. This list of integers will determine a possible matrix $\sigma$ as follows:
\be \label{Matrix from Continued Fraction}
    \sigma = \prod_{i = 1}^n T^{(-1)^{i+1}p_i}S = T^{p_1}S T^{-p_2}S \ldots T^{(-1)^{n+1}p_n}S ~.
\ee
Note that this matrix is not unique, with the same monodromy being reproduced by $\sigma T^n$, for some integer $n\in \mathbb{Z}$. Let us also mention that the continued fraction representation of the cusp-positions can be used to find accumulation rays of the BPS quivers of the KK theories, which, as opposed to 4d SQCD theories do not necessarily lie along the real axis of the central charge plane. We hope to come back to this issue in the future. For now, we would like to understand how changes of the fundamental domains affect the BPS quivers, in light of the above identifications.

\medskip

\noindent
\paragraph{Constant shifts of $\cF$.} As previously discussed, fundamental domains are subsets of the upper half-plane $\mathbb{H}$ whose closure contain the images of all points of $\mathbb{H}$ under the action of some $\Gamma \subset {\rm PSL}(2,\mathbb{Z})$. As a result, there is an infinite number of possibilities of drawing a fundamental domain for any such subgroup $\Gamma$. One of our goals is to understand how different choices of fundamental domains are reflected in the physical theory, under the identifications introduced above.

We first point out that the choice of a fundamental domain is arbitrary and, usually, there is no notion of a `correct' choice. Based on the identification \eqref{Identification}, different fundamental domains can be also seen as a change of conventions or normalization of the BPS states; indeed, the charge of the monopole, for instance, is chosen to be $(1,0)$ by convention. What we aim to do, however, is to fix a fundamental domain, \textit{i.e.} fix a basis of BPS states, and see how changes of this basis are reflected on the fundamental domains. Naturally, we will find that in certain cases such basis changes can be interpreted as quiver mutations.

The simplest way of changing a fundamental domain is by a constant shift $\tau \rightarrow \tau +w_{\infty}$, for $w_{\infty} \in \mathbb{Z}$ the width of the cusp at infinity. In this case, the monodromies around all singularities in the bulk get conjugated by $\mathbb{M}_{\infty}$ - the monodromy around the cusp at infinity - such that we might assign new electromagnetic charges:
\be
    \mathbb{M}_{(m,q)} \longrightarrow \mathbb{M}_{(m',q')} = \mathbb{M}_{\infty} \mathbb{M}_{(m,q)} \mathbb{M}_{\infty}^{-1}~, 
\ee
to all light BPS states. Then, for $\mathbb{M}_{\infty} \in {\rm SL}(2,\mathbb{Z})$ of the form:
\be
    \mathbb{M}_{\infty} = \left(  \begin{matrix}  a & b \\ c & d \end{matrix} \right)~,
\ee
we have:
\bea
    m_j' = d m_j - c q_j~, \qquad q_j' = -b m_j + a q_j~,
\eea
and thus the Dirac pairing is preserved:
\be
    n_{ij}' = det(\mathbb{M}_{\infty})~ n_{ij} = n_{ij}~.
\ee
This thus shows that the BPS quiver does not change under such shifts of the fundamental domain. However, since the electromagnetic charges do change, one can think of this as a sequence of mutations that preserves the number of arrows but change the basis of BPS states.

\medskip

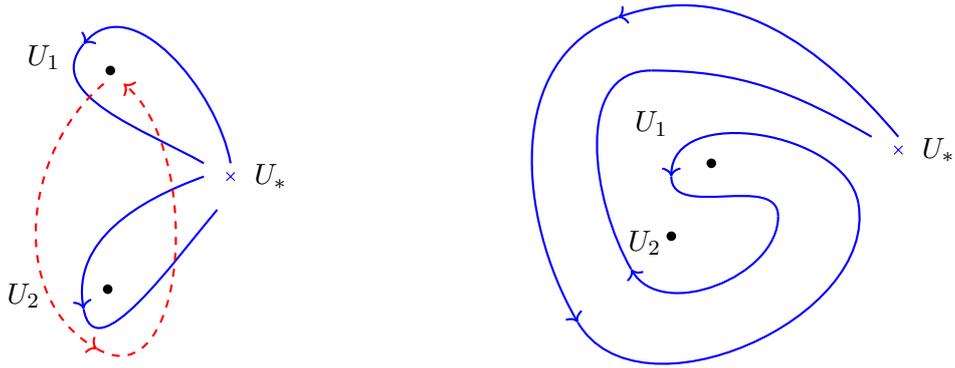
\begin{figure}[t]
\centering
\tikzset{cross/.style={cross out, draw=black, fill=none, minimum size=2*(#1-\pgflinewidth), inner sep=0pt, outer sep=0pt}, cross/.default={2pt}}
\begin{tikzpicture}[x=50pt,y=50pt]
\begin{scope}[shift={(0,0)}]
    \draw (0.1,0) node[cross, blue] {};
    \fill (-0.82,-0.85)  circle[radius=1.8pt] {};
    \fill (-0.8,0.8)  circle[radius=1.8pt] {};
    \draw[blue, thick, ->] ($(-0.1, 0)$) to[out=200, in=100] ($(-1, -1)$);
    \draw[blue, thick] ($(-1, -1)$) to[out=280, in=230] ($(0, -0.25)$);

    \draw[blue, thick, ->] ($(0.1, 0.1)$) to[out=100, in=50] ($(-1, 1)$);
    \draw[blue, thick] ($(-1, 1)$) to[out=50+180, in=150] ($(-0.1, 0.1)$);
    
    \draw[red, thick, dashed, ->] ($(-0.85, 0.7)$) to[out=220, in=150] ($(-0.9, -1.3)$);
    \draw[red, thick, dashed, ->] ($(-0.9,-1.3)$) to[out=150+180, in=320] ($(-0.7, 0.7)$);
    
    \node at (0.4,0) {$U_*$};
    \node at (-1.45,-0.9) {$U_2$};
    \node at (-1.3,0.9) {$U_1$};
\end{scope}

\begin{scope}[shift={(4,0.2)}]
    \draw (1+0.1,0) node[cross, blue] {};
    \fill (-0.6,-0.65)  circle[radius=1.8pt] {};
    \fill (-0.3,-0.1)  circle[radius=1.8pt] {};

    \draw[blue, thick, ->] ($(1+0.1, 0.1)$) to[out=130, in=20] ($(-1, 1)$);
    \draw[blue, thick, ->] ($(-1, 1)$) to[out=20+180, in=120] ($(-1.3, -1.3)$);
    \draw[blue, thick] ($(-1.3, -1.3)$) to[out=120+180, in=280] ($(0.8, -0.4)$);
    \draw[blue, thick, ->] ($(0.8, -0.4)$) to[out=280+180, in=90] ($(-0.6, -0.2)$);
    \draw[blue, thick] ($(-0.6, -0.2)$) to[out=90+180, in=90] ($(0.2, -0.5)$);
    \draw[blue, thick, ->] ($(0.2, -0.5)$) to[out=90+180, in=270 + 30] ($(-0.9, -0.9)$);
    \draw[blue, thick] ($(-0.9, -0.9)$) to[out=90+30, in=180] ($(-0.75, 0.6)$);
    \draw[blue, thick] ($(-0.75, 0.6)$) to[out=0, in=150] ($(-0.1+1, 0.1)$);
    
    \node at (1+0.4,0) {$U_*$};
    \node at (-0.8,-0.7) {$U_2$};
    \node at (-0.75,0.2) {$U_1$};
\end{scope}

\end{tikzpicture}
    \caption{Composition of monodromies for non-trivial paths of the singularities on the $U$-plane.}
    \label{fig:paths U plane}
\end{figure}

\noindent
\paragraph{Non-trivial paths around singularities.} Let us consider the more general scenario depicted in figure~\ref{fig:paths U plane}, with two $U$-plane singularities denoted by $U_1$ and $U_2$. Then, we imagine a situation in which by turning on some deformation pattern, the singularity $U_1$ makes a closed path around $U_2$, marked in the figure by the dashed red line. In this case, the monodromies with base point $U_*$ get mixed up, with the monodromy around $U_1$ becoming:
\be \label{monodromy change}
    \mathbb{M}_{U_1} \longrightarrow \mathbb{M}_{U_1}' = \mathbb{M}_{U_2} \mathbb{M}_{U_1} \mathbb{M}_{U_2}^{-1}~.
\ee %
%
This can be read from figure~\ref{fig:paths U plane} by following the continuous blue path on the right diagram. Note that in this qualitative picture we ignored the branch cut structure of the periods of the Seiberg-Witten differential. In fact, for a path crossing a branch cut, one automatically changes from the principal Riemann sheet of the periods to another sheet. Following the analysis of \cite{Bilal:1996sk, Ferrari:1996sv} for the spectrum of the 4d SQCD theories, the above monodromy change can be interpreted as moving on a different Riemann sheet, where a new light BPS state becomes massless. While this picture might appear rather simplistic, it is a direct application of Picard-Lefshetz theory, which guarantees a consistent physical picture \cite{Klemm:1995wp}.

We will see momentarily that such monodromy changes appear rather naturally at the level of the fundamental domains. The question we want to answer is how these changes affect the BPS quiver. In this case, from \eqref{monodromy change}, the new monodromy can be thought of as being associated to the electromagnetic charges:
\be \label{BPS charge changes}
    \gamma'_1 = \gamma_1 - \langle\gamma_1 , \gamma_2\rangle \gamma_2~,
\ee
where $\gamma_j = (m_j, q_j)$ and $\langle\gamma_i, \gamma_j\rangle$ is the usual Dirac pairing. It is clear that the intersection matrix $n_{ij}$ doesn't change for $i,j \neq 1$, in the above notation. Hence, we only need to see how $n_{1j}$ change. One finds:
\bea
    n_{1j}' = \begin{cases} n_{12}~, & j = 2~, \\
    n_{1j} - n_{12}n_{2j}~, & j\neq 2~.
    \end{cases}
\eea
which will generally lead to a new BPS quiver. In fact, this could be interpreted as a quiver mutation on the node $\gamma_2$, if $n_{j2} = \langle\gamma_j, \gamma_2\rangle  \geq 0$, for $j\neq 1$ and $n_{j2} = \langle \gamma_j, \gamma_2 \rangle <0$ for $j = 1$, as the other BPS states should remain unchanged. Note, that such a quiver mutation on $\gamma_2$ is also accompanied by a change $\gamma_2 \rightarrow -\gamma_2$, which would need to be imposed manually in this case. Thus, we have shown that in certain scenarios, simple paths around singularities on the Coulomb branch can be interpreted as quiver mutations. This will serve as our main tool in trying to relate different modular configurations of the same theory.


\medskip

\noindent \paragraph{Elliptic points.} Let us also comment on the other special points of the fundamental domain which correspond to $U$-plane singularities, namely the elliptic points. These lie on the boundary of the fundamental domain and appear when mutually non-local BPS states become massless simultaneously. For this reason, there will be more than one way of `reconstructing' such singularities from the underlying BPS states.

In general, we will try to use configurations of singular fibers that only involve the multiplicative $I_n$ fibers, and make an ansatz for configurations that involve additive fibers in terms of merging multiplicative fibers. The most common positions of the elliptic points studied throughout this work are:
\bea    \label{III, II Notation}
    II^{(n)}~: & \qquad \tau = n + e^{2\pi i \ov 3}~, & \qquad \mathbb{M}_{II^{(n)}} = T^n(ST)^{-1} T^{-n}~,  \\
    III^{(n)}~: & \qquad \tau = n + i~, &  \mathbb{M}_{III^{(n)}} = T^nS^{-1} T^{-n}~, \\
\eea
for some integer $n \in \mathbb{Z}$. In such cases, the possible `decompositions' of the type $II$ singularity are into the BPS states given by:
\be \label{II Decomposition}
    II^{(n)}~:~ (1, 1-n) \oplus (1,-n)~, \qquad (1,-n) \oplus (0,1) ~, \qquad (0,1) \oplus (1,1-n)~,
\ee
where these are obtained only from monodromy considerations. Note that due to this construction, the order in which these states appear is important for determining the correct monodromy around the corresponding elliptic point. For instance, for the first decomposition in \eqref{II Decomposition}, the correct monodromy is $\mathbb{M}_{II} = \mathbb{M}_{(1,1-n)} \mathbb{M}_{(1,-n)}$.

While this decomposition might appear somewhat artificial, it is in fact in line with the identification \eqref{Identification}. In particular, the $II$ singularity for instance corresponds to the simplest Argyres-Douglas (AD) SCFT \cite{Argyres:1995jj, Argyres:1995xn}, with the above decompositions corresponding to the deformation $II \rightarrow 2I_1$. While for this case the decomposition does not lead to a modular configuration of the $AD$ theory, it turns out that all possibilities \eqref{II Decomposition} do in fact reproduce the $A_1$ quiver. Additionally, while the `charges' of the states forming these singularities might appear arbitrary, this is due to the fact that we can embed multiple AD theories on the Coulomb branch of some other larger theory, allowing thus for different charge normalizations. 

Similarly, for the $H_1$ AD theory, for the split $III \rightarrow I_2 \oplus I_1$ we have the following possibilities:
\bea    \label{III Decomposition}
    III^{(n)}~:~ & \qquad 2(1,-n) \oplus (1,-1-n)~, \qquad 2(0,1) \oplus (1,1-n)~, \\
    & \qquad (1,1-n) \oplus 2(1,-n)~, \qquad (1,-1-n) \oplus 2(0,1)~.
\eea
as well as:
\bea \label{III Decomposition v2}
    III^{(n)}~:~&  \qquad (1,-n) \oplus (0,1) \oplus (1,-n)~,\\
        & \qquad (0,1) \oplus (1,-n) \oplus (0,1)~.
\eea
We postpone the details of this identification to section \ref{Rank-one SCFTs}, where we will discuss 4d rank-one SCFTs. Astute readers may have noticed that the cusps of the BPS states $(0,1)$ used in the above description would correspond to cusps at infinity in view of \eqref{Identification}. In fact, we immediately notice that:
\be
    \mathbb{M}_{(0,1)} = T = T^n T T^{-n} ~.
\ee
Thus, such cusps do not lie on the real axis, but are instead more similar to the `fixed' singularity at infinity, corresponding to $T^n \CF_{\Gamma(1)}$ copies of the fundamental domain. In the context of 4d SQCD, such states correspond to quarks, which can appear as nodes of the BPS quivers \cite{Alim:2011kw, Alim:2011ae}. These should not be confused with the W-bosons, which are in fact accumulation rays of the 4d quivers and cannot appear as nodes of the BPS quiver. 

Let us finally mention that from the point of view of the monodromy change \eqref{monodromy change}, the states $(0,1)$ are not at all unconventional. In this spirit, one has:
\be
    \mathbb{M}_{(1,-k+1)} \mathbb{M}_{(1,-k)} \left(\mathbb{M}_{(1,-k+1)}\right)^{-1} = \mathbb{M}_{(0,1)}~, \qquad k\in \mathbb{Z}~,
\ee
for example, which shows how $(1,-k)$ states can turn into $(0,1)$ sates. Similar expressions can be worked out for generic $(m, -q)$ charges.


\subsection{Normalization of $U(\tau)$}

A fundamental domain for a given modular configuration on the extended Coulomb branch can be drawn by first identifying the monodromy group. In practice, this is done by starting with the Weierstrass form of the SW geometry and solving the equation:
\be \label{J(U) = J(t)}
    J(U) = J(\tau) = 1728 j(\tau) ~,
\ee
for $U = U(\tau)$, for fixed mass parameters. This leads to a polynomial equation $P(U) = 0$, where:
\be \label{Polynomial J(U)=J(t)}
    P(U) = \left( g_2(U, \boldsymbol{M})^3 - 27 g_3(U, \boldsymbol{M})^2\right) j(\tau) - 1728 g_2(U, \boldsymbol{M})^3~.
\ee
For generic mass parameters, this is a degree 6 equation in $U$ for 4d SQCD theories \cite{Aspman:2020lmf, Aspman:2021evt, Aspman:2021vhs}, or a degree 12 equation in $U$ for the KK theories. If the configuration is modular, any root of this polynomial will be a modular function of the monodromy group $\Gamma \subset \rm{PSL}(2,\mathbb{Z})$, with the other roots corresponding to transformations under coset representatives of $\Gamma$. Thus, choosing one of these roots $U_0(\tau)$, a fundamental domain can be drawn by finding a list of coset representatives $(\alpha_i)$ such that $U_0(\alpha_i \tau)$ reproduces the $U$-plane singularities. This can be achieved by using the modular transformations of this function.

A rather subtle point discussed in \cite{Aspman:2021evt} in the context of 4d SQCD theories concerns the `correct' choice of the  solution for $U_0(\tau)$.  This choice, in particular, needs to be made in such a way that it is consistent with the choices for the other theories in the same class. 

For the KK theories, this consistency requirement can be satisfied, in principle, by starting with $\KK E_8$ and subsequently decoupling the flavours, as follows. Recall first that the $\KK E_n$ theories have the fixed fiber at infinity $F_{\infty} = I_{9-n}$, which is mapped to a cusp of width $w_{\infty} = 9-n$ on the upper half-plane. On the maximally deformed Coulomb branch of $\KK E_8$, the $12$ solutions of \eqref{J(U) = J(t)} would then be in one-to-one correspondence with the $12$ simple $I_1$ cusps. However, since only one of these describes the $I_1$ cusp at infinity, there is a `unique' solution with asymptotics $U \approx \infty$ as $\tau \approx i\infty$. This solution and its descendants obtained from decoupling limits would thus be a consistent choice for the whole $\KK E_n$ family. Computationally, this is a non-trivial task, as one requires closed-form expressions of the solutions for this approach to be feasible, which we leave for future work.

In this paper, for each of the $\KK E_n$ theories we will use the choice of $U(\tau)$ with the following asymptotics:
\be     \label{U choice}
    \t U(\tau) = q^{-{1 \ov 9-n}} + \ldots ~,
\ee
in the limit $\tau \rightarrow i\infty$, where $\t U$ is the parameter appearing in the Seiberg-Witten curves of Eguchi and Sakai \cite{Eguchi:2002fc, Eguchi:2002nx}, \textit{i.e.} the curves expressed in terms of the $E_n$ characters. As pointed out in \cite{Closset:2021lhd}, this differs from the `physical' $U$-parameter by a normalization factor:
\be
    \t U = {1 \ov \alpha} U~,
\ee
with $\alpha$ being a function of the gauge theory parameters which becomes unity in the massless limit. We make the choice \eqref{U choice} for all $\KK E_n$ theories other than $\KK E_2$ and $\KK \t E_1$, for which the expression involves an additional normalization factor. We stress that despite not proving that this choice is `correct', for our purposes it suffices to have a consistent choice for each individual $\KK E_n$ theory, ensuring thus that the quivers obtained from the fundamental domains are related by quiver mutations. For the non-toric theories (\textit{i.e.} $n\geq 4$), we will also use $U$ and $\t U$ interchangeably, but, one should be aware that the expressions do in fact involve the $\t U$ parameter in general. 

Meanwhile, for the toric geometries (\textit{i.e.} $n\leq 3$), we will predominantly use the SW curves expressed in terms of gauge theory parameters of \cite{Closset:2021lhd}, for which we have:
\bea    \label{alpha E1, E3}
    \alpha_{E_1} = \sqrt{\lambda}~, \qquad \quad \alpha_{E_3} = \left( M_1 M_2 \lambda^2\right)^{1\ov 6}~,
\eea
with $\lambda$ and $M_i$ the (exponentiated) inverse gauge coupling and flavour mass parameters.


\subsection{BPS states from 2-block quivers} \label{subsection: BPS spectra}

Let us consider some simple examples of 4d theories and their associated BPS quivers. We limit ourselves to the massless SQCD theories with $N_f \leq 3$, which, apart from $N_f = 1$, turn out to be modular. Fundamental domains for these theories can be obtained as discussed in \cite{Aspman:2020lmf, Closset:2021lhd, Aspman:2021vhs}. The aim of this section is to show through examples how changes of the fundamental domains are reflected at the level of the quivers, when one makes the identification \eqref{Identification}. This ultimately leads to a different interpretation of the quiver mutation method \cite{Alim:2011kw, Alim:2011ae}. Recall also that 4d SQCD theories have the fixed fiber at infinity $F_{\infty} = I_{4-N_f}^*$, and thus the index of the monodromy groups $\Gamma \in \rm{PSL}(2,\mathbb{Z})$ cannot be larger than 6. This is due to the Euler number constraint in \eqref{euler number sum}.

Additionally, note that since the massless SQCD theories have only two BPS chambers \cite{Bilal:1996sk, Ferrari:1996sv, Bilal:1997st}, the correspondence between the BPS states and cusps on the upper half-plane introduced in section \ref{sec: BPS states - Singularities identification} does hold. The analysis presented in this section does hold in fact for any 2-block quiver.


\subsubsection{Pure $SU(2)$ theory.} 

The pure $SU(2)$ theory in four-dimensions has no free parameters, having the `unique' configuration $(I_4^*, 2I_1)$, where the fixed fiber at infinity is $F_{\infty} = I_4^*$. The BPS quiver is given by the usual Kronecker quiver:
\bea \label{4dpureSU2 quiver}
 \begin{tikzpicture}[baseline=1mm]
\node[] (1) []{$\CE_{\gamma_1=(1,0)}$};
\node[] (2) [right = of 1]{$\CE_{\gamma_2=(-1,2)}$};
\draw[->>-=0.5] (1) to   (2);
\end{tikzpicture}
\eea
There are two BPS chambers, corresponding to strong $(S)$ and weak $(W)$ coupling which are given by:
\be
     (S)~: \quad arg~\cZ(\gamma_2) >  arg~\cZ(\gamma_1)~, \qquad  (W)~: \quad arg~\cZ(\gamma_1) >  arg~\cZ(\gamma_2)~,
\ee
This quiver has been discussed in \textit{e.g.} \cite{Alim:2011kw}. We will briefly summarise the mutation algorithm for this quiver and analyze the link to the fundamental domain of the underlying theory.
\medskip

\noindent
\paragraph{Strong-coupling chamber.} In the strong-coupling chamber, rotating the half-plane $\cH_{\cZ}$ boils down to the mutations: $\gamma_2$ or $\hat{\gamma}_1$. From the perspective of the central charges, a mutation on $\gamma_2$ can only be followed by either a left-mutation on the new $\gamma_1'$ node or by a right mutation on the new $\hat{\gamma}_2'$ node. By a slight abuse of notation, we will drop the prime symbol on the new nodes. The sequence of mutations $\gamma_2 - \gamma_1 - \gamma_2$ leads to the successive quivers:
\bea \label{4dpureSU2 S quivers}
 \begin{tikzpicture}[baseline=1mm]
\node[] (1) []{$\CE_{\gamma_1}$};
\node[] (2) [right = of 1]{$\CE_{-\gamma_2}$};
\draw[->>-=0.5] (2) to   (1);
\end{tikzpicture} \qquad \qquad
 \begin{tikzpicture}[baseline=1mm]
\node[] (1) []{$\CE_{-\gamma_1}$};
\node[] (2) [right = of 1]{$\CE_{-\gamma_2}$};
\draw[->>-=0.5] (1) to   (2);
\end{tikzpicture} \qquad \qquad
 \begin{tikzpicture}[baseline=1mm]
\node[] (1) []{$\CE_{-\gamma_1}$};
\node[] (2) [right = of 1]{$\CE_{\gamma_2}$};
\draw[->>-=0.5] (2) to   (1);
\end{tikzpicture}
\eea
with an additional mutation on the $\gamma_1$ node leading back to \eqref{4dpureSU2 quiver}.

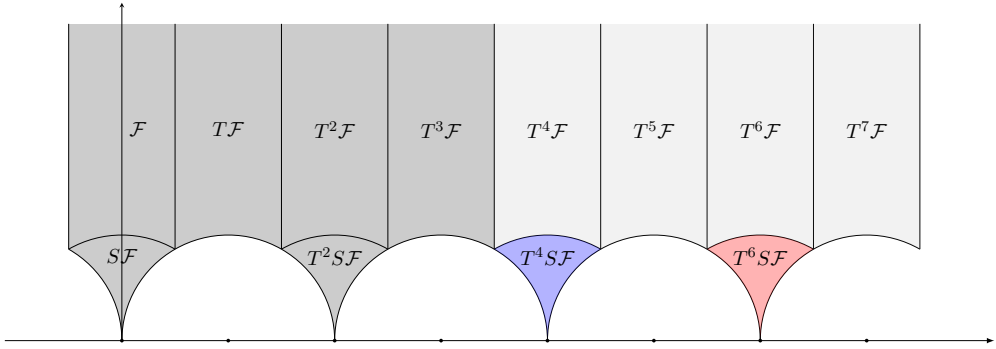
\begin{figure}[t]
    \centering
    \scalebox{0.7}{


\begin{tikzpicture}[scale=2]

    \pgfdeclarelayer{background}
    \pgfsetlayers{background,main}

    \pgfmathsetmacro{\myxlow}{-1}
    \pgfmathsetmacro{\myxhigh}{8}
    \pgfmathsetmacro{\myiterations}{1}
    
    \draw[-latex](\myxlow-0.1,0) -- (\myxhigh+0.2,0);
    \pgfmathsetmacro{\succofmyxlow}{\myxlow+0.5}
    
    \draw[-stealth, very thin] (0,0)--(0,3.2) node[above] {}; 
    \begin{scope}
    
    \foreach \i in {-0.5,0.5,1.5,2.5,3.5,4.5,5.5,6.5}
            {\draw[very thin, black] (\i,0.866) arc(120:60:1);
            \draw[very thin, black] (\i,0.866) -- (\i,3);
            }

            \draw[very thin, black] (7.5,0.866) -- (7.5,3);
            
            \draw[very thin, black] (-0.5,0.866) arc(60:0:1);
            \draw[very thin, black] (0,0) arc(180:120:1);
            
            \draw[very thin, black] (+2,0) arc(180:120:1);
            \draw[very thin, black] (+1.5,0.866) arc(60:0:1);
    
    \end{scope}
    
    \draw[very thin, black] (4-0.5,0.866) arc(60:0:1);
    \draw[very thin, black] (4,0) arc(180:120:1);

        \draw[very thin, black] (4+2,0) arc(180:120:1);
        \draw[very thin, black] (4+1.5,0.866) arc(60:0:1);

    \begin{scope}
        \begin{pgfonlayer}{background}
            \clip
                (-0.5,3) 
                { -- (-0.5,0.866) arc(60:0:1)
             -- (0,0) arc(180:0:1)
             -- (2,0) arc(180:60:1)
             }
                -- (3.5, 3) -- cycle
            ;
            \fill[gray,opacity=0.4] (-1,-1) rectangle (4.5,3);
        \end{pgfonlayer}
    \end{scope}

    \begin{scope}
            \clip
                (3.5,3) 
                { -- (3.5,0.866) arc(120:60:1)
             -- (4.5,0.866) arc(120:60:1)
             -- (5.5,0.866) arc(120:60:1)
             -- (6.5,0.866) arc(120:60:1)
             }
                -- (7.5, 3) -- cycle;
            \fill[gray,opacity=0.1] (4-1,-1) rectangle (4+4.5,3);
    \end{scope}

    \begin{scope}
            \clip
                (4-0.5,0.866) 
                { -- (4-0.5,0.866) arc(60:0:1)
             -- (4+0,0) arc(180:120:1)
             -- (4+0.5,0.866) arc(60:120:1)
             }
                -- (4-0.5, 0.866) -- cycle
            ;
            \fill[blue,opacity=0.3] (4-1,0) rectangle (4+1,1);
    \end{scope}

    \begin{scope}
            \clip
                (4+1.5,0.866) 
                { -- (4+1.5,0.866) arc(60:0:1)
             -- (4+2,0) arc(180:120:1)
             -- (4+2.5,0.866) arc(60:120:1)
             }
                -- (4+1.5, 0.866) -- cycle
            ;
            \fill[red,opacity=0.3] (4+0,0) rectangle (4+4,1);
    \end{scope}

    \begin{scope}
        \node at (0.15,2) {$\mathcal{F}$};
        \node at (1,2) {$T\mathcal{F}$};
        
        \foreach \i in {2,3,4,5,6,7}
            {\node at (\i,2) {$T^{\i}\mathcal{F}$};
            }

        \node at (0,0.8) {$S\mathcal{F}$};
        \node at (2,0.8) {$T^2S\mathcal{F}$};
        \node at (4,0.8) {$T^4S\mathcal{F}$};
        \node at (6,0.8) {$T^6S\mathcal{F}$};
        
         \foreach \i in {0,1,2,3,4,5,6,7}
            {\fill (\i,0)  circle[radius=0.5pt];
            }
            
    \end{scope}
    
\end{tikzpicture}

}
    \caption{Fundamental domain for $\Gamma^0(4)$ in dark gray. The regions in blue $(T^4S\cF)$ and red $(T^6S\cF)$ are identified with $S\cF$ and $T^2S\cF$, respectively, while the regions in light gray are identified with the corresponding $T^{k-4}\cF$ regions.}
    \label{fig: 4dPureSU2}
\end{figure}

The fundamental domain for this configuration is that of the $\Gamma^0(4)$ congruence subgroup, with the two width-one cusps at $\tau_0 = 0$ and $\tau_2 = 2$, respectively, as shown in dark grey in figure~\ref{fig: 4dPureSU2}. For more details about this subgroup, we refer to table~\ref{tab: Modular Functions Summary}. Recall from \eqref{summary Gamma - RES relation} that the $I_4^*$ singular fiber is mapped to a width-four cusp. Note again that there is some ambiguity in choosing the overall signs of the light BPS states, as the states $\pm(m,q)$ produce the same monodromy. We notice that the above mutations do not change the choice of the fundamental domain. Thus, under the identification \eqref{Identification}, we would like to emphasise that there is a unique way of choosing the cusps of the fundamental domain that reproduces the strong-coupling BPS spectrum. The latter only includes the monopole and dyon, which become massless at the two $I_1$ singularities on the $u$-plane.
\medskip

\noindent
\paragraph{Weak-coupling chamber.} In the weak coupling chamber we can instead mutate on the node $\gamma_1$ (or, alternatively, on $\hat{\gamma}_2$). This chamber has one accumulation ray associated to the W-boson. As a result, in order to explore the whole BPS spectrum, one needs to consider both clockwise and counter-clockwise rotations of the  $\cH_{\cZ}$ half-plane. 

At the level of the fundamental domain, there are two operations that we can perform which preserve the width $w_{\infty}$ of the cusp at infinity. The first one is a shift of the fundamental domain by:
\be
    \tau \longrightarrow \tau + w_{\infty}~,
\ee %
where $w_{\infty} = 4$. This is also depicted in figure~\ref{fig: 4dPureSU2}. In this case, the monodromies are conjugated by the monodromy around the cusp at infinity, $\mathbb{M}_{\infty} = PT^4$, as discussed before, and transform as:
\bea
    \mathbb{M}_{(1,0)} & \longrightarrow (PT^4) \mathbb{M}_{(1,0)} (PT^4)^{-1} = \mathbb{M}_{(1,-4)}~, \\ \mathbb{M}_{(1,-2)} & \longrightarrow (PT^4) \mathbb{M}_{(1,-2)} (PT^4)^{-1} = \mathbb{M}_{(1,-6)}~,
\eea
which is in agreement with the new fundamental domain. Note that, as expected, the intersection form of the BPS quiver remains the same. In fact, since the charges actually change, we can consider the sequence of quiver mutations: $\hat{\gamma}_2 - \hat{\gamma}_1$, leading to:
\bea
 \begin{tikzpicture}[baseline=1mm]
\node[] (1) []{$\CE_{\gamma_1=(1,-4)}$};
\node[] (2) [right = of 1]{$\CE_{\gamma_2=(-1,6)}$};
\draw[->>-=0.5] (1) to   (2);
\end{tikzpicture}
\eea
The second possibility of having a connected fundamental domain is performing the change $I_1(\tau = 0) \rightarrow I_1(\tau = 4)$, without changing the other $I_1$ cusp at $\tau = 2$, and adjusting the other copies of $\cF_{{\rm PSL}(2,\mathbb{Z})}$ such that the fundamental domain is connected. This can be thought of as taking a closed path around the other $I_1$ cusp, or, equivalently, around the cusp at infinity (since there are only three singular fibers), with:
\be
    \mathbb{M}_{(1,0)} \longrightarrow  \mathbb{M}_{(1,-2)}^{-1} \mathbb{M}_{(1,0)} \mathbb{M}_{(1,-2)} = (PT^4) \mathbb{M}_{(1,0)} (PT^4)^{-1}  =  \mathbb{M}_{(1,-4)}~.
\ee
This is, of course, the monodromy change observed in \eqref{monodromy change}, which can be interpreted as a mutation on node $\hat{\gamma}_2$, leading to the new quiver description:
\bea
 \begin{tikzpicture}[baseline=1mm]
\node[] (1) []{$\CE_{\gamma_1=(-1,4)}$};
\node[] (2) [right = of 1]{$\CE_{\gamma_2=(1,-2)}$};
\draw[->>-=0.5] (2) to   (1);
\end{tikzpicture}
\eea
Finally, the left mutations can be treated in a similar fashion, with the only difference being that the value of $\tau$ for the two cusps decreases, indicating a `motion' in the negative ${\rm Re}(\tau)$-axis direction. Of course, this mutation method has been used in \cite{Alim:2011kw} to derive the full BPS spectrum of the theory. Here we only give an interpretation of this algorithm in terms of fundamental domains and paths on the $U$-plane.

Finally, repeating the above mutations a large number of times one quickly realizes that this quiver has an accumulation ray, which is given by the states at $\tau \rightarrow \pm \infty$. From the correspondence introduced before, these are, of course, the W-bosons, with charges $\pm (0,2)$ in our conventions.

Let us also note that the above results can be generalised to other 4d theories for which the $U$-plane contains only three singularities. It is a general feature of such theories that a closed path around one of the bulk singularities is equivalent to a path around the cusp at infinity, and, thus, the allowed changes of the fundamental domain are the trivial ones.


\subsubsection{$4d~SU(2)~N_f = 2$ Theory}

For completeness, we will also briefly discuss the $4d$ $SU(2)$ theories with $N_f \leq 3$ flavours. In \cite{Aspman:2021vhs}, the map from the $U$-plane to the upper half-plane $\mathbb{H}$ was derived for generic values of the mass parameters for these theories. The corresponding fundamental domains have branch points and cuts generically, but, for certain values of the mass parameters $U(\tau)$ becomes a modular function for a congruence subgroup of ${\rm PSL}(2,\mathbb{Z})$.

\medskip

\noindent
\paragraph{The massless configuration.} The massless configuration $(I_2^*, 2I_2)$ is extremal and has monodromy group $\Gamma(2)$ \cite{Nahm:1996di}. The associated BPS quiver has been discussed in \textit{e.g.} \cite{Alim:2011kw}. At the bulk singularities, there are two $(1,0)$ and two $(-1,1)$ light BPS states, respectively, transforming as doublets of the $SU(2)$ factors of the $SU(2)\times SU(2)$ flavour symmetry (algebra), in agreement with the fundamental domain shown in figure~\ref{fig: 4dSU2Nf=2}. The BPS quiver that describes this scenario is given below:
\bea
 \begin{tikzpicture}[baseline=1mm]
\node[] (1) []{$\CE_{\gamma_{1,2}=(1,0)}$};
\node[] (2) [right = of 1]{$\CE_{\gamma_{3,4}=(-1,1)}$};
\draw[->-=0.5] (1) to   (2);
\end{tikzpicture}
\eea
There are again two BPS chambers, given by the ordering of $\cZ(\gamma_1) = \cZ(\gamma_2)$ and $\cZ(\gamma_3) = \cZ(\gamma_4)$. Additionally, we stress out that it is not possible to rotate the central charge half-plane $\cH_{\cZ}$ such that the central charge of a single node in a 2-block node leaves $\cH_{\cZ}$ -- \textit{i.e.} performing a mutation on a single node from a 2-block node -- in a way that still reproduces a quiver for the massless theory. We will see, however, that such mutations will lead to massive quivers. 

In the strong-coupling chamber, the admissible mutations are $\gamma_{3,4}$ or $\hat{\gamma}_{1,2}$, where we indicate by $\gamma_{i,j}$ the fact that one needs to mutate on nodes $\gamma_i$ and $\gamma_j$ simultaneously. Note that in practice this should be equivalent to successive mutations. One quickly notices that these mutations do not lead to changes of the fundamental domain, with the only changes being the signs of the basis BPS charges \cite{Alim:2011kw}. Thus, the only states in this chamber are the $\pm(1,0)$ and $\pm(1,-1)$ dyons, similarly to the strong-coupling chamber of the pure $SU(2)$ theory.

In the weak-coupling chamber, the possible mutations are $\gamma_{1,2}$ or $\hat{\gamma}_{3,4}$. The first change of fundamental domain that we would like to consider consists of mapping the whole $I_2$ cusp at $\tau = 0$ to $\tau = 2$. In this case, both $(1,0)$ singularities follow a closed path around the other two states, with the monodromies transforming according to:
\be \label{4dSU2Nf2 Monodromies}
    \mathbb{M}_{(1,0)}^k \longrightarrow \mathbb{M}_{(-1,1)}^{-2} \mathbb{M}_{(1,0)}^k \mathbb{M}_{(-1,1)}^2 = \mathbb{M}_{(1,-2)}^k~,
\ee
In fact, since the configuration only has three cusps, we can also view this as a path around the cusp at infinity, with:
\be
    \mathbb{M}_{(1,0)}^k \longrightarrow (PT^2) \mathbb{M}_{(1,0)}^k (PT^2)^{-1} = \mathbb{M}_{(1,-2)}^k~.
\ee
At the level of the BPS quiver, the mutations $ \hat{\gamma}_{3,4} = \hat{\gamma}_3 - \hat{\gamma}_4$ lead to:
\bea
 \begin{tikzpicture}[baseline=1mm]
\node[] (1) []{$\CE_{\gamma_{1,2}=(-1,2)}$};
\node[] (2) [right = of 1]{$\CE_{\gamma_{3,4}=(1,-1)}$};
\draw[->-=0.5] (2) to   (1);
\end{tikzpicture}
\eea
One can keep rotating the central charge half-plane $\cH_{\cZ}$ in this direction, leading to successive shifts of the cusps to $\tau \rightarrow \tau + 2$, as described above. Counter-clockwise rotations can be treated in a similar way. We emphasise that the copies of $\cF$ must remain in strips of width-two, for the $I_2$ cusps, for the associated quiver to correspond to the massless theory.

\medskip

\noindent
\paragraph{Massive configurations.} The question that arises is what happens when the copies of $\cF$ are not in strips of width-two for one of the $I_2$ cusps. The answer to this question was given in \cite{Aspman:2021vhs}, where a fundamental domain for generic (equal) mass parameters was derived. Thus, we expect that such changes lead to massive BPS quivers. As a result, we might relax the constraint that two nodes need to mutate simultaneously. 

We first note that the regions $TST$ and $TST^{-1}$ are in the same $\Gamma(2)$ orbit, both being in the width-two strip of the cusp at $\tau = 1$. As a result, choosing either of these two regions will not change neither the BPS states, nor the BPS quiver. The cusp at $\tau = 0$ is identified with that at $\tau = 2$, and the region $ST^{-1}\cF$ is in the same $\Gamma(2)$ orbit with $T^2ST\cF$. %
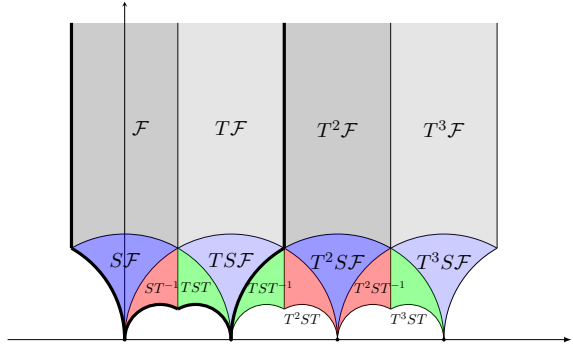
\begin{figure}[t]
    \centering
    \scalebox{0.7}{


\begin{tikzpicture}[scale=2]
    
    \pgfdeclarelayer{background}
    \pgfsetlayers{background,main}
    \pgfmathsetmacro{\myxlow}{-1}
    \pgfmathsetmacro{\myxhigh}{4}
    \pgfmathsetmacro{\myiterations}{1}

    \draw[-latex](\myxlow-0.1,0) -- (\myxhigh+0.2,0);
    \pgfmathsetmacro{\succofmyxlow}{\myxlow+0.5}
    
    \draw[-stealth, very thin] (0,0)--(0,3.2) node[above] {}; 
    \begin{scope}
        \draw[very thin, black] (0,0) arc(180:60:1);
        \draw[very thin, black] (1,0) arc(0:120:1);

        \draw[very thin, black] (0.5,0.29) -- (0.5, 3);
        \draw[very thin, black] (2.5,0.29) -- (2.5, 3);
        \draw[very thin, black] (3.5,0.866) -- (3.5, 3);
        
        \draw[very thin, black] (1.5,0.29) -- (1.5, 0.866);
        
        \draw[very thin, black] (1.5,0.866) arc(120:60:1);
        \draw[very thin, black] (1.5,0.866) arc(60:0:1);
        \draw[very thin, black] (2,0) arc(180:60:1);
        \draw[very thin, black] (2.5,0.866) arc(60:0:1);
        \draw[very thin, black] (3,0) arc(180:120:1);
        
        \draw[very thin, black] (1,0) arc(180:60:1/3);
        \draw[very thin, black] (2,0) arc(0:120:1/3);
        \draw[very thin, black] (2,0) arc(180:60:1/3);
        \draw[very thin, black] (3,0) arc(0:120:1/3);
        
    \end{scope}

    \begin{scope}
        \begin{pgfonlayer}{background}
            \clip
                (-0.5,3) 
                { -- (-0.5,0.866) arc(120:60:1)
                -- (0.5, 3)
                -- (1.5, 3)
                -- (1.5,0.866) arc(120:60:1)
                -- (2.5, 3)
             }
                -- (0.5, 3) -- cycle;
            \fill[gray,opacity=0.4] (-1,-1) rectangle (4.5,3);
        \end{pgfonlayer}
    \end{scope}
    
    \begin{scope}
            \clip
                (0.5,3) 
                { -- (0.5,0.866) arc(120:60:1)
                -- (1.5, 3)
                -- (2.5, 3)
                -- (2.5,0.866) arc(120:60:1)
                -- (3.5, 3)
             }
                -- (1.5, 3) -- cycle;
            \fill[gray,opacity=0.2] (-1,-1) rectangle (4.5,3);
    \end{scope}

    \begin{scope}
            \clip
                (-0.5,0.866) 
                { -- (-0.5,0.866) arc(60:0:1)
                -- (0,0) arc(180:120:1)
                -- (0.5,0.866) arc(60:120:1)
             }
                -- (-0.5,0.866) -- cycle;
            \fill[blue,opacity=0.4] (-1,-1) rectangle (4.5,3);
    \end{scope}
    
    \begin{scope}
            \clip
                (2-0.5,0.866) 
                { -- (2-0.5,0.866) arc(60:0:1)
                -- (2,0) arc(180:120:1)
                -- (2.5,0.866) arc(60:120:1)
             }
                -- (2-0.5,0.866) -- cycle;
            \fill[blue,opacity=0.4] (-1,-1) rectangle (4.5,3);
    \end{scope}
    
    \begin{scope}
            \clip
                (1-0.5,0.866) 
                { -- (1-0.5,0.866) arc(60:0:1)
                -- (1,0) arc(180:120:1)
                -- (1.5,0.866) arc(60:120:1)
             }
                -- (1-0.5,0.866) -- cycle;
            \fill[blue,opacity=0.2] (-1,-1) rectangle (4.5,3);
    \end{scope}
    
    \begin{scope}
            \clip
                (3-0.5,0.866) 
                { -- (3-0.5,0.866) arc(60:0:1)
                -- (3,0) arc(180:120:1)
                -- (3.5,0.866) arc(60:120:1)
             }
                -- (3-0.5,0.866) -- cycle;
            \fill[blue,opacity=0.2] (-1,-1) rectangle (4.5,3);
    \end{scope}
    
    \begin{scope}
            \clip
                (0,0) 
                { -- (0,0) arc(180:60:1/3)
                -- (0.5,0.866) arc(120:180:1)
             }
                -- (0,0) -- cycle;
            \fill[red,opacity=0.4] (-1,-1) rectangle (4.5,3);
    \end{scope}
    
    \begin{scope}
            \clip
                (1.5,0.866) 
                { -- (1.5,0.866) arc(60:0:1)
                -- (2,0) arc(180:120:1)
                -- (2.5, 0.29) arc(60:180:1/3)
                -- (2,0) arc(0:120:1/3)
             }
                -- (1.5, 0.866) -- cycle;
            \fill[red,opacity=0.4] (-1,-1) rectangle (4.5,3);
    \end{scope}
    
    \begin{scope}
            \clip
                (1.5-1,0.866) 
                { -- (1.5-1,0.866) arc(60:0:1)
                -- (2-1,0) arc(180:120:1)
                -- (2.5-1, 0.29) arc(60:180:1/3)
                -- (2-1,0) arc(0:120:1/3)
             }
                -- (1.5-1, 0.866) -- cycle;
            \fill[green,opacity=0.4] (-1,-1) rectangle (4.5,3);
    \end{scope}
    
    \begin{scope}
            \clip
                (2.5,0.866) 
                { -- (2.5,0.866) arc(60:0:1)
                -- (3,0) arc(0:120:1/3)
             }
                -- (2.5, 0.866) -- cycle;
            \fill[green,opacity=0.4] (-1,-1) rectangle (4.5,3);
    \end{scope}
    
    \begin{scope}
        \draw[line width=0.6mm, black] (-0.5,0.866) arc(60:0:1);
        \draw[line width=0.6mm, black] (1,0) arc(180:120:1);
        \draw[line width=0.6mm, black] (0,0) arc(180:60:1/3);
        \draw[line width=0.6mm, black] (1,0) arc(0:120:1/3);
        \draw[line width=0.6mm, black] (1.5,0.866) -- (1.5, 3);
        \draw[line width=0.6mm, black] (-0.5,0.866) -- (-0.5, 3);
    \end{scope}
    
    \begin{scope}
        \node at (0.15,2) {$\mathcal{F}$};
        \node at (1,2) {$T\mathcal{F}$};
        \node at (2,2) {$T^2\mathcal{F}$};
        \node at (3,2) {$T^3\mathcal{F}$};
        
        \node at (0,0.75) {$S\mathcal{F}$};
        \node at (1,0.75) {$TS\mathcal{F}$};
        \node at (2,0.75) {$T^2S\mathcal{F}$};
        \node at (3,0.75) {$T^3S\mathcal{F}$};
        
        \node at (0.35,0.5) [scale=0.7] {$ST^{-1}$};
        \node at (0.67,0.5) [scale=0.7] {$TST$};
        
        \node at (1+0.37,0.5) [scale=0.7] {$TST^{-1}$};
        \node at (1+0.67,0.2) [scale=0.7] {$T^2ST$};
        
        \node at (2+0.4,0.5) [scale=0.7] {$T^2ST^{-1}$};
        \node at (2+0.67,0.2) [scale=0.7] {$T^3ST$};

        \fill (0,0)  circle[radius=0.5pt];
        \fill (1,0)  circle[radius=0.5pt];
        \fill (2,0)  circle[radius=0.5pt];
        \fill (3,0)  circle[radius=0.5pt];
        
    \end{scope}
    
\end{tikzpicture}

}
    \caption{Fundamental domain for $\Gamma(2)$ delimited by the dark thick line. The coloured regions are identified under $\Gamma(2)$.}
    \label{fig: 4dSU2Nf=2}
\end{figure}%
In fact, we notice that this again corresponds to:
\be
    \mathbb{M}_{(1,0)} \longrightarrow \mathbb{M}_{(-1,1)}^{-2} \mathbb{M}_{(1,0)} \mathbb{M}_{(-1,1)}^2 = \mathbb{M}_{(1,-2)}~,
\ee
namely one of the $(1,0)$ singularities makes a close path around the two $(-1,1)$ singularities. At the level of the quiver, the mutation sequence $\hat{\gamma}_1 - \hat{\gamma}_{3,4}$ leads to the quiver: %
\bea
 \begin{tikzpicture}[baseline=1mm]
\node[] (1) []{$\CE_{\gamma_1=(-1,0)}$};
\node[] (2) [right = of 1]{$\CE_{\gamma_2=(-1,2)}$};
\node[] (3) [below = of 1]{$\CE_{\gamma_{3,4}=(1,-1)}$};
\draw[->>-=0.5] (2) to   (1);
\draw[->-=0.5] (3) to   (2);
\draw[->-=0.5] (1) to   (3);
\end{tikzpicture}
\eea %
This sequence can be interpreted as follows. The first mutation on $\hat{\gamma}_1$ is done in order to separate the two $(1,0)$ states, which can be achieved by turning on some mass parameter. Thus, $\gamma_1$ leaves the central charge half-plane $\cH_{\cZ}$, while $\gamma_2$ does not. Then, the mutations on the nodes $\hat{\gamma}_3$ and $\hat{\gamma}_4$ correspond to the action described above, namely one of the $(1,0)$ states making a path around the $I_2$ cusp. The final configuration is $(I_2^*, I_2, 2I_1)$, with the $I_2$ cusp being generated by $m_1 = m_2 \neq 0$.

We expect that a similar argument can be used to split the remaining $I_2$ singularity, leading to the generic configuration $(I_2^*, 4I_1)$. As a result, we effectively introduce a different type of quiver mutation, which corresponds to mass deformations of the original quiver.

\subsubsection{$4d~SU(2)~N_f = 3$ Theory}

Finally, the last 4d theory of interest is the $N_f = 3$ theory, which we discuss in this subsection.

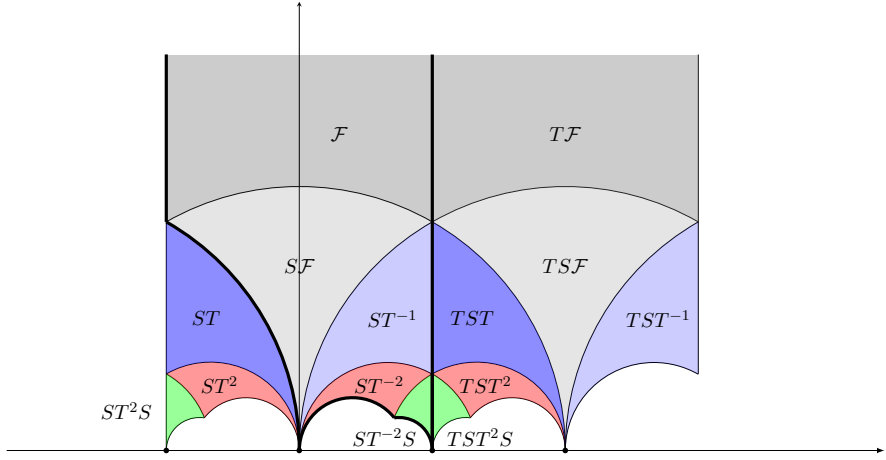
\begin{figure}[t]
    \centering
    \scalebox{0.7}{


\begin{tikzpicture}[scale=5]

    \pgfdeclarelayer{background}
    \pgfsetlayers{background,main}
    \pgfmathsetmacro{\myxlow}{-1}
    \pgfmathsetmacro{\myxhigh}{2}
    \pgfmathsetmacro{\myiterations}{1}

    \draw[-latex](\myxlow-0.1,0) -- (\myxhigh+0.2,0);
    \pgfmathsetmacro{\succofmyxlow}{\myxlow+0.5}

    \draw[-stealth, very thin] (0,0)--(0,1.5+0.2) node[above] {}; 
    \begin{scope}
        \draw[very thin, black] (0,0) arc(180:120:1);
        \draw[very thin, black] (-0.5,0.866) arc(120:60:1);
        
         \draw[very thin, black] (1-0.5,0.866) arc(60:0:1);
        \draw[very thin, black] (1,0) arc(180:120:1);
        \draw[very thin, black] (1-0.5,0.866) arc(120:60:1);

        \draw[very thin, black] (0,0) arc(180:60:1/3);
        \draw[very thin, black] (-0.5,0.29) arc(120:0:1/3);
        
        \draw[very thin, black] (-0.5, 0) arc(180:180-98.3:1/8); 
        \draw[very thin, black] (-0.357, 0.124) arc(180-38.3:0:1/5);
             
        \draw[very thin, black] (-0.5,0) -- (-0.5, 1.5);
        \draw[very thin, black] (-0.5, 0.29) arc(60:60-38.4:1/3); 
        
        \draw[very thin, black] (0.5, 0.29) arc(120:120+38.4:1/3); 
        
        \draw[very thin, black] (1,0) arc(180:60:1/3);
        \draw[very thin, black] (1-0.5,0.29) arc(120:0:1/3);
        
        \draw[very thin, black] (1-0.5, 0) arc(180:180-98.3:1/8);
        \draw[very thin, black] (1-0.357, 0.124) arc(180-38.3:0:1/5); 
             
        \draw[very thin, black] (1+0.5,0.29) -- (1+0.5, 1.5);
        \draw[very thin, black] (1-0.5, 0.29) arc(60:60-38.4:1/3);
        
    \end{scope}

    \begin{scope}
        \begin{pgfonlayer}{background}
            \clip
                (-0.5,1.5) 
                { -- (-0.5, 0.866) arc(120:60:1)
             }
                -- (0.5, 1.5) -- cycle;
            \fill[gray,opacity=0.4] (-1,-1) rectangle (2.5, 1.5);
        \end{pgfonlayer}
    \end{scope}

    \begin{scope}
            \clip
                (1-0.5,1.5) 
                { -- (1-0.5, 0.866) arc(120:60:1)
             }
                -- (1+0.5, 1.5) -- cycle
            ;
            \fill[gray,opacity=0.4] (-1,-1) rectangle (2.5, 1.5);
    \end{scope}

    \begin{scope}
            \clip
                (-0.5,0.866) 
                { -- (-0.5, 0.29) arc(120:0:1/3)
             -- (0,0) arc(0:60:1)
             }
                -- (-0.5, 0.866) -- cycle
            ;
            \fill[blue,opacity=0.45] (-1,-1) rectangle (2.5,2);
    \end{scope}
    
    \begin{scope}
            \clip
                (1-0.5,0.866) 
                { -- (1-0.5, 0.29) arc(120:0:1/3)
             -- (1,0) arc(0:60:1)
             }
                -- (1-0.5, 0.866) -- cycle
            ;
            \fill[blue,opacity=0.45] (-1,-1) rectangle (2.5,2);
    \end{scope}
    
    \begin{scope}
            \clip
                (-0.5,0.866) 
                { -- (-0.5, 0.866) arc(60:0:1)
             -- (0,0) arc(180:120:1)
             -- (0.5, 0.866) arc(60:120:1)
             }
                -- (-0.5, 0.866) -- cycle
            ;
            \fill[gray,opacity=0.2] (-1,-1) rectangle (2.5,2);
    \end{scope}
    
    \begin{scope}
            \clip
                (1-0.5,0.866) 
                { -- (1-0.5, 0.866) arc(60:0:1)
             -- (1,0) arc(180:120:1)
             -- (1+0.5, 0.866) arc(60:120:1)
             }
                -- (1-0.5, 0.866) -- cycle
            ;
            \fill[gray,opacity=0.2] (-1,-1) rectangle (2.5,2);
    \end{scope}
    
        \begin{scope}
            \clip
                (0,0) 
                { -- (0,0) arc(180:60:1/3)
             -- (0.5,0.866) arc(120:180:1)
             }
                -- (0,0) -- cycle
            ;
            \fill[blue,opacity=0.2] (-1,-1) rectangle (2.5,2);
    \end{scope}
    
    \begin{scope}
            \clip
                (1+0,0) 
                { -- (1+0,0) arc(180:60:1/3)
             -- (1+0.5,0.866) arc(120:180:1)
             }
                -- (1+0,0) -- cycle
            ;
            \fill[blue,opacity=0.2] (-1,-1) rectangle (2.5,2);
    \end{scope}

    \begin{scope}
            \clip
                (-0.5, 0.29) 
                { -- (-0.5, 0.29) arc(60:60-38.4:1/3)
                -- (-0.357, 0.124) arc(180-38.3:0:1/5)
                -- (0,0) arc(0:120:1/3)
             }
                -- (-0.5, 0.29) -- cycle
            ;
            \fill[red,opacity=0.4] (-1,-1) rectangle (2.5,2);
    \end{scope}
    
    \begin{scope}
            \clip
                (1-0.5, 0.29) 
                { -- (1-0.5, 0.29) arc(60:60-38.4:1/3)
                -- (1-0.357, 0.124) arc(180-38.3:0:1/5)
                -- (1,0) arc(0:120+38.4:1/3)
                -- (0.357, 0.124) arc(38.3:180:1/5)
                -- (0, 0) arc(180:60:1/3)
             }
                -- (1-0.5, 0.29) -- cycle
            ;
            \fill[red,opacity=0.4] (-1,-1) rectangle (2.5,2);
    \end{scope}

    \begin{scope}
            \clip
                (-0.5, 0) 
                { -- (-0.5, 0) arc(180:180-98.3:1/8) 
                -- (-0.357, 0.124) arc(60-38.4:60:1/3)
             }
                -- (-0.5, 0) -- cycle
            ;
            \fill[green,opacity=0.4] (-1,-1) rectangle (2.5,2);
    \end{scope}

    \begin{scope}
            \clip
                (1-0.5, 0) 
                { -- (1-0.5, 0) arc(180:180-98.3:1/8)
                -- (1-0.357, 0.124) arc(60-38.3:60:1/3)
                -- (0.5, 0.29) arc(120:120+38.3:1/3)
                -- (0.357, 0.124) arc(98.3:0:1/8)
             }
                -- (1-0.5, 0) -- cycle;
            \fill[green,opacity=0.4] (-1,-1) rectangle (2.5,2);
    \end{scope}
    
    \begin{scope}
        \draw[line width=0.6mm, black] (-0.5,0.866) arc(60:0:1);
        \draw[line width=0.6mm, black] (0.5,0) -- (0.5, 1.5);
        \draw[line width=0.6mm, black] (-0.5,0.866) -- (-0.5, 1.5);
        \draw[line width=0.6mm, black] (0.5, 0) arc(0:98.3:1/8); 
        \draw[line width=0.6mm, black] (0.357, 0.124) arc(38.3:180:1/5);
        \draw[line width=0.6mm, black] (1-0.5,0) -- (1-0.5, 0.29);
    \end{scope}
        
    \begin{scope}
        \node at (0.15,1.2) {$\mathcal{F}$};
        \node at (1,1.2) {$T\mathcal{F}$};
        
        \node at (0,0.7) {$S\mathcal{F}$};
        \node at (1,0.7) {$TS\mathcal{F}$};
        
        \node at (0.35,0.5)  {$ST^{-1}$};
        \node at (-0.35,0.5)  {$ST$}; 
        \node at (1+0.35,0.5)  {$TST^{-1}$};
        \node at (1-0.35,0.5)  {$TST$};
        
        \node at (-0.3,0.25)  {$ST^2$}; 
        \node at (0.3,0.25)  {$ST^{-2}$};
        \node at (1-0.3,0.25)  {$TST^2$};
        
        \node at (-0.65,0.15)  {$ST^2S$};
        \node at (0.32,0.05)  {$ST^{-2}S$};
        \node at (1-0.32,0.05)  {$TST^{2}S$};

        \fill (0,0)  circle[radius=0.3pt];
        \fill (-0.5,0)  circle[radius=0.3pt];
        \fill (0.5,0)  circle[radius=0.3pt];
        \fill (1,0)  circle[radius=0.3pt];
        
    \end{scope}
    
\end{tikzpicture}

}
    \caption{Fundamental domain for $\Gamma_0(4)$ delimited by the dark thick line. The coloured regions are identified under $\Gamma_0(4)$.}
    \label{fig: 4dSU2Nf=3}
\end{figure}

\paragraph{Massless Configuration.} The massless configuration of this theory is $(I_1^*, I_4, I_1)$, with the monodromy group being $\Gamma_0(4)$, whose modular properties have been recently discussed in \cite{Closset:2021lhd, Aspman:2021vhs}. For the fundamental domain shown in figure~\ref{fig: 4dSU2Nf=3}, the light BPS states are $4(1,0)$ and $(2,-1)$. The BPS quiver for this configuration is given by \cite{Alim:2011kw}:
\bea    \label{Nf=3 quiver}
 \begin{tikzpicture}[baseline=1mm]
\node[] (2) []{$\CE_{\gamma_5=(-2,1)}$};
\node[] (1) [left = of 2]{$\CE_{\gamma_{1,2,3,4}=(1,0)}$};;
\draw[->-=0.5] (1) to   (2);
\end{tikzpicture}
\eea
To make the following discussion clearer, let us note that:
\bea
    (ST^2S) T (ST^2S)^{-1} = \mathbb{M}_{\pm(2,1)}~, \qquad \qquad \qquad \qquad \\
    (ST^{-2}S) T (ST^{-2}S)^{-1} = (TST^{2}S) T (TST^{2}S)^{-1} =  \mathbb{M}_{\pm(2,-1)}~,
\eea %
which is equivalent to saying that the cusps at $\tau = \pm {1\ov 2}$ correspond to the light BPS states $(2, \mp 1)$, according to \eqref{Identification}. %
There are again two chambers for the massless configuration, with $arg~\cZ(\gamma_5) > arg~\cZ(\gamma_1)$ for the strong-coupling regime and $arg~\cZ(\gamma_5) < arg~\cZ(\gamma_1)$ for weak-coupling. As before, mutations in the strong-coupling chamber (\textit{i.e.} sequences of the type $\gamma_5 - \gamma_{1,2,3,4} - \gamma_5$ etc.) only change the signs of the basis states and leave the fundamental domain invariant. The BPS spectrum is, however, richer for the weak-coupling chamber.

The first change of fundamental domain that we consider is the one that involves moving the $I_1$ cusp from $\tau = {1\ov 2}$ to $\tau = -{1\ov 2}$, which amounts to moving the $I_1$ cusp past the $I_4$ cusp. In fact, we find that:
\be
    \mathbb{M}_{(2,1)} = (PT)^{-1} \mathbb{M}_{(2,-1)} (PT) = \mathbb{M}_{(1,0)}^4\mathbb{M}_{(2,-1)} \mathbb{M}_{(1,0)}^{-4}~,
\ee
in agreement with the previous comments. At the level of the BPS quiver, one can follow the mutation sequence $\gamma_{1,2,3,4}$, leading to the BPS quiver:
\bea
 \begin{tikzpicture}[baseline=1mm]
\node[] (1) []{$\CE_{\gamma_5=(2,1)}$};
\node[] (2) [left = of 1]{$\CE_{\gamma_{1,2,3,4}=(-1,0)}$};;
\draw[->-=0.5] (1) to   (2);
\end{tikzpicture}
\eea
Similarly, starting with the quiver \eqref{Nf=3 quiver} we can instead move the $I_4$ cusp past the $I_1$ singularity, with:
\be
    \mathbb{M}_{(1,-1)}^k = (PT) \mathbb{M}_{(1,0)}^k (PT)^{-1} = \mathbb{M}_{(2,-1)}^{-1}\mathbb{M}_{(1,0)}^k \mathbb{M}_{(2,-1)}~,
\ee
with the corresponding quiver mutation $\hat{\gamma}_5$ leading to the new quiver:
\bea
 \begin{tikzpicture}[baseline=1mm]
\node[] (1) []{$\CE_{\gamma_5=(2,-1)}$};
\node[] (2) [left = of 1]{$\CE_{\gamma_{1,2,3,4}=(-1,1)}$};;
\draw[->-=0.5] (1) to   (2);
\end{tikzpicture}
\eea
One can keep mutating these quivers and recover the full BPS spectrum in the weak coupling chamber, namely:
\bea
    I_1~:~ \pm(2, 2n + 1)~, \qquad I_4~:~ \pm(1,n)~, \qquad n\in\mathbb{Z}~,
\eea
with the W-boson $(0,2)$ being an accumulation ray. For more details about the spectrum, including the representations under the $SU(4)$ flavour symmetry, see \cite{Alim:2011kw} for instance. Note that, as for $N_f = 0, 2$, the weak coupling chamber only allows `trivial' changes of the fundamental domain, while the domain for the strong coupling chamber is fixed.
\medskip

\noindent
\paragraph{Massive configurations.} The other changes that we consider involve splitting the $I_4$ cusp and taking a closed path of the underlying singularities around the other $I_1$ cusp. The monodromies for the $(1,0)$ states become:
\be
    \mathbb{M}_{(1,0)} \longrightarrow  \mathbb{M}_{(2,-1)}^{-1} \mathbb{M}_{(1,0)} \mathbb{M}_{(2,-1)} = \mathbb{M}_{(1,-1)}~.
\ee
Note that, this is also equivalent to:
\be
    (\mathbb{M}_{(1,0)}^k \mathbb{M}_{(2,-1)} )^{-1} \mathbb{M}_{(1,0)} (\mathbb{M}_{(1,0)}^k\mathbb{M}_{(2,-1)}) = \mathbb{M}_{(1,-1)}~,
\ee
for any integer $k\in \mathbb{Z}$. Thus, for only one $(1,0)$ state being changed, the mutation sequence $\hat{\gamma}_{2,3,4} - \hat{\gamma}_{5}$ leads to the quiver:
\bea
 \begin{tikzpicture}[baseline=1mm]
\node[] (1) []{$\CE_{\gamma_1=(-1,1)}$};
\node[] (2) [right = of 1]{$\CE_{\gamma_{2,3,4}=(-1,0)}$};
\node[] (3) [below = of 1]{$\CE_{\gamma_{5}=(2,-1)}$};
\draw[->-=0.5] (3) to   (1);
\draw[->-=0.5] (2) to   (3);
\draw[->-=0.5] (1) to   (2);
\end{tikzpicture}
\eea %
This mutation sequence can be interpreted as moving one $I_1$ cusp from the $I_4$ singularity of the massless theory around the $I_3 \subset I_4$ and the remaining $I_1$ cusps. This can be done by turning on $m_1 \neq 0$, while leaving the other three masses to zero. Similar arguments hold for moving $k(1,0)$ states around the $(2,-1)$ singularity. For instance, for $k=2$, the mutation sequence $\hat{\gamma}_{3,4} - \hat{\gamma}_5$ leads to the massive BPS quiver:
\bea
 \begin{tikzpicture}[baseline=1mm]
\node[] (1) []{$\CE_{\gamma_{1,2}=(-1,1)}$};
\node[] (2) [right = of 1]{$\CE_{\gamma_{3,4}=(-1,0)}$};
\node[] (3) [below = of 1]{$\CE_{\gamma_{5}=(2,-1)}$};
\draw[->-=0.5] (3) to   (1);
\draw[->-=0.5] (2) to   (3);
\draw[->-=0.5] (1) to   (2);
\end{tikzpicture}
\eea %


\section{Modular Coulomb branches}   \label{section: modular RES classification}


In this section, we review all modular rational elliptic surfaces. As shown in \cite{Doran:1998hm}, this is a set of $33$ surfaces, up to quadratic twists, which then extends to $47$ different surfaces in Persson's classification \cite{Persson:1990, Miranda:1990}. The subgroups of ${\rm PSL}(2,\mathbb{Z})$ associated to each modular rational elliptic surface can be deduced from the list of genus 0 subgroups of ${\rm PSL}(2,\mathbb{Z})$ of index less or equal to $12$ \cite{Sebbar:2001, Cummins2003, STROMBERG2019436}, together with the map \eqref{summary Gamma - RES relation}. 

Let us also briefly comment on the notation used for these subgroups. Apart from the well-established notation for $\Gamma_0(N)$, $\Gamma_1(N)$ and the principal congruence subgroup $\Gamma(N)$, as well as their conjugates, we follow the notation of \cite{STROMBERG2019436}, denoting conjugacy classes by the index, the genus of associated Riemann surface, the number of cusps and elliptic elements, as well as the level:\footnote{The level of a congruence subgroup is the largest $N$ for which $\Gamma(N)$ is a subgroup. For non-congruence subgroups, this is the least common multiple of the widths of the cusps. For more details see appendix \ref{Appendix Permutations}.}
\be
    \Gamma \left( n_{\Gamma}; g, c, e_2, e_3; N\right)~.
\ee
At times, we also use the more compact notation of \cite{Cummins2003}, namely:
\be
    N\,(label)\,^g~,
\ee
with $N, g$ the level and genus, as before, with the groups labelled by a capital letter.

We order the modular configurations by the rank of the MW group, also indicating the CB of the field theories on which these can appear. 
Note that we only discuss the theories with trivial Coulomb branch, but the modular properties hold for any rank-one theory for which the given configuration is allowed. 

Consider any theory $\cT$ in this class. An essential tool in finding its quiver description is the fundamental domain, which, for a given conjugacy class in ${\rm PSL}(2,\mathbb{Z})$ can be determined from the isomorphism between these conjugacy classes and permutations of the symmetric group $S_n$, with $n$ the index in ${\rm PSL}(2,\mathbb{Z})$ \cite{10.1112/jlms/s2-1.1.351}. We review these aspects in appendix~\ref{Appendix Permutations} and refer to \cite{STROMBERG2019436} for the data needed to draw these domains. The representative of the conjugacy class that correctly describes the theory $\cT$ can be determined from consistency with the massless configuration, as well as from the asymptotics \eqref{U choice} in the limit $\tau \rightarrow i \infty$. This will usually lead to new quivers of $\cT$, which we relate to the massless BPS quivers by a sequence of quiver mutations.

It is worth pointing out that many of the modular rational elliptic surfaces have singular fibers that are not of $I_n$ type. Thus, these correspond to AD theories on the $U$-plane, which are mapped to the elliptic points of the monodromy groups. However, in order to use the BPS quiver, we will implicitly deform these singularities to $I_n$ cusps, as outlined in the previous section. On the upper half-plane $\mathbb{H}$, this operation corresponds to copies of the fundamental domain of ${\rm PSL}(2,\mathbb{Z})$ that are `absorbed' into the elliptic points. Since this splitting is not always unique, there is no `a priori' reason for why a specific choice should be preferred. However, we will pick the decomposition that is in agreement with the massless configuration.

Let us mention again that we will use the conventions of \cite{Closset:2021lhd} for the Seiberg-Witten curves of SQCD and $\KK E_n$ theories. Note that the curves expressed in terms of the $E_n$ characters differ by those in the original work of Eguchi and Sakai \cite{Eguchi:2002fc, Eguchi:2002nx, Huang:2013yta, Kim:2014nqa} by the change $\chi_2 \leftrightarrow \chi_3$.


\subsection{Proper field extensions}

Before presenting the lists of modular rational elliptic surfaces, let us emphasize an important aspect of our analysis. We will find closed-form expressions for the modular functions of all congruence subgroups of interest, as well as for certain non-congruence subgroups. This is done by solving \eqref{J(U) = J(t)}:
\be \label{J(U) = J(t) v2}
    J(U) = J(\tau)~,    
\ee
for $U(\tau)$ explicitly, where the LHS is the expression obtained from the Weierstrass form of the elliptic fibration while the RHS refers to the modular $J$-function. For generic mass parameters, this is a degree $12$ equation in $U$ and, thus, closed-form solutions do not always exist. However, an important tool that we apply in certain cases follows a more general result from \cite{2006math.....12100K}, which we shall comment on momentarily. This result emphasizes the interplay between Galois theory and four-dimensional supersymmetric quantum field theories \cite{Aspman:2021vhs, Cecotti:2015qha, Bourget:2017goy}. This interplay was also emphasized using \textit{dessins d'enfants} in, \textit{e.g.} \cite{Ashok:2006br, He:2020eva, Bao:2021vxt}.
 
When the mass parameters, or alternatively, the characters are fixed to some complex numbers, the polynomial \eqref{Polynomial J(U)=J(t)} obtained from \eqref{J(U) = J(t) v2} has coefficients in the field  $\mathbb{C}(\Gamma(1))$ of (meromorphic) modular functions of $\Gamma(1)$ over $\mathbb{C}$, where $\Gamma(1) \cong {\rm PSL}(2,\mathbb{Z})$. Turning on mass parameters, the new polynomial defines a field extension over $\mathbb{C}(\Gamma(1))$. In certain cases, there exist intermediate fields for which this polynomial partially splits, \textit{i.e.} it factors into polynomials of lower degree, as explained in \cite{Aspman:2021evt}.\footnote{We are very grateful to Elias Furrer and Johannes Aspman for explaining this to us.} Schematically, one has:
\be
    \mathbb{C}(\Gamma(1)) \subset \mathbb{C}(\Gamma') \subset \mathbb{C}(\Gamma(1), \boldsymbol{M})~,
\ee
where by $\boldsymbol{M}$ we indicate any mass parameters, while $\Gamma'$ is some subgroup of ${\rm PSL}(2,\mathbb{Z})$. In the scenarios obtained by fixing the mass parameters such that the configuration is modular, there exists a field extension where the polynomial splits, called the splitting field of the polynomial \eqref{Polynomial J(U)=J(t)}. This is, of course, $\mathbb{C}(\Gamma)$, for $\Gamma$ the monodromy group of the modular configuration. It was shown in \cite{2006math.....12100K} that for a sequence of inclusions of the type:
\be \label{Subgroups inclusions}
    \Gamma(1) \supset \Gamma' \supset \Gamma~,
\ee
there is a one-to-one correspondence between the subgroups $\Gamma'$ and the field extensions over $\mathbb{C}(\Gamma(1))$ which are contained in $\mathbb{C}(\Gamma)$, so there does exist an intermediate field $\mathbb{C}(\Gamma')$ of this field extension, with:
\be
    \mathbb{C}(\Gamma(1)) \subset \mathbb{C}(\Gamma') \subset \mathbb{C}(\Gamma)~.
\ee
Thus, the polynomial factors over the intermediate field, leading to simpler polynomial equations. Note that these field extensions are Galois since they are splitting fields for certain polynomials in $\mathbb{C}(\Gamma(1))$. The proof of the above result follows from the fact that the Galois group ${\rm Gal}(\mathbb{C}(\Gamma)/\mathbb{C}(\Gamma'))$ for $\Gamma$ a normal subgroup of $\Gamma'$ is isomorphic to $\Gamma'/\Gamma$, combined with with the fundamental theorem of Galois theory.

In practice, given the subgroup $\Gamma'$ in the inclusion sequence \eqref{Subgroups inclusions}, the solution modular function of $\Gamma$ can be determined from the simplified equation:
\be
    J(U) = J_{\Gamma'}(f)~,
\ee
where $f$ is the Hauptmodul of $\Gamma'$. This has the effect of reducing the degree $12$ polynomial equation in $U$ \eqref{J(U) = J(t)} to an equation of degree $[\Gamma':\Gamma]$ in $U$. Note that this index satisfies:
\be
    [\Gamma(1):\Gamma] = [\Gamma(1):\Gamma'] \times [\Gamma':\Gamma]~.
\ee
Let us finally mention that non-trivial inclusions \eqref{Subgroups inclusions} always exist congruence subgroups $\Gamma'$, as they must contain the principal congruence subgroup of level $N$, for some positive integer $N \in \mathbb{N}$.


\subsection{Congruence subgroups}

Many of the modular rational elliptic surfaces associated to congruence subgroups have been discussed in \cite{Closset:2021lhd}. Some of their properties and their modular functions are summarised in table~\ref{tab: Modular Functions Summary}. Note that for other subgroups in the same ${\rm PSL}(2,\mathbb{Z})$ conjugacy class, these Hauptmoduln can be found by shifting and/or rescaling the modular parameter $\tau$. Moreover, it turns out that the modular configurations have ${\rm rk}(\Phi) \leq 3$.

\medskip

\noindent
\subsubsection{${\rm rk}(\Phi) = 0$ Configurations.} Let us first mention the extremal modular surfaces, which are listed in table~\ref{tab:Extremal2}.%
\renewcommand{\arraystretch}{1}
\begin{table}[h]
\small
\centering
\begin{tabular}{ |c||c|c|c|c|} 
 \hline
$\{F_v\}$ & $\Phi_{\rm tor}$& \textit{Field theory} & $\Fg_F$ & $\Gamma$ 
\\
\hline 
 \hline
\multirow{3}{*}{$III^*, I_2, I_1$} & \multirow{3}{*}{$\mathbb{Z}_2$} & $\KK E_8$ & $E_7 \oplus A_1$ &  \multirow{3}{*}{$\Gamma_0(2)$}   \\ \cline{3-4}
&  & $\KK E_7$ & $E_7$  & \\ \cline{3-4}
&  & \text{AD} $H_1$ & $A_1$  &  \\ \hline
 \hline
\multirow{3}{*}{$IV^*, I_3, I_1$}  & \multirow{3}{*}{$\mathbb{Z}_3$} & $\KK E_8$ & $E_6 \oplus A_2$ & \multirow{3}{*}{$\Gamma_0(3)$}  \\ \cline{3-4}
 & & $\KK E_6$ & $E_6$  &  \\  \cline{3-4}
 & &  \text{AD} $H_2$ & $A_2$  &  \\ \hline
 \hline
\multirow{2}{*}{$I_4^*, I_1, I_1$}  &\multirow{2}{*}{ $\mathbb{Z}_2$} & $\KK E_8$  & $D_8$ & \multirow{2}{*}{$\Gamma_0(4)$ } \\ \cline{3-4}
 & & \text{4d pure} $SU(2)$  &  $-$  &   \\ \hline \hline
 
\multirow{3}{*}{$I_1^*, I_4, I_1$}  & \multirow{3}{*}{$\mathbb{Z}_4$} & $\KK E_8$ & $D_5 \oplus A_3$ &  \multirow{3}{*}{$\Gamma_0(4)$}  \\ \cline{3-4}
 & & $\KK E_5$  & $D_5$  &   \\\cline{3-4}
 & & \text{4d} $SU(2)\, N_f=3$  & $A_3$  &    \\ \hline
 \hline
\multirow{2}{*}{$I_2^*, 2I_2$}  & \multirow{2}{*}{$\mathbb{Z}_2\times \mathbb{Z}_2$ }& $\KK E_7$  & $D_6 \oplus A_1 $ &\multirow{2}{*}{ $\Gamma(2)$ } \\\cline{3-4}
& & \text{4d} $SU(2)\, N_f=2$  & $A_1\oplus A_1$  &  \\ \hline
 \hline
\multirow{2}{*}{$2I_5, 2I_1$}  & \multirow{2}{*}{$\mathbb{Z}_5$} & $\KK E_8$ & $A_4 \oplus A_4$ & \multirow{2}{*}{$\Gamma_1(5)$}  \\ \cline{3-4}
 & & $\KK E_4$  & $A_4$ &  \\\hline
 \hline
\multirow{4}{*}{$I_6, I_3, I_2, I_1$}  & \multirow{4}{*}{$\mathbb{Z}_6$} & $\KK E_8$ & $A_5 \oplus A_2 \oplus A_1$ & \multirow{4}{*}{$\Gamma_0(6)$}   \\ \cline{3-4}
 & & $\KK E_7$ & $A_5 \oplus A_2 $ &    \\ \cline{3-4}
 & & $\KK E_6$ & $A_5 \oplus A_1 $ &   \\ \cline{3-4}
 & & $\KK E_3$ & $A_2 \oplus A_1$ &   \\ \hline
 \hline
\multirow{3}{*}{$I_8, I_2, 2I_1 $}  & \multirow{3}{*}{$\mathbb{Z}_4$} & $\KK E_8$ & $A_7 \oplus A_1$ & \multirow{3}{*}{$\Gamma_0(8)$}  \\ \cline{3-4}
 & & $\KK E_7$ & $A_7$ &  \\\cline{3-4}
 & & $\KK E_1$  & $A_1$ &  \\\hline
 \hline
\multirow{2}{*}{$2I_4, 2I_2$} & \multirow{2}{*}{$\mathbb{Z}_4\times \mathbb{Z}_2$} & $\KK E_7$ & $2A_3 \oplus A_1 $&  \multirow{2}{*}{$\Gamma_0(4) \cap \Gamma(2)$}   \\ \cline{3-4}
 & & $\KK E_5$  & $A_3 \oplus 2A_1$ &   \\ \hline
 \hline
\multirow{2}{*}{$I_9, 3I_1$} & \multirow{2}{*}{$\mathbb{Z}_3$} & $\KK E_8$ & $A_8$ & \multirow{2}{*}{$\Gamma_0(9)$}   \\ \cline{3-4}
 & & $\KK E_0$  & $-$ &   \\ \hline
 \hline
$4I_3$ &  $\mathbb{Z}_3 \times \mathbb{Z}_3$ & $\KK E_6$ & $3A_2$  & $\Gamma(3)$ \\
 \hline
\end{tabular}
    \caption{Extremal modular RES and corresponding field theories.}
    \label{tab:Extremal2}
\end{table}
%
As pointed out in \cite{Closset:2021lhd}, these are all the extremal rational elliptic surfaces with three or four singular fibers, excluding the configuration $(II^*, 2I_1)$, which is not modular. Additionally, in all cases other than the $(2I_5,2I_1)$ configuration, the associated congruence subgroup is in the same conjugacy class as $\Gamma_0(N)$, for some integer value of $N$ \cite{Sebbar:2001}. As a result, their Hauptmoduln can be expressed in terms of $\eta$-quotients, whose $q$-series also appear in the monstrous moonshine \cite{Conway:1979qga}. Note also that most of the configurations can be understood as the massless limit of the $\KK E_n$ theories, for $n\neq \t 1, 2,8$ \cite{Alim:2013eja}. For a more detailed analysis of the BPS states of these configurations and the global aspects of the flavour symmetry, see \cite{Closset:2021lhd}.

It is worth pointing out that the Hauptmodul for $\Gamma_1(5)$ can be obtained from that of $\Gamma(5)$ or $\Gamma^0(5)$, as discussed in \cite{Sebbar:2001}, which is a first instance of \eqref{Subgroups inclusions}. Its expression is showed in table~\ref{tab: Modular Functions Summary}, where $\left( n \ov p \right)$ denotes the Legendre symbol. The configurations $(4I_3)$ and $(2I_4,2I_2)$ do not correspond to massless $\KK E_n$ theories, but can be used to determine the BPS states becoming massless at the additive singularities of the $E_{6}$ and $E_7$ theories \cite{Closset:2021lhd}. 

Finally, before moving on to non-extremal configurations, let us review some of the extremal configurations which we will make use of later.
\medskip

\begin{figure}[t]
    \centering
    \begin{subfigure}{0.5\textwidth}
    \centering
    \scalebox{0.7}{


\begin{tikzpicture}[scale=2]

    \pgfdeclarelayer{background}
    \pgfsetlayers{background,main}

    \pgfmathsetmacro{\myxlow}{-1}
    \pgfmathsetmacro{\myxhigh}{4}
    \pgfmathsetmacro{\myiterations}{1}
    
    \draw[-latex](\myxlow-0.1,0) -- (\myxhigh+0.2,0);
    \pgfmathsetmacro{\succofmyxlow}{\myxlow+0.5}
    
    \draw[-stealth, very thin] (0,0)--(0,3.2) node[above] {}; 
    \begin{scope}
        \draw[very thin, black] (-0.5,0.8666) arc(60:0:1);
        \draw[very thin, black] (0,0) arc(180:0:1);
        \draw[very thin, black] (1,0) arc(0:120:1);
        \draw[very thin, black] (1,0) arc(180:60:1);
        \draw[very thin, black] (2,0) arc(180:60:1);
        
        \draw[very thin, black] (0,0) arc(180:60:1/3);
        \draw[very thin, black] (1,0) arc(0:120:1/3);
        \draw[very thin, black] (1,0) arc(180:60:1/3);
        \draw[very thin, black] (2,0) arc(0:120:1/3);
        
        \draw[very thin, black] (1-0.357, 0.124) arc(180-38.3:0:1/5); 
        \draw[very thin, black] (1-0.5, 0.29) arc(60:60-38.3:1/3); 
        
        \draw[very thin, black] (0.5,0.29) -- (0.5, 3);
        \draw[very thin, black] (-0.5,0.866) -- (-0.5, 3);
        \draw[very thin, black] (1.5,0.29) -- (1.5, 3);
        \draw[very thin, black] (2.5,0.866) -- (2.5, 3);
        \draw[very thin, black] (3.5,0.866) -- (3.5, 3);
    \end{scope}

    \begin{scope}
        \begin{pgfonlayer}{background}
            \clip
                (-0.5,3) 
                { -- (-0.5,0.866) arc(60:0:1)
             -- (0,0) arc(180:60:1/3)
             -- (1-0.5, 0.29) arc(60:60-38.3:1/3)
             -- (1-0.357, 0.124) arc(180-38.3:0:1/5)
             -- (1,0) arc(180:60:1/3)
             -- (1.5, 0.29) arc(120:0:1/3)
             -- (2,0) arc(180:60:1)
             }
                -- (3.5, 3) -- cycle
            ;
            \fill[gray,opacity=0.4] (-1,-1) rectangle (4.5,3);
        \end{pgfonlayer}
    \end{scope}

    \begin{scope}
        \node at (0.15,2) {$\mathcal{F}$};
        \node at (1,2) {$T\mathcal{F}$};
        \node at (2,2) {$T^2\mathcal{F}$};
        \node at (3,2) {$T^3\mathcal{F}$};
        
        \node at (0,0.75) {$S\mathcal{F}$};
        \node at (1,0.75) {$TS\mathcal{F}$};
        \node at (2,0.75) {$T^2S\mathcal{F}$};
        
        \node at (-0.3,0.13) {$I_2$};
        \node at (1+0.3,0.13) {$I_4$};
        \node at (2+0.3,0.13) {$I_2$};

        \fill (0,0)  circle[radius=0.5pt];
        \fill (1,0)  circle[radius=0.5pt];
        \fill (2,0)  circle[radius=0.5pt];
        \fill (3,0)  circle[radius=0.5pt];
        
        \node at (0,-0.2) {$(1,0)$};
        \node at (1,-0.2) {$(-1,1)$};
        \node at (2,-0.2) {$(1,-2)$};
        
    \end{scope}
    
\end{tikzpicture}

}
    \caption{ }
    \end{subfigure}%
    \begin{subfigure}{0.5\textwidth}
    \centering
    \scalebox{0.7}{


\begin{tikzpicture}[scale=2]

    \pgfdeclarelayer{background}
    \pgfsetlayers{background,main}

    \pgfmathsetmacro{\myxlow}{-1}
    \pgfmathsetmacro{\myxhigh}{4}
    \pgfmathsetmacro{\myiterations}{1}
    
    \draw[-latex](\myxlow-0.1,0) -- (\myxhigh+0.2,0);
    \pgfmathsetmacro{\succofmyxlow}{\myxlow+0.5}
    
    \draw[-stealth, very thin] (0,0)--(0,3.2) node[above] {}; 
    \begin{scope}
        \draw[very thin, black] (-0.5,0.866) arc(120:60:1);
        \draw[very thin, black] (-0.5,0.866) arc(60:0:1);
        \draw[very thin, black] (0,0) arc(180:0:1);
        \draw[very thin, black] (1+2,0) arc(0:120:1);
        \draw[very thin, black] (1+2,0) arc(180:120:1);
        \draw[very thin, black] (2,0) arc(180:60:1);
        
        \draw[very thin, black] (0,0) arc(180:60:1/3);
        \draw[very thin, black] (1+2,0) arc(0:120:1/3);
        \draw[very thin, black] (1+2,0) arc(180:60:1/3);
        \draw[very thin, black] (2,0) arc(0:120:1/3);
        
        \draw[very thin, black] (2+1-0.357, 0.124) arc(180-38.3:0:1/5); 
        \draw[very thin, black] (2+1-0.5, 0.29) arc(60:60-38.3:1/3); 
        
        \draw[very thin, black] (0.5,0.29) -- (0.5, 3);
        \draw[very thin, black] (-0.5,0.866) -- (-0.5, 3);
        \draw[very thin, black] (1.5,0.29) -- (1.5, 3);
        \draw[very thin, black] (2.5,0.29) -- (2.5, 3);
        \draw[very thin, black] (3.5,0.29) -- (3.5, 3);
    \end{scope}

    \begin{scope}
        \begin{pgfonlayer}{background}
            \clip
                (-0.5,3) 
                { -- (-0.5,0.866) arc(60:0:1)
             -- (0,0) arc(180:60:1/3)
             -- (0.5, 0.866) arc(120:60:1)
             -- (1.5, 0.29) arc(120:0:1/3)
             -- (2,0) arc(180:120:1)
             -- (2+1-0.5, 0.29) arc(60:60-38.3:1/3)
             -- (2+1-0.357, 0.124) arc(180-38.3:0:1/5)
             -- (2+1,0) arc(180:60:1/3)
             }
                -- (3.5, 3) -- cycle;
            \fill[gray,opacity=0.4] (-1,-1) rectangle (4.5,3);
        \end{pgfonlayer}
    \end{scope}

    \begin{scope}
        \node at (0.15,2) {$\mathcal{F}$};
        \node at (1,2) {$T\mathcal{F}$};
        \node at (2,2) {$T^2\mathcal{F}$};
        \node at (3,2) {$T^3\mathcal{F}$};
        
        \node at (0,0.75) {$S\mathcal{F}$};
        \node at (1+2,0.75) {$T^3S\mathcal{F}$};
        \node at (2,0.75) {$T^2S\mathcal{F}$};
        
        \node at (-0.3,0.13) {$I_2$};
        \node at (2+1+0.3,0.13) {$I_4$};
        \node at (2+0.3,0.13) {$I_2$};

        \fill (0,0)  circle[radius=0.5pt];
        \fill (1,0)  circle[radius=0.5pt];
        \fill (2+1,0)  circle[radius=0.5pt];
        \fill (2,0)  circle[radius=0.5pt];
        
        \node at (0,-0.2) {$(1,0)$};
        \node at (1+2,-0.2) {$(-1,3)$};
        \node at (2,-0.2) {$(1,-2)$};
        
    \end{scope}
    
\end{tikzpicture}

}
    \caption{ }
    \end{subfigure}
    \caption{Possible fundamental domains for $\Gamma^0(4)\cap\Gamma(2)$, with BPS states corresponding to each cusp indicated.}
    \label{fig: Gamma04Gamma2}
\end{figure}
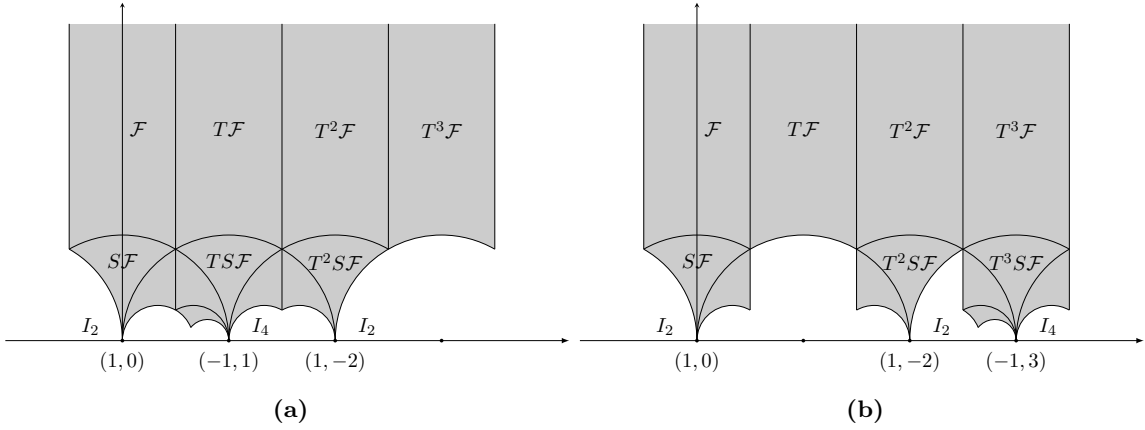

\noindent
\paragraph{$\boldsymbol{(2I_4, 2I_2)}$ configuration for $\KK E_5$.} Consider first the $(2I_4, 2I_2)$ configuration. Fixing the $I_4$ fibre at infinity the rational elliptic surface corresponds to an allowed configuration on the Coulomb branch of the $\KK E_5$ theory, with the Weierstrass form given by:
\be
    g_2(U) = {4\ov 3}(U^4 + U^2 + 16)~, \qquad g_3(U) = -{8\ov 27}(U^6 + 6U^4 - 24U^2 - 64)~,
\ee
and discriminant:
\be
    \Delta(U) = 256U^4(U^2 + 4)^2~.
\ee
In particular, this is obtained for:
\be
\boldsymbol{\chi}^{E_5} = \left( -2, -3, 0, 8, 0 \right)~.
\ee
The BPS quiver associated to this configuration was found in \cite{Closset:2021lhd} and reads:
\bea\label{quiver E5 orbi}
 \begin{tikzpicture}[baseline=1mm]
\node[] (1) []{$\CE_{\gamma_{1,2}=(1,0)}$};
\node[] (2) [right = of 1]{$\CE_{\gamma_{3,4,5,6}=(-1,1)}$};
\node[] (3) [below = of 1]{$\CE_{\gamma_{7,8}=(1,-2)}$};
\draw[->-=0.5] (3) to   (1);
\draw[->-=0.5] (1) to   (2);
\draw[->>-=0.5] (2) to   (3);
\end{tikzpicture}
\eea
which is a known quiver for local $dP_5$ -- see \textit{e.g.} `Model 4d' in \cite{Hanany:2012hi}. The fundamental domain for this configuration is shown in figure~\ref{fig: Gamma04Gamma2}. This can be found by solving the polynomial equation \eqref{Polynomial J(U)=J(t)} for $U = U(\tau)$ and analysing the modular transformations of this function \cite{Aspman:2020lmf, Aspman:2021evt, Aspman:2021vhs, Closset:2021lhd}.

At the level of the fundamental domain, we notice that the $I_4$ cusp at $\tau = 1$ can be moved to $\tau = 3$, transforming thus all $(-1,1)$ BPS states to $(1,-3)$ states. This is equivalent to the $I_4$ singularity making a closed path around the $I_2$ singularity generated by the $(1,-2)$ states, with:
\be
    \mathbb{M}_{(-1,1)} \longrightarrow \mathbb{M}_{(1,-2)}^{-2} \mathbb{M}_{(1,-1)} \mathbb{M}_{(1,-2)}^2 = \mathbb{M}_{(1,-3)}~.
\ee
At the level of the BPS quiver, the mutation $\hat{\gamma}_{7,8}$ applied to \eqref{quiver E5 orbi} leads to:
\bea
 \begin{tikzpicture}[baseline=1mm]
\node[] (1) []{$\CE_{\gamma_{1,2}=(1,0)}$};
\node[] (2) [right = of 1]{$\CE_{\gamma_{3,4,5,6}=(1,-3)}$};
\node[] (3) [below = of 1]{$\CE_{\gamma_{7,8}=(-1,2)}$};
\draw[->>>-=0.5] (2) to   (1);
\draw[->-=0.5] (3) to   (2);
\draw[->>-=0.5] (1) to   (3);
\end{tikzpicture}
\eea %
This is simply a feature of the $\Gamma^0(4) \cap \Gamma(2)$ congruence subgroup which allows to move the $I_4$ cusp from $\tau = 1$ to $\tau = 3$. The two fundamental domains are shown in figure \ref{fig: Gamma04Gamma2}.
As before, one can consider splitting the $I_4$ cusp, leading to configurations that are no longer modular and involve branch cuts and branch points in general.

Let us also mention that this configuration can be used to understand the origin of the massless configuration of the $\KK E_5$ theory, namely $(I_4, I_1, I_1^*)$, which was already discussed in \cite{Closset:2021lhd}. To summarise the argument, the monodromy group of the latter is $\Gamma^0(4)$, which can be chosen with the $I_1$ cusp at $\tau = 0$ and the $I_1^*$ cusp at $\tau = 2$. Thus, the origin of the $I_1^*$ cusp can be understood from the light BPS states:
\be
    I_1^*~:~ \mathbb{M}_{(1,0)} \mathbb{M}_{(1,-1)}^4 \mathbb{M}_{(1,-2)}^2 = (T^2 S)(PT)(T^2S)^{-1}~.
\ee
We will make use of this configuration throughout this section to understand other modular configurations on the Coulomb branch of the $\KK E_5$ theory.


\subsubsection{${\rm rk}(\Phi) = 1$ Configurations} The modular configurations with ${\rm rk}(\Phi) = 1$ and the associated modular group being a congruence subgroup are listed in table~\ref{tab:Congruence Rank1}.%
\renewcommand{\arraystretch}{1}
\begin{table}[h]
\small
\centering
\begin{tabular}{ |c||c|c|c|c|} 
 \hline
$\{F_v\}$ & $\Phi_{\rm tor}$& \textit{Field theory} & $\Fg_F$ & $\Gamma$ 
\\
\hline 
 \hline
\multirow{3}{*}{$III^*, II, I_1$} & \multirow{3}{*}{$-$} & $\KK E_8$ & $E_7 \oplus \frak{u}(1)$ &  \multirow{3}{*}{$\Gamma(1)$} \\ \cline{3-4}
&  & $MN~ E_8$ & $E_7 \oplus \frak{u}(1)$  & \\ \cline{3-4}
&  & \text{AD} $H_1$ & $\frak{u}(1)$  &  \\ \hline
 \hline
\multirow{3}{*}{$IV^*, III, I_1$} & \multirow{3}{*}{$-$} & $\KK E_8$ & $E_6 \oplus A_1 \oplus \frak{u}(1)$ &  \multirow{3}{*}{$\Gamma(1)$} \\ \cline{3-4}
&  & $MN~ E_7$ & $E_6 \oplus \frak{u}(1)$  &  \\ \cline{3-4}
&  & \text{AD} $H_2$ & $A_1 \oplus \frak{u}(1)$  &  \\ \hline
 \hline
\multirow{3}{*}{$IV^*, I_2, II$} & \multirow{3}{*}{$-$} & $\KK E_7$ & $E_6 \oplus \frak{u}(1)$ &  \multirow{3}{*}{$\Gamma^2$}  \\ \cline{3-4}
&  & $MN~ E_8$ & $E_6 \oplus A_1 \oplus \frak{u}(1)$  &   \\ \cline{3-4}
&  & \text{AD} $H_2$ & $A_1 \oplus \frak{u}(1)$  &  \\ \hline
 \hline
\multirow{3}{*}{$I_1^*, III, I_2$} & \multirow{3}{*}{$\mathbb{Z}_2$} & $\KK E_7$ & $D_5 \oplus A_1 \oplus \frak{u}(1)$ &  \multirow{3}{*}{$\Gamma_0(2)$} \\ \cline{3-4}
&  & \text{MN} $E_7$ & $D_5 \oplus A_1 \oplus \frak{u}(1)$  &  \\ \cline{3-4}
&  & \text{4d} $SU(2)\, N_f=3$ & $2A_1 \oplus \frak{u}(1)$  &  \\ \hline
 \hline
\multirow{3}{*}{$I_2^*, III, I_1$} & \multirow{3}{*}{$\mathbb{Z}_2$} & $\KK E_8$ & $D_6 \oplus A_1 \oplus \frak{u}(1)$ &  \multirow{3}{*}{$\Gamma_0(2)$}  \\ \cline{3-4}
&  & \text{MN} $E_7$ & $D_6 \oplus \frak{u}(1)$  &   \\ \cline{3-4}
&  & \text{4d} $SU(2)\, N_f=2$ & $A_1 \oplus \frak{u}(1)$  &  \\ \hline
 \hline
\multirow{3}{*}{$I_1^*, I_3, II$} & \multirow{3}{*}{$-$} & $\KK E_6$ & $D_5 \oplus \frak{u}(1)$ &  \multirow{3}{*}{$\Gamma_0(3)$}  \\ \cline{3-4}
&  & \text{MN} $E_8$ & $D_5 \oplus A_2 \oplus \frak{u}(1)$  &   \\ \cline{3-4}
&  & \text{4d} $SU(2)\, N_f=3$ & $A_2 \oplus \frak{u}(1)$  &  \\ \hline
 \hline
\multirow{3}{*}{$I_3^*, II, I_1$} & \multirow{3}{*}{$-$} & $\KK E_8$ & $D_7 \oplus \frak{u}(1)$ &  \multirow{3}{*}{$\Gamma_0(3)$}  \\ \cline{3-4}
&  & \text{MN} $E_8$ & $D_7 \oplus \frak{u}(1)$  &  \\ \cline{3-4}
&  & \text{4d} $SU(2)\, N_f=1$ & $\frak{u}(1)$  &   \\ \hline
\hline
\multirow{3}{*}{$I_0^*, I_4, 2I_1$} & \multirow{3}{*}{$\mathbb{Z}_2$} & \text{4d} $SU(2)\, N_f=4$ & $A_3 \oplus \frak{u}(1)$ &  \multirow{3}{*}{$\Gamma_0(4)$}  \\ \cline{3-4}
&  &$\KK E_5$ & $D_4 \oplus \frak{u}(1)$  &   \\ \cline{3-4}
&  &$\KK E_8$ & $D_4 \oplus A_3 \oplus \frak{u}(1)$  & \\ \hline
\hline
\multirow{2}{*}{$I_0^*, 3I_2$} & \multirow{2}{*}{$\mathbb{Z}_2 \times \mathbb{Z}_2$} & \text{4d} $SU(2)\, N_f=4$ & $3A_1 \oplus \frak{u}(1)$ &  \multirow{2}{*}{$\Gamma(2)$} \\ \cline{3-4}
&  &$\KK E_7$ & $D_4 \oplus 2A_1 \oplus \frak{u}(1)$  & \\ \hline

\end{tabular}
    \caption{Modular RES with ${\rm rk}(\Phi)=1$ and corresponding field theories.}
    \label{tab:Congruence Rank1}
\end{table}%
Some configurations are simply obtained from extremal configurations by quadratic twists. As a result, they have the same $j$-invariant as their corresponding extremal configurations, and, thus, the same modular properties. This is not the case, however, for the first three configurations, which we briefly discuss now.
\medskip

\noindent
\paragraph{$\boldsymbol{(III^*, II, I_1)}$ and $\boldsymbol{(IV^*, III, I_1)}$ configurations for $\KK E_8$.} These configurations are related by a quadratic twist, so let us focus on the former configuration. This can be obtained on the Coulomb branch of the $\KK E_8$ theory, by fixing the $I_1$ fibre at infinity. As discussed in \cite{Closset:2021lhd}, it is convenient to find the map between the $E_8$ characters and the gauge theory parameters $(\boldsymbol{M},\lambda)$, where $\lambda$ and $M_i$ are the exponentiated (inverse) gauge coupling and flavour masses. One can then check that for the $\KK E_8$ theory, for equal masses $M_i = m$, we have the configuration $(I_7, 4I_1)$, which, by setting $\lambda = {1\ov m^4}$ becomes the $(III^*, 3I_1)$ configuration. Then, for certain values of the masses, we have:
\bea
    (III^*, II, I_1)~: \qquad M_i = m~, \quad \lambda = {1\ov m^4}~, \quad m_{\pm} = e^{\pm{\pi i \ov 2}}(3\pm2\sqrt{2})~.
\eea
Focusing on\footnote{The $(III^*, II, I_1)$ configuration can also found for $\boldsymbol{\chi}^{E_8} = \{ 122243, \,7432962624,$ $-25018478,\, 299178594518496,\, -641511136266,\, 902309696,\, -507892,\, -616 \}$.} the  $m_+$ solutions, the curve becomes:
\be
    g_2(U) = {1 \ov 12} (U-432)^3 (U+1296) ~, \qquad g_3(U) = -{1 \ov 216} (U-432)^5 (U+1296) ~, 
\ee
with the discriminant and $J$-invariant given by:
\be
    \Delta(U) = (U-432)^9 (U+1296)^2~, \qquad J(U) = {3\ov 4} + {U \ov 1728}~.
\ee
Rearranging the last expression, we find that:
\be
    U(\tau) = j(\tau) - 1296~.
\ee
As a result, the modular group of this configuration is $\Gamma(1) = {\rm PSL}(2,\mathbb{Z})$, with the two elliptic elements at $\tau = e^{2\pi i \ov 3}$ and $\tau = i$ being the type $II$ and type $III^*$ singular fibers. A similar argument holds for the $(IV^*, III, I_1)$, on the Coulomb branch of any of the theories indicated in table~\ref{tab:Congruence Rank1}. However, note that the latter configuration can no longer be obtained for equal masses $M_i$ on the CB of the $\KK E_8$ theory. By working with the characters of $E_8$ and following the branching rules for $E_8 \rightarrow E_6 \oplus A_1 \oplus \frak{u}(1)$, one can find the $(IV^*, III, I_1)$ configuration\footnote{This can be obtained for $\boldsymbol{\chi}^{E_8} = \{143843, 9014103360, 25481458,  312413106260832, -48762273066$, $-863671744, -1042708, 248 \}$.} with the Weierstrass form:
\be
    g_2(U) = {1 \ov 12} (U+432)^3 (U-1296) ~, \qquad g_3(U) = -{1 \ov 216} (U+432)^4 (U-1296)^2 ~, 
\ee
with the discriminant and $J$-invariant given by:
\be
    \Delta(U) = (U+432)^8 (U-1296)^3~, \qquad J(U) = {1\ov 4} + {U \ov 1728}~.
\ee
Note that since this configuration was not directly obtained by a quadratic twist from the $(III^*, II, I_1)$ configuration, the modular functions differ by a constant shift. 
\medskip

\noindent
\paragraph{$\boldsymbol{(IV^*, I_2, II)}$ configuration for $\KK E_7$.} The next configuration is the $(IV^*, I_2, II)$ configuration, which can be obtained on the CB of the $E_7$ theory. One finds that the $j$-invariant is of the form:
\be
    j(U) = U^2 - \alpha U  - {3\ov 4}\alpha^2~.
\ee
for $\alpha \in \mathbb{C}$ some constant.\footnote{An example is $\boldsymbol{\chi}^{E_7} = \{ -299, 42341, -66\alpha, -3774250, 2366\alpha, 675, \alpha \}$ where $\alpha = 24i\sqrt{3}~$.} Following the analysis of \cite{Aspman:2021vhs}, we find that:
\be
    U(\tau) = -{\alpha\ov 2} \, \pm \, \sqrt{j(\tau) - 1728} = -{\alpha\ov 2}\,\pm\, 8{\left(\vartheta_2^4 + \vartheta_3^4\right) \left(\vartheta_3^4 + \vartheta_4^4\right)\left(\vartheta_4^4 -\vartheta_2^4 \right) \ov \vartheta_2^4 \vartheta_3^4 \vartheta_4^4 }~.
\ee
Consequently, $U(\tau)$ is a modular function for the unique index 2 subgroup of ${\rm PSL}(2,\mathbb{Z})$. This is usually denoted by $\Gamma^2$, being the subgroup generated by the squares of the elements of the modular group.

\medskip

\noindent
\paragraph{$\boldsymbol{(I_0^*, I_4, 2I_1)}$ configuration for $\KK E_5$.} The remaining configurations are all obtained from the extremal configurations, upon quadratic twists and can be thus analyzed in a similar way. Perhaps rather more curious are the configurations involving the $I_0^*$ singularity. In the context of the $4d\, SU(2)\, N_f=4$ theory these have been recently discussed in \cite{Aspman:2021evt}, where the $I_0^*$ fiber is the singular fiber at infinity. Note, however, that since at the level of subgroups of the modular group $I_n^*$ and $I_n$ singular fibers both correspond to cusps of width $n$, $I_0^*$ singular fibers are not distinguished from the smooth generic fiber $I_0$. Thus, these do not appear as cusps of the fundamental domain and, rather interestingly, the cusp at infinity `disappears', with the remaining cusps being mapped to the real line \cite{Aspman:2021evt}.

Consider now the $(I_0^*, I_4, 2I_1)$ configuration on the CB of the $\KK E_5$ theory, found for:
\be
    \boldsymbol{\chi}^{E_5} = \left(  6, 13, 0, 8, 0 \right)~,
\ee
for instance. In this case the discriminant and $j$-invariant read:
\bea
    \Delta(U) = U^6 (U^2 - 64) ~, \qquad \quad j(U) = { (U^2 - 48)^3 \ov (U^2 - 64)}~,
\eea
with a possible solution for $U=U(\tau)$ given by the usual Hauptmodul for $\Gamma^0(4)$:
\be
    U(\tau) = \left( {\eta(\tau) \ov \eta(4\tau)} \right)^8~.
\ee
related to the Hauptmodul of $\Gamma_0(4)$ shown in table~\ref{tab: Modular Functions Summary} by $\tau \rightarrow \tau/4$. This is then consistent with the previous $\KK E_5$ configurations as follows. From the $(2I_4, 2I_2)$ configuration with the BPS states $2(1,0)$, $4(-1,1)$ and $2(1,-2)$, the $I_0^*$ singular fiber can be formed as:
\be
    \mathbb{M}_{I_0^*} = \mathbb{M}_{(1,0)} \mathbb{M}_{(-1,1)}^4 \mathbb{M}_{1,-2} = P~,
\ee
and thus the BPS quiver used to describe the $(2I_4, 2I_2)$ configuration still holds here as well.

\subsubsection{${\rm rk}(\Phi) = 2$ Configurations}

The modular configurations having ${\rm rk}(\Phi) =2$ are listed in table~\ref{tab:Congruence Rank2}.%
\renewcommand{\arraystretch}{1}
\begin{table}[t]
\small
\centering
\begin{tabular}{ |c||c|c|c|c|} 
 \hline
$\{F_v\}$ & $\Phi_{\rm tor}$& \textit{Field theory} & $\Fg_F$ & $\Gamma$ 
\\
\hline 
 \hline
\multirow{3}{*}{$I_1^*, III, II$} & \multirow{3}{*}{$-$} & \text{MN} $E_8$ & $D_5 \oplus A_1 \oplus 2\frak{u}(1)$ &  \multirow{3}{*}{$\Gamma(1)$}  \\ \cline{3-4}
&  & \text{MN} $E_7$ & $D_5 \oplus 2\frak{u}(1)$  &   \\ \cline{3-4}
&  & \text{4d} $SU(2)\, N_f=3$ & $A_1\oplus 2\frak{u}(1)$  &  \\ \hline
 \hline
\multirow{2}{*}{$I_2^*, 2II$} & \multirow{2}{*}{$-$} & \text{MN} $E_8$ & $D_6 \oplus 2\frak{u}(1)$ &  \multirow{2}{*}{$\Gamma^2$}  \\ \cline{3-4}
&  & \text{4d} $SU(2)\, N_f=2$ & $2\frak{u}(1)$  & \\ \hline
\hline
\multirow{4}{*}{$I_0^*, III, I_2, I_1$} & \multirow{4}{*}{$\mathbb{Z}_2$}  &\text{4d} $SU(2)\, N_f=4$ & $A_2 \oplus 2\frak{u}(1)$ &  \multirow{4}{*}{$\Gamma_0(2)$} \\ \cline{3-4}
&  &\text{MN} $E_6$ & $D_4 \oplus A_1 \oplus 2\frak{u}(1)$  &  \\ \cline{3-4}
&  &$\KK E_7$ & $D_4 \oplus A_1 \oplus 2\frak{u}(1)$  &   \\ \cline{3-4}
&  & $\KK E_8$ & $D_4 \oplus 2A_1 \oplus 2\frak{u}(1)$  &   \\ \hline
\hline
\multirow{4}{*}{$I_0^*, I_3, II, I_1$} & \multirow{4}{*}{$-$} & \text{4d} $SU(2)\, N_f=4$ & $A_2 \oplus 2\frak{u}(1)$ &  \multirow{4}{*}{$\Gamma_0(3)$}   \\ \cline{3-4}
&  &$\KK E_8$ & $D_4 \oplus A_2 \oplus 2\frak{u}(1)$  &   \\ \cline{3-4}
&  &$\KK E_6$ & $D_4 \oplus 2\frak{u}(1)$  &  \\ \cline{3-4}
&  & $\KK E_8$ & $D_4 \oplus A_2 \oplus 2\frak{u}(1)$  &  \\ \hline
 \hline
\multirow{2}{*}{$2I_3, 2III$} & \multirow{2}{*}{$-$} & $\KK E_6$ & $A_2 \oplus 2A_1 \oplus 2\frak{u}(1)$ &  \multirow{2}{*}{$3C^0$}  \\ \cline{3-4}
&  & \text{MN} $E_7$ & $2A_2 \oplus A_1 \oplus 2\frak{u}(1)$  &   \\ \hline
 \hline
\multirow{2}{*}{$2I_4, 2II$} & \multirow{2}{*}{$-$} & $\KK E_5$ & $A_3 \oplus 2\frak{u}(1)$ &  \multirow{2}{*}{$4D^0$}  \\ \cline{3-4}
&  & \text{MN} $E_8$ & $2A_3 \oplus 2\frak{u}(1)$  &   \\ \hline
 \hline
\multirow{3}{*}{$I_4, 2III, I_2$} & \multirow{3}{*}{$\mathbb{Z}_2$} & $\KK E_7$ & $A_3 \oplus 2A_1 \oplus 2\frak{u}(1)$ &  \multirow{3}{*}{$4C^0$}   \\ \cline{3-4}
&  & $\KK E_5$ & $3A_1 \oplus 2\frak{u}(1)$  &   \\ \cline{3-4}
&  & \text{MN} $E_7$ & $A_3 \oplus 2A_1 \oplus 2\frak{u}(1)$  &   \\ \hline
 \hline
\multirow{3}{*}{$I_5, 2III, I_1$} & \multirow{3}{*}{$-$} & $\KK E_8$ & $A_4 \oplus 2A_1 \oplus 2\frak{u}(1)$ &  \multirow{3}{*}{$\Gamma_0(5)$}  \\ \cline{3-4}
&  & $\KK E_4$ & $2A_1 \oplus 2\frak{u}(1)$  &   \\ \cline{3-4}
&  & \text{MN} $E_7$ & $A_4 \oplus A_1 \oplus 2\frak{u}(1)$  &   \\ \hline
 \hline
\multirow{3}{*}{$I_6, I_2, 2II$} & \multirow{3}{*}{$-$} & $\KK E_7$ & $A_5 \oplus 2\frak{u}(1)$ &  \multirow{3}{*}{$6C^0$}  \\ \cline{3-4}
&  & $\KK E_3$ & $A_1 \oplus 2\frak{u}(1)$  &  \\ \cline{3-4}
&  & \text{MN} $E_8$ & $A_5 \oplus A_1 \oplus 2\frak{u}(1)$  &   \\ \hline
 \hline 
\multirow{3}{*}{$I_7, 2II, I_1$} & \multirow{3}{*}{$-$} & $\KK E_8$ & $A_6 \oplus 2\frak{u}(1)$ &  \multirow{3}{*}{$\Gamma_0(7)$}   \\ \cline{3-4}
&  & $\KK E_2$ & $2\frak{u}(1)$  &  \\ \cline{3-4}
&  & \text{MN} $E_8$ & $A_6 \oplus 2\frak{u}(1)$  & \\ \hline 
\end{tabular}
    \caption{Modular RES with ${\rm rk}(\Phi)=2$ and corresponding field theories.}
    \label{tab:Congruence Rank2}
\end{table}%
Some of these configurations have already been discussed in \cite{Closset:2021lhd, Aspman:2021evt}. We will only discuss here those for which the congruence subgroup is not one of the `standard' groups introduced so far.

\medskip

\begin{figure}[t]
    \centering
    \begin{subfigure}{0.5\textwidth}
    \centering
    \scalebox{0.7}{


\begin{tikzpicture}[scale=2]

    \pgfdeclarelayer{background}
    \pgfsetlayers{background,main}

    \pgfmathsetmacro{\myxlow}{-1}
    \pgfmathsetmacro{\myxhigh}{3}
    \pgfmathsetmacro{\myiterations}{1}

    \draw[-latex](\myxlow-0.1,0) -- (\myxhigh+0.2,0);
    \pgfmathsetmacro{\succofmyxlow}{\myxlow+0.5}
    
    \begin{scope}
        \draw[-stealth, very thin] (0,0)--(0,3.2) node[above] {}; 
        
        \foreach \i in {-0.5,0.5,1.5}
            {\draw[very thin, black] (\i,0.866) arc(120:60:1);
            \draw[very thin, black] (\i,0.866) -- (\i,3);
            }
        \draw[very thin, black] (2.5,0.866) -- (2.5,3);  
        \draw[very thin, black] (-0.5,0.866) arc(60:0:1);
        \draw[very thin, black] (0,0) arc(180:120:1);
        \draw[very thin, black] (0.5,0.866) arc(60:0:1);
        \draw[very thin, black] (1,0) arc(180:120:1);
        
        \draw[very thin, black] (0,0) arc(180:60:1/3);
        \draw[very thin, black] (0.5,0.29) arc(120:0:1/3);
        \draw[very thin, black] (1,0) arc(180:60:1/3);
        
        \draw[very thin, black] (1-0.357, 0.124) arc(180-38.3:0:1/5);
        \draw[very thin, black] (1-0.5, 0.29) arc(60:60-38.3:1/3); 
        
        \draw[very thin, black] (1, 0) arc(180:38.3:1/5); 
        \draw[very thin, black] (2-0.5, 0.29) arc(120:120+38.3:1/3); 
        
        \draw[very thin, black] (2-0.5, 0) arc(0:99:1/8); 
        
        \draw[very thin, black] (1, 0) arc(180:27.7:1/7);
        \draw[very thin, black] (1.357, 0.124) arc(98.3:148:1/8); 

        \draw[very thin, black] (0.5,0.29) -- (0.5,3);
        \draw[very thin, black] (1.5,0) -- (1.5,3);
    
        \clip
                (-0.5,3) 
                {-- (-0.5,0.866) arc(60:0:1)
             -- (0,0) arc(180:60-38.3:1/3)
             -- (1-0.357, 0.124) arc(180-38.3:0:1/5) 
             -- (1,0) arc(180:27.7:1/7)
             -- (1+0.269,0.0666) arc(148:0:1/8)
              -- (1.5,0.866) arc(120:60:1)
             }
                -- (2.5, 3) -- cycle;
            \fill[gray,opacity=0.4] (-1,-1) rectangle (3.5,3);
            
    \end{scope}

    \begin{scope}
        \node at (0.15,2) {$\mathcal{F}$};
        \node at (1,2) {$T\mathcal{F}$};
        \node at (2,2) {$T^2\mathcal{F}$};

        \node at (0,0.75) {$S\mathcal{F}$};
        \node at (1,0.75) {$TS\mathcal{F}$};
        
        \node at (2,0.45) [scale=0.8] {$T^2ST^{-2}S$};
        \draw [->] (1.9,0.3) {} -- (1.55,0.15) {};  

        \fill (0,0)  circle[radius=0.5pt] {};
        \fill (1,0)  circle[radius=0.5pt] {};
        \fill (1.5,0)  circle[radius=0.5pt] {};
        \fill (2,0)  circle[radius=0.5pt];
        
        \node at (0,-0.2) {$2(1,0)$};
        \node at (0.85,-0.2) {$6(-1,1)$};
        \node at (1.7,-0.2) {$(2,-3)$};
    \end{scope}
    
\end{tikzpicture}

}
    \caption{ }
    \end{subfigure}%
    \begin{subfigure}{0.5\textwidth}
    \centering
    \scalebox{0.7}{


\begin{tikzpicture}[scale=2]

    \pgfdeclarelayer{background}
    \pgfsetlayers{background,main}

    \pgfmathsetmacro{\myxlow}{-1}
    \pgfmathsetmacro{\myxhigh}{3}
    \pgfmathsetmacro{\myiterations}{1}

    \draw[-latex](\myxlow-0.1,0) -- (\myxhigh+0.2,0);
    \pgfmathsetmacro{\succofmyxlow}{\myxlow+0.5}
    
    \begin{scope}
        \draw[-stealth, very thin] (0,0)--(0,3.2) node[above] {}; 
    
        \foreach \i in {-0.5,0.5,1.5}
            {\draw[very thin, black] (\i,0.866) arc(120:60:1);
            \draw[very thin, black] (\i,0.866) -- (\i,3);
            }
        \draw[very thin, black] (2.5,0.866) -- (2.5,3);  
        
        \draw[very thin, black] (0.5,0.866) arc(60:0:1);
        \draw[very thin, black] (1,0) arc(180:120:1);
        
        \draw[very thin, black] (0.5,0.29) arc(120:0:1/3);
        \draw[very thin, black] (1,0) arc(180:60:1/3);
        
        \draw[very thin, black] (0.5,0.29) -- (0.5,3);
        \draw[very thin, black] (1.5,0.29) -- (1.5,3);
        
        \draw[very thin, black, dashed] (-0.5,0.866) arc(60:0:1);
        \draw[very thin, black, dashed] (0,0) arc(180:120:1);
        \draw[very thin, black, dashed] (0,0) arc(180:60-38.3:1/3);
        \draw[very thin, black, dashed] (1-0.357, 0.124) arc(180-38.3:0:1/5);
        
        \draw[very thin, black, dash pattern=on 1.5pt off 1pt] (1,0) arc(180:27.7:1/7);
        \draw[very thin, black, dash pattern=on 1.5pt off 1pt] (1+0.269,0.0666) arc(148:0:1/8);
        \draw[very thin, black, dash pattern=on 1.5pt off 1pt] (1.5, 0) -- (1.5, 0.29);
        \draw[very thin, black, dash pattern=on 1.5pt off 1pt] (1, 0) arc(180:38.3:1/5);
        \draw[very thin, black, dash pattern=on 1.5pt off 1pt] (2-0.5, 0.29) arc(120:120+38.3:1/3);

        \clip
                (-0.5,3) 
                {-- (-0.5,0.866) arc(120:60:1)
             -- (0.5, 0.29) arc(120:0:1/3) 
             -- (1,0) arc(180:60:1/3)
              -- (1.5,0.866) arc(120:60:1)
             }
                -- (2.5, 3) -- cycle
            ;
            \fill[gray,opacity=0.4] (-1,-1) rectangle (3.5,3);
            
    \end{scope}
    
    \begin{scope}
            \clip
                (-0.5,0.866) 
                { -- (-0.5,0.866) arc(60:0:1)
             -- (0,0) arc(180:60-38.3:1/3)
             -- (1-0.357, 0.124) arc(180-38.3:0:1/5)
             -- (1,0) arc(0:120:1/3)
             -- (0.5, 0.866) arc (60:120:1)
             }
                -- (-0.5, 0.866) -- cycle;
            \fill[blue, opacity=0.25] (-1,0) rectangle (1,1);
    \end{scope}
    
        \begin{scope}
            \clip
                (1,0) 
                { -- (1,0) arc(180:27.7:1/7)
             -- (1+0.269,0.0666) arc(148:0:1/8)
             -- (1.5, 0.29) arc(60:180:1/3)
             }
                -- (1,0) -- cycle;
           \fill[red, opacity=0.2] (1,0) rectangle (1.5,1);
    \end{scope}
    
    \begin{scope}
        \node at (0.15,2) {$\mathcal{F}$};
        \node at (1,2) {$T\mathcal{F}$};
        \node at (2,2) {$T^2\mathcal{F}$};
        \node at (1,0.75) {$TS\mathcal{F}$};
        
        \fill (0,1)  circle[radius=0.8pt] {};
        \fill (1,0)  circle[radius=0.5pt] {};
        \fill (0,0)  circle[radius=0.5pt] {};
        \fill (1.5,0)  circle[radius=0.5pt] {};
        \fill (2,0)  circle[radius=0.5pt] {};
        \fill (1.5,0.5)  circle[radius=0.8pt] {};
        
        \node at (-0.05,1.2) {$III^{(0)}$};
        \node at (1,-0.2) {$3(-1,1)$};
        \node at (1.85,0.5) {$III_{(1)}$};
        
    \end{scope}
    
\end{tikzpicture}

}
    \caption{ }
    \end{subfigure}
    \caption{Fundamental domains for configurations on the CB of the $\KK E_6$ theory. (a)~The $(I_6,I_3,I_2,I_1)$ configuration, with the BPS states indicated and the modular group $\Gamma^0(3)\cap\Gamma_0(2)$. (b)~The $(2I_3,2III)$ configuration with the  fundamental domain of the modular group $3C^0$ in gray. The coloured regions are `absorbed' into the AD points and can be used to visualize the light BPS states.}
    \label{fig: E6 Configurations}
\end{figure}
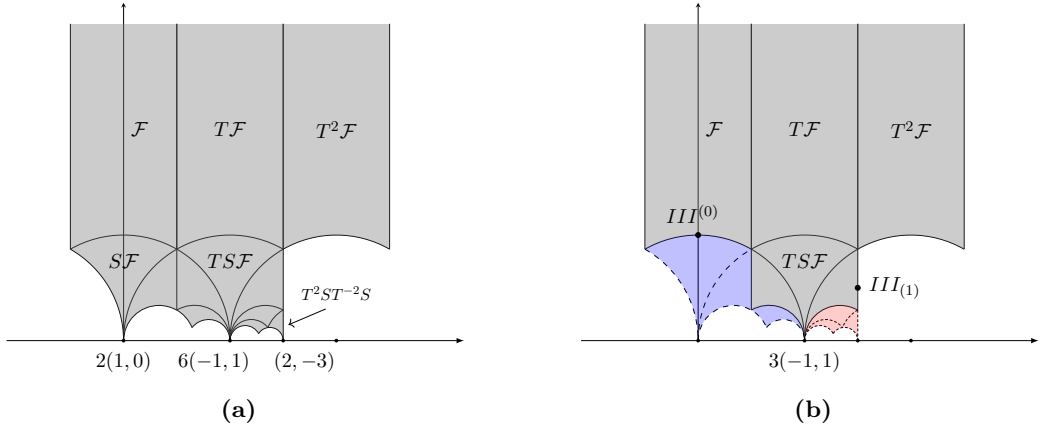

\noindent
\paragraph{$\boldsymbol{(2I_3,2III)}$ configuration for $\KK E_6$.} This configuration appears on the CB of the $\KK E_6$ theory and can be found by considering the branching of the characters $E_6 \rightarrow A_2 \oplus 2A_1 \oplus 2\frak{u}(1)$, with a possibility being given by:
\be
    \boldsymbol{\chi}^{E_6} = \left( {81 \ov 4}, 81, 24, -639, -{567\ov 4}, 0 \right)~,
\ee
In this case, the $j$-invariant becomes:
\be
    j(U) = {(U-3)^3(U+9)^3 \ov (U+6)^3}~,
\ee
with the $I_3$ cusp at $U_0 = -6$ and the type $III$ singularities at $U_{\pm} = 3(1\pm 2\sqrt{3})$. One of the roots of the above equation is:
\be
    U(\tau) = -6 + \left({\eta(\tau)^2 \ov  \eta\left({\tau\ov 3}\right) \eta(3\tau)}\right)^6~,
\ee
whose series expansion reproduces the McKay-Thompson series of class $9A$. For this solution, using the modular properties of the Dedekind eta-function, one can check that the $I_3$ cusp corresponds to $\tau_0 = 1$, while the elliptic points of type $III$ are in the orbits of $\tau_{\pm} = i$ and ${3+i\ov 5}$ (or, equivalently, of ${3+i \ov 2}$). In particular, we see that:
\be
    TST^2 (i) = {3+i\ov 5}~, \qquad TST^{-1} (i) =  {3+i \ov 2}~.
\ee
which then leads to a choice of a fundamental domain for the group $3C^0$, where we follow the notation of \cite{Cummins2003}. We denote the elliptic point at $\tau = i$ by $III^{(0)}$, following \eqref{III, II Notation}, while the remaining elliptic point will be referred to as $III_{(1)}$. The fundamental domains of the subgroups of ${\rm PSL}(2,\mathbb{Z})$ can be equivalently understood using permutations, as discussed in \cite{STROMBERG2019436}, for example, and in appendix \ref{Appendix Permutations}. In their notation, this subgroup is:
\be
    3C^0 \cong \Gamma(6; 0, 2, 2, 0; 3)~.
\ee
In fact, from the data available in \cite{STROMBERG2019436}, one can easily see that the fundamental domain agrees with our findings. This domain is shown in figure \ref{fig: E6 Configurations}. To show that this is indeed a congruence subgroup one needs to check that $\Gamma(3) \subset 3C^0$. To do this, consider the monodromies around the $U$-plane singularities, namely:
\be
    M_{\infty} = T^3~, \quad M_{I_3} = T^5S T^3 (T^5S)^{-1}~,\quad M_{III^{(0)}} = S^{-1}~, \quad M_{III_{(1)}} = TST^{-1}(S^{-1})(TST^{-1})^{-1}~,
\ee
which generate the group $3C^0$. Then, the subgroup generated by:
\be
    M_{\infty}~, \qquad M_{\infty}^{-3} M_{I_3}M_{\infty}^{3}~, \qquad M_{III^{(0)}}^{-1}M_{\infty}M_{III^{(0)}}~,
\ee
is indeed the principal congruence subgroup of level three $\Gamma(3)$. %
We can understand the BPS states of this configuration from the $(I_6, I_3, I_2, I_1)$ configuration of $\KK E_6$, discussed in \cite{Closset:2021lhd}. Recall that the BPS states becoming massless at the cusps of the latter configuration were found to be:
\be
    I_1~:~ 2(1,0)~, \qquad I_6~:~ 6(-1,1)~, \qquad I_1~:~(2,-3)~,
\ee
with the monodromies generating the group $\Gamma^0(3)\cap \Gamma_0(2) \cong \Gamma^0(6)$, and the fundamental domain shown in figure~\ref{fig: E6 Configurations}. From these states, we notice that:
\be
    M_{III^{(0)}} = S^{-1} = M_{(1,0)}^2 M_{(-1,1)}~, \qquad M_{III_{(1)}} = (TST^{-1})S^{-1}(TST^{-1})^{-1} = M_{(-1,1)}^2 M_{(2,-3)}~,
\ee
and, thus, these states (and the corresponding BPS quiver) offer a good description for the $(2I_3, 2III)$ configuration.  %
\medskip

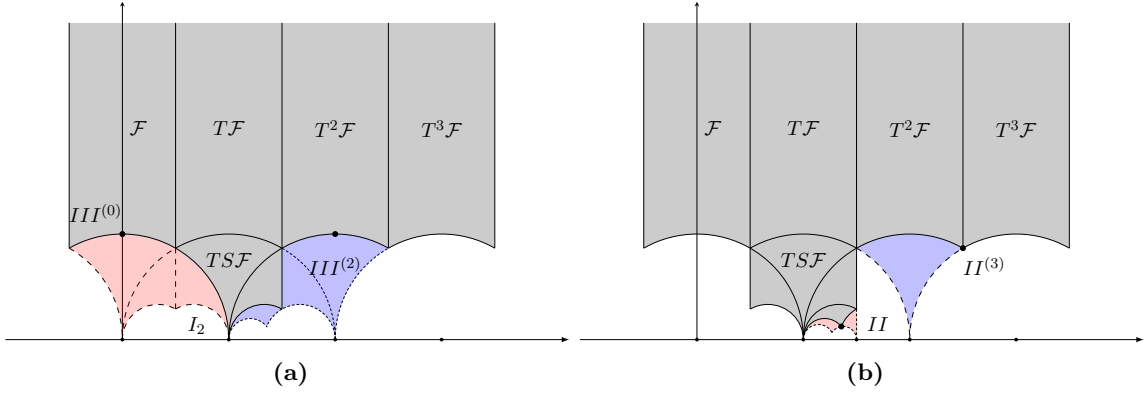
\begin{figure}[t]
    \centering
    \begin{subfigure}{0.5\textwidth}
    \centering
    \scalebox{0.7}{


\begin{tikzpicture}[scale=2]

    \pgfdeclarelayer{background}
    \pgfsetlayers{background,main}

    \pgfmathsetmacro{\myxlow}{-1}
    \pgfmathsetmacro{\myxhigh}{4}
    \pgfmathsetmacro{\myiterations}{1}
    
    \draw[-latex](\myxlow-0.1,0) -- (\myxhigh+0.2,0);
    \pgfmathsetmacro{\succofmyxlow}{\myxlow+0.5}
    
    \draw[-stealth, very thin] (0,0)--(0,3.2) node[above] {}; 
    
    \begin{scope}
        \draw[very thin, black, dashed] (-0.5,0.866) arc(60:0:1);
        \draw[very thin, black, dashed] (0,0) arc(180:120:1);
        \draw[very thin, black, dashed] (0,0) arc(180:60:1/3);
        \draw[very thin, black, dashed] (0.5,0.29) arc(120:0:1/3);
        \draw[very thin, black, dashed] (0.5,0.29) -- (0.5, 0.866);
        
            \clip
                (-0.5,0.866) 
                { -- (-0.5,0.866) arc(60:0:1)
             -- (0,0) arc(180:60:1/3)
             -- (0.5,0.29) arc(120:0:1/3)
             -- (1, 0) arc(0:120:1)
             }
                -- (-0.5, 0.866) -- cycle;
            \fill[red,opacity=0.2]  (-1,-1) rectangle (4.5,3);
    \end{scope}

    \begin{scope}
        \draw[very thin, black, dash pattern=on 1.5pt off 1pt] (2-0.5,0.866) arc(60:0:1);
        \draw[very thin, black, dash pattern=on 1.5pt off 1pt] (2,0) arc(180:120:1);
        \draw[very thin, black, dash pattern=on 1.5pt off 1pt] (2,0) arc(0:120:1/3);
        \draw[very thin, black, dash pattern=on 1.5pt off 1pt] (1,0) arc(180:180-144:0.2);
        
        \draw[very thin, black, dash pattern=on 1.5pt off 1pt] (1+0.5, 0.29) arc(120:120+38.3:1/3);
        
        \clip
                (1,0) 
                { -- (1,0) arc(180:180-144:0.2)
             -- (1+0.357, 0.124) arc(120+38.3:0:1/3)
             -- (2,0) arc(180:120:1)
             -- (2.5, 0.866) arc(60:120:1)
             -- (1.5, 0.29) arc(60:180:1/3)
             }
                -- (1,0) -- cycle;
        \fill[blue,opacity=0.25]  (-1,-1) rectangle (4.5,3);
    \end{scope}
    
    \begin{scope}
        \draw[very thin, black] (0.5,0.866) arc(120:60:1);
        \draw[very thin, black] (1,0) arc(0:120:1);
        \draw[very thin, black] (1,0) arc(180:60:1);
        \draw[very thin, black] (2.5,0.866) arc(120:60:1);
        
        \draw[very thin, black] (1,0) arc(180:60:1/3);
        
        \draw[very thin, black] (0.5,0.866) -- (0.5, 3);
        \draw[very thin, black] (-0.5,0.866) -- (-0.5, 3);
        \draw[very thin, black] (1.5,0.29) -- (1.5, 3);
        \draw[very thin, black] (2.5,0.866) -- (2.5, 3);
        \draw[very thin, black] (3.5,0.866) -- (3.5, 3);
        
    \end{scope}

    \begin{scope}
        \begin{pgfonlayer}{background}
            \clip
                (-0.5,3) 
                { -- (-0.5,0.866) arc(120:0:1)
             -- (1,0) arc(180:60:1/3)
             -- (1.5,0.866) arc(120:60:1)
             -- (2.5, 0.866) arc(120:60:1)
             }
                -- (3.5, 3) -- cycle;
            \fill[gray,opacity=0.4] (-1,-1) rectangle (4.5,3);
        \end{pgfonlayer}
    \end{scope}

    \begin{scope}
        \node at (0.15,2) {$\mathcal{F}$};
        \node at (1,2) {$T\mathcal{F}$};
        \node at (2,2) {$T^2\mathcal{F}$};
        \node at (3,2) {$T^3\mathcal{F}$};
        
        \node at (1,0.75) {$TS\mathcal{F}$};
        
        \node at (0-0.25,1.2) {$III^{(0)}$};
        \node at (1-0.3,0.13) {$I_2$};
        \node at (2+0,0.7) {$III^{(2)}$};

        \fill (0,1)  circle[radius=0.8pt];
        \fill (1,0)  circle[radius=0.5pt];
        \fill (2,0)  circle[radius=0.5pt];
        \fill (0,0)  circle[radius=0.5pt];
        \fill (3,0)  circle[radius=0.5pt];
        \fill (2,1)  circle[radius=0.8pt];

    \end{scope}
    
\end{tikzpicture}

}
    \caption{ }
    \label{fig: E5 I4-2III-I2}
    \end{subfigure}%
    \begin{subfigure}{0.5\textwidth}
    \centering
    \scalebox{0.7}{


\begin{tikzpicture}[scale=2]

    \pgfdeclarelayer{background}
    \pgfsetlayers{background,main}

    \pgfmathsetmacro{\myxlow}{-1}
    \pgfmathsetmacro{\myxhigh}{4}
    \pgfmathsetmacro{\myiterations}{1}
    
    \draw[-latex](\myxlow-0.1,0) -- (\myxhigh+0.2,0);
    \pgfmathsetmacro{\succofmyxlow}{\myxlow+0.5}
    
    \draw[-stealth, very thin] (0,0)--(0,3.2) node[above] {}; 
    
    \begin{scope}
            \clip
                (1,0) 
               { -- (1,0) arc(180:38.3:1/5)
               -- (1+0.357, 0.124) arc(120+38.3:120:1/3)
               -- (1.5, 0) arc(0:148:1/8)
               -- (1.269,0.0666) arc(27.7:180:1/7)
           }
              -- (1, 0) -- cycle;
            \fill[red,opacity=0.2] (-1,-1) rectangle (4.5,3);
    \end{scope}

    \begin{scope}
            \clip
                (2-0.5,3) 
                { -- (2-0.5,0.866) arc(60:0:1)
             -- (2,0) arc (180:120:1)
             -- (2.5,0.866) arc(60:120:1)
             }
                -- (1.5, 3) -- cycle;
            \fill[blue,opacity=0.25] (-1,-1) rectangle (4.5,3);
    \end{scope}
    
    \begin{scope}
    
    \foreach \i in {-0.5,0.5,1.5,2.5}
            {\draw[very thin, black] (\i,0.866) arc(120:60:1);
            \draw[very thin, black] (\i,0.866) -- (\i,3);
            }

            \draw[very thin, black] (3.5,0.866) -- (3.5,3);
            \draw[very thin, black] (1-0.5,0.866) arc(60:0:1);
            \draw[very thin, black] (1,0) arc(180:120:1);
            \draw[very thin, black] (1,0) arc(180:60:1/3);
            \draw[very thin, black] (2-0.5,0.29) -- (2-0.5, 0.866);
            \draw[very thin, black] (1-0.5,0.29) -- (1-0.5, 0.866);
            \draw[very thin, black] (1,0) arc(0:120:1/3);
            
            \draw[very thin, black] (1, 0) arc(180:38.3:1/5); 
            \draw[very thin, black] (-0.5+2, 0.29) arc(120:120+38.4:1/3); 
            
            \draw[very thin, black, dashed] (2-0.5,0.866) arc(60:0:1);
            \draw[very thin, black, dashed] (2,0) arc(180:120:1);
            
            \draw[very thin, black, dash pattern=on 1.5pt off 1pt] (2-0.5,0) -- (2-0.5, 0.29);
            \draw[very thin, black, dash pattern=on 1.5pt off 1pt] (2-0.5,0) arc(0:148:1/8);
            \draw[very thin, black, dash pattern=on 1.5pt off 1pt] (1,0) arc(180:27.7:1/7);
            
    \end{scope}

    \begin{scope}
        \begin{pgfonlayer}{background}
            \pgfmathsetmacro{\myradius}{pow(1/3,1)}
            \clip
                (-0.5,3) 
                { -- (-0.5,0.866) arc(120:60:1)
             -- (0.5,0.29) arc(120:0:1/3)
             -- (1, 0) arc(180:38.3:1/5)
             -- (1+0.357, 0.124) arc(120+38.3:120:1/3)
             -- (1.5,0.866) arc(120:60:1)
             -- (2.5,0.866) arc(120:60:1)
             }
                -- (3.5, 3) -- cycle
            ;
            \fill[gray,opacity=0.4] (-1,-1) rectangle (3.5,3);
        \end{pgfonlayer}
    \end{scope}

    \begin{scope}
        \node at (0.15,2) {$\mathcal{F}$};
        \node at (1,2) {$T\mathcal{F}$};
        \node at (1,0.75) {$TS\mathcal{F}$};

        \foreach \i in {2,3}
            {\node at (\i,2) {$T^{\i}\mathcal{F}$};
            \fill (\i,0)  circle[radius=0.5pt];
            }
        
        \node at (1.7,0.15) {$II$};
        \node at (2.7,0.7) {$II^{(3)}$};

        \fill (0,0)  circle[radius=0.5pt];
        \fill (1,0)  circle[radius=0.5pt];
        \fill (1.5,0)  circle[radius=0.5pt];
        \fill (1+0.357, 0.124)  circle[radius=0.8pt];
        \fill (2.5,0.866)  circle[radius=0.8pt];
        
    \end{scope}
    
\end{tikzpicture}

}
    \caption{ }
    \label{fig: E5 2I4-2II}
    \end{subfigure}
    \caption{Modular configurations of $\KK E_5$ theory. $(a)~(I_4,2III,I_2)$ configuration with the monodromy group $4C^0$. $(a)~(I_4,2I_2, 2II)$ configuration with monodromy group $4D^0$. Fundamental domains are drawn in gray, while coloured regions are absorbed in the AD points.}
\end{figure}

\noindent
\paragraph{$\boldsymbol{(I_4,2III, I_2)}$ configuration for $\KK E_5$.} This configuration appears on the CB of the $\KK E_5$ theory, as well as on the CB of $\KK E_7$. Focusing on the former case, this can be seen by setting the characters to:
\be
    \boldsymbol{\chi}^{E_5} = \left( 10, 13, 0, -72, 0 \right)~.
\ee
In this case the discriminant and $j$-invariant become:
\be
    \Delta(U) = U^2 (U^2 - 64)^3~, \qquad \quad j(U) = {(U^2 - 16)^3 \ov U^2}~.
\ee
One root of the equation $j = j(\tau)$ is then given by:
\be
    U(\tau) = \left({ \eta(\tau)^2 \ov \eta\left( {\tau \ov 2}\right) \eta(2\tau)}\right)^{12}~,
\ee
whose series expansion reproduces the McKay-Thompson series of class $8B$ of the Monster group. Additionally, under a $T$ transformation, we recover the $4D$ series. One can then check that the type $III$ singularities of the $U$-plane at $U_{\pm} = \pm 8$ correspond to $\tau_+ = i$ and $\tau_- = 2+ i$. Hence, we will denote them by $III^{(0)}$ and $III^{(2)}$, respectively, following \eqref{III, II Notation}. Moreover, the $I_2$ cusp lies at $\tau = 1$. The monodromy group in this case is referred to as $4C^0$ in \cite{Cummins2003}, with the fundamental domain drawn in figure~\ref{fig: E5 I4-2III-I2}. Note that in the notation of \cite{STROMBERG2019436}, this group is:
\be
    4C^0 \cong \Gamma(6; 0,2, 2,0; 4)~.
\ee
Let us also note that the representative chosen in figure~\ref{fig: E5 I4-2III-I2} from this conjugacy class is an index $2$ subgroup of $\Gamma_{\theta} \cong \Gamma^0(2)$. This can also be seen at the level of the $J$-invariant, from which one can show explicitly that:
\be
    U(\tau)^2 = - \left( { \eta\left( {\tau + 1 \ov 2}\right) \ov \eta(\tau + 1) } \right)^{24}~,
\ee
where we recognise on the RHS the Hauptmodul for $\Gamma_{\theta}$. It is also straightforward to show that $\Gamma(4) \subset 4C^0$. The BPS states becoming massless at the different singularities can be understood from the $(2I_4,2I_2)$ configuration described in the previous section, namely:
\be
    I_2~:~2(1,0)~, \qquad I_4~:~ 4(-1,1)~, \qquad I_2~:~2(1,-2)~,
\ee
with the fundamental domain of $\Gamma^0(4) \cap \Gamma(2)$ drawn in figure~\ref{fig: Gamma04Gamma2}. We then note that:
\be
    M_{III^{(0)}} = S^{-1} = M_{(1,0)}^2 M_{(-1,1)}~, \qquad M_{III^{(2)}} = T^2S^{-1}T^{-2}= M_{(-1,1)}M_{(1,-2)}^2~,
\ee
which is also graphically depicted in figure~\ref{fig: E5 I4-2III-I2}. %
Note that this identification holds since $\Gamma(4) \subset \Gamma^0(4) \cap \Gamma(2) \subset 4C^0$, as also pointed out in \cite{Bourget:2017goy}.

\medskip

\noindent
\paragraph{$\boldsymbol{(2I_4,2II)}$ configuration for $\KK E_5$.} A different configuration appearing on the CB of the $\KK E_5$ theory is $(2I_4, 2II)$, found for the following values of the characters: 
\be
    \boldsymbol{\chi}^{E_5} = \left( -{19 i\ov2}, -31, 8(1+i), 30i, -4(1-i)\right)~.
\ee
In this case the bulk $I_4$ singularity resides at $U_{I_4} = 3(1-i)$, while the two type $II$ singularities are at $U_{II_{(1,2)}} = (1-i)( -2 \pm 3\sqrt{3})$. The monodromy group is referred to as $4D^0$ in \cite{Cummins2003} or as $\Gamma(8; 0,2,0,2; 4)$ in \cite{STROMBERG2019436}.

This configuration is related to the $(I_4, 2III, II)$ configuration studied in the next subsection, which corresponds to the congruence subgroup $4A^0$. In particular, for the values of the characters used in \eqref{E5 characters I4 2III II}, we find:
\be
    U_{4A^0} = {i (4i - 3(1-i)U_{4D^0} + U_{4D^0}^2) \ov -3(1-i) + U_{4D^0}}~.
\ee
This quadratic equation can be solved explicitly, leading to:
\be
    U_{4D^0} = -{i\ov 2} \left(3(1+i) + U_{4A^0} \pm \sqrt{34i - 6(1+i)U_{4A^0} + U_{4A^0}^2} \right)~.
\ee
It is in fact not difficult to show that $4D^0 \subset 4A^0$, which is what the above computation suggests. Using the expression for the modular function of $4A^0$, the square root resolves, with the solution that has the correct asymptotics reading:
\be \label{4D0}
    U_{4D^0} = -(1+i) {\vartheta_2^6 - i\vartheta_3^6 + \vartheta_4^6 - i\vartheta_3^4 \vartheta_4^2 + \vartheta_3^2 \vartheta_4^4 - \vartheta_2^4(\vartheta_3^2 + i \vartheta_4^2) + i\vartheta_2^2(\vartheta_3^4 - \vartheta_3^2 \vartheta_4^2 +\vartheta_4^4) \ov \vartheta_2^2 \vartheta_3^2 \vartheta_4^2}~.
\ee
The choice of fundamental domain consistent with the previous configurations is shown in figure~\ref{fig: E5 2I4-2II}. Here, the two $II$ elliptic points lie at $T^3 e^{2\pi i \ov 3}$ and $TST^2 e^{2\pi i \ov 3}$, with the former denoted by $II^{(3)}$ in figure~\ref{fig: E5 2I4-2II}, following the notation introduced in \eqref{III, II Notation}. It is again a non-trivial task to find a mutation sequence that reproduces this configuration. In particular, starting with the $(2I_4, 2I_2)$ configuration, it becomes clear that the bulk $I_4$ singularity needs to be split in order to obtain the type $II$ singularities, as there is no physical way to realise a transition of the type $I_2 \rightarrow II$. Thus, we propose the mutation sequence: $\hat{\gamma}_3 - \hat{\gamma}_1 - \hat{\gamma}_{4,5}$ on the quiver \eqref{quiver E5 orbi}. This has the effect of splitting the original width-four cusp into $I_4 \rightarrow I_3 \oplus I_1 \rightarrow I_2 \oplus 2I_1$, with the resulting $I_2$ cusp merging with that corresponding to the $\gamma_{4,5}$ states. The resulting BPS quiver becomes:
\bea\label{quiver E4 2I4 2II}
 \begin{tikzpicture}[baseline=1mm]
\node[] (6) []{$\CE_{\gamma_{6}=(-1,1)}$};
\node[] (9) [left = of 6] {};
\node[] (10) [right = of 6] {};
\node[] (1) [below = of 9]{$\CE_{\gamma_{1}=(0,-1)}$};
\node[] (2) [below = of 10]{$\CE_{\gamma_{2}=(-2,3)}$};
\node[] (8) [below = of 1]{$\CE_{\gamma_{4,5,7,8}=(1,-1)}$};
\node[] (3) [below = of 2]{$\CE_{\gamma_{3}=(-1,2)}$};
\draw[->-=0.5] (1) to   (8);
\draw[->-=0.5] (8) to   (3);
\draw[->-=0.5] (3) to   (2);
\draw[->>-=0.5] (2) to   (1);
\draw[->-=0.3] (8) to   (2);
\draw[->-=0.3] (3) to   (1);
\draw[->-=0.7] (3) to   (6);
\draw[->-=0.5] (2) to   (6);
\draw[->-=0.5] (6) to   (1);
\end{tikzpicture}
\eea
Using these basis states and the logic of \eqref{II Decomposition}, at the type $II$ singularities the following mutually non-local states become massless:
\bea
    T^3 e^{2\pi i\ov 3} \qquad    & II^{(3)}~:~ & (0, -1) \oplus (-1,2)~,\\
    TST^2 e^{2\pi i\ov 3} \qquad    & II~:~ &  (-2, 3) \oplus (-1,1) ~.\\
\eea

\medskip

\noindent
\paragraph{$\boldsymbol{(I_5,2III, I_1)}$ configuration for $\KK E_4$.} We will now consider some configurations appearing on the CB of the $\KK E_4$ theory. The massless configuration has already been studied in \cite{Closset:2021lhd}. It was shown that its monodromy group is $\Gamma^1(5)$, with the BPS quiver given by:
\bea\label{quiver E4}
 \begin{tikzpicture}[baseline=1mm]
\node[] (1) []{$\CE_{\gamma_1=(1,0)}$};
\node[] (2) [right = of 1]{$\CE_{\gamma_{2}=(-2,5)}$};
\node[] (3) [below = of 1]{$\CE_{\gamma_{3,4,5,6,7}=(1,-3)}$};
\draw[->>>>>-=0.5] (1) to   (2);
\draw[->-=0.3] (2) to   (3);
\draw[->>>-=0.5] (3) to   (1);
\end{tikzpicture}
\eea
We first consider the $(I_5,2III, I_1)$ configuration, which can be obtained for:
\be
    \boldsymbol{\chi}^{E_4} = \left( {17\ov 4}, 9, 6, 10\right)~,
\ee
with the discriminant and $j$-invariant:
\be
    \Delta(U) = (U-8)(U^2 + 6U + 13)^2~, \qquad j(U) = { (U^2 - 6U - 11)^2 \ov U-8}~.
\ee
Hence, the $I_1$ cusp resides at $U = 8$, while the two $III$ singularities are at $U_{\pm} = -3 \pm 2i$. One finds that:
\be
    U(\tau) = 8 + \left( {\eta\left( {\tau\ov 5}\right) \ov \eta(\tau)}\right)^6~,
\ee
with the $q$-series reproducing the McKay-Thompson series of class $5B$ of the Monster group. It is then straightforward to check that the $I_1$ cusp corresponds to $\tau = 0$, while the two type $III$ singularities are the elliptic points $\tau_- = 2 + i$ and $\tau_+ = 3+i$, respectively. Here $U(\tau)$ is chosen in agreement with the massless configuration. It is in fact not difficult to see by following the $U$-plane singularities that there is a deformation pattern from the massless configuration to $(I_5, 2III, I_1)$ that leaves the $I_1$ cusp at $\tau = 0$ invariant. 

\begin{figure}[t]
    \centering
    \begin{subfigure}{0.5\textwidth}
    \centering
    \scalebox{0.6}{


\begin{tikzpicture}[scale=2]

    \pgfdeclarelayer{background}
    \pgfsetlayers{background,main}

    \pgfmathsetmacro{\myxlow}{-1}
    \pgfmathsetmacro{\myxhigh}{5}
    \pgfmathsetmacro{\myiterations}{1}

    \draw[-latex](\myxlow-0.1,0) -- (\myxhigh+0.2,0);
    \pgfmathsetmacro{\succofmyxlow}{\myxlow+0.5}
    
    \draw[-stealth, very thin] (0,0)--(0,3.2) node[above] {}; 
    \begin{scope}
        
        \draw[very thin, black] (-0.5,0.866) arc(60:0:1);
        \draw[very thin, black] (-0.5,0.866) arc(120:60:1);
        \draw[very thin, black] (0,0) arc(180:60:1);
        
        \draw[very thin, black] (2.5,0.866) arc(120:60:1);
        
        \draw[very thin, black] (3,0) arc(0:120:1);
        \draw[very thin, black] (3,0) arc(180:60:1);   
        
        \draw[very thin, black] (3,0) arc(180:60:1/3);
        \draw[very thin, black] (2.5,0.29) arc(120:0:1/3);
        
        \draw[very thin, black] (-0.5+3, 0) arc(180:180-98.3:1/8); 
        \draw[very thin, black] (3-0.357, 0.124) arc(180-38.3:0:1/5); 
        \draw[very thin, black] (3, 0) arc(180:38.3:1/5);
        
        \draw[very thin, black] (1.5,0.866) -- (1.5, 3);
        \draw[very thin, black] (0.5,0.866) -- (0.5, 3);
        \draw[very thin, black] (-0.5,0.866) -- (-0.5, 3);
        \draw[very thin, black] (2.5,0) -- (2.5, 3);
        \draw[very thin, black] (3.5,0.29) -- (3.5, 3);
        \draw[very thin, black] (4.5,0.866) -- (4.5, 3);
        
        \draw[very thin, black] (-0.5+3, 0.29) arc(60:60-38.3:1/3); 
        \draw[very thin, black] (-0.5+4, 0.29) arc(120:180-60+38.3:1/3);
        
    \end{scope}

    \begin{scope}
        \begin{pgfonlayer}{background}
            \clip
                (-0.5,3) 
                { -- (-0.5,0.866)
             -- (-0.5,0.866) arc(60:0:1)
             -- (0,0) arc(180:60:1)
             -- (1.5,0.866) arc(120:60:1)
             -- (2.5,0) arc(180:180-98.3:1/8)
             -- (3-0.357, 0.124) arc(180-38.3:0:1/5)
             -- (3,0) arc(180:38.3:1/5)
             -- (3+0.357, 0.124) arc(180-60+38.3:120:1/3)
             -- (3.5, 0.866) arc(120:60:1)
             }
                -- (4.5, 3) -- cycle
            ;
            \fill[gray,opacity=0.4] (-1,-1) rectangle (4.5,3);
        \end{pgfonlayer}
    \end{scope}

    \begin{scope}
        \node at (0.15,2) {$\mathcal{F}$};
        \node at (1,2) {$T\mathcal{F}$};
        \node at (2,2) {$T^2\mathcal{F}$};
        \node at (3,2) {$T^3\mathcal{F}$};
        \node at (4,2) {$T^4\mathcal{F}$};
        
        \node at (0,0.75) {$S\mathcal{F}$};
        \node at (3,0.75) {$T^3S\mathcal{F}$};

        \node at (2,0.37) [scale=0.7] {$T^3ST^{2}S$};
        \draw [->] (2.25,0.28) {} -- (2.47,0.15) {};  
        
        \fill (0,0)  circle[radius=0.5pt];
        \fill (1,0)  circle[radius=0.5pt];
        \fill (2,0)  circle[radius=0.5pt];
        \fill (3,0)  circle[radius=0.5pt];
        \fill (4,0)  circle[radius=0.5pt];
        \fill (2.5,0)  circle[radius=0.5pt];
        
    \end{scope}
    
\end{tikzpicture}

}
    \caption{ }
    \label{fig: E4 Gamma15}
    \end{subfigure}%
    \begin{subfigure}{0.5\textwidth}
    \centering
    \scalebox{0.6}{


\begin{tikzpicture}[scale=2]

    \pgfdeclarelayer{background}
    \pgfsetlayers{background,main}

    \pgfmathsetmacro{\myxlow}{-1}
    \pgfmathsetmacro{\myxhigh}{5}
    \pgfmathsetmacro{\myiterations}{1}
    
    \draw[-latex](\myxlow-0.1,0) -- (\myxhigh+0.2,0);
    \pgfmathsetmacro{\succofmyxlow}{\myxlow+0.5}
    
    \draw[-stealth, very thin] (0,0)--(0,3.2) node[above] {}; 
    
    \begin{scope}
            \clip
                (1.5,0.866) 
                { -- (1.5,0.866) arc(60:0:1)
             -- (2,0) arc(180:60:1/3)
             -- (2.5,0.29) arc(120:0:1/3)
             -- (3,0) arc(0:60:1)
             -- (2.5, 0.866) arc(60:120:1)
             }
                -- (1.5, 0.866) -- cycle
            ;
            \fill[blue,opacity=0.25] (0,0) rectangle (4,1);
    \end{scope}

        \begin{scope}
            \clip
                (1+1.5,0.866) 
                { -- (1+1.5,0.866) arc(60:0:1)
             -- (1+2,0) arc(180:60:1/3)
             -- (1+2.5, 0.866) arc(60:120:1)
             }
                -- (1+1.5, 0.866) -- cycle
            ;
            \fill[red,opacity=0.2] (0,0) rectangle (4,1);
    \end{scope}
    
    \begin{scope}
    
    \foreach \i in {-0.5,0.5,1.5,2.5,3.5}
            {\draw[very thin, black] (\i,0.866) arc(120:60:1);
            \draw[very thin, black] (\i,0.866) -- (\i,3);
            }

            \draw[very thin, black] (4.5,0.866) -- (4.5,3);
            
            \draw[very thin, black] (-0.5,0.866) arc(60:0:1);
            \draw[very thin, black] (0,0) arc(180:120:1);
            
            \draw[very thin, black, dashed] (2-0.5,0.866) arc(60:0:1);
            \draw[very thin, black, dashed] (2,0) arc(180:120:1);
            \draw[very thin, black, dashed] (2,0) arc(180:60:1/3);
            \draw[very thin, black, dashed] (3-0.5,0.29) -- (3-0.5, 0.866);
            
            \draw[very thin, dash pattern=on 1.5pt off 1pt] (3,0) arc(180:60:1/3);
            \draw[very thin, dash pattern=on 1.5pt off 1pt] (4-0.5,0.29) -- (4-0.5, 0.866);
            
            \draw[very thin, dash pattern=on 1.5pt off 1pt] (3-0.5,0.866) arc(60:0:1);
            \draw[very thin, dash pattern=on 1.5pt off 1pt] (3,0) arc(180:120:1);
            \draw[very thin, dashed] (3,0) arc(0:120:1/3);

    \end{scope}

    \begin{scope}
        \begin{pgfonlayer}{background}
            \pgfmathsetmacro{\myradius}{pow(1/3,1)}
            \clip
                (-0.5,3) 
                { -- (-0.5,0.866) arc(60:0:1)
             -- (0,0) arc(180:60:1)
             -- (1.5,0.866) arc(120:60:1)
             -- (2.5,0.866) arc(120:60:1)
             -- (3.5,0.866) arc(120:60:1)
             }
                -- (4.5, 3) -- cycle
            ;
            \fill[gray,opacity=0.4] (-1,-1) rectangle (4.5,3);
        \end{pgfonlayer}
    \end{scope}
    
    \begin{scope}
        \node at (0.15,2) {$\mathcal{F}$};
        \node at (1,2) {$T\mathcal{F}$};
        \node at (0,0.75) {$S\mathcal{F}$};

        \foreach \i in {2,3,4}
            {\node at (\i,2) {$T^{\i}\mathcal{F}$};
            \fill (\i,0)  circle[radius=0.5pt];
            }
        
        \node at (2,1.3) {$III^{(2)}$};
        \node at (3,1.3) {$III^{(3)}$};

        \fill (0,0)  circle[radius=0.5pt];
        \fill (1,0)  circle[radius=0.5pt];
        
        \fill (2,1)  circle[radius=1pt];
        \fill (3,1)  circle[radius=1pt];

    \end{scope}
    
\end{tikzpicture}

}
    \caption{ }
    \label{fig: E4 Gamma05}
    \end{subfigure}
    \caption{Fundamental domains for configurations on the CB of the $\KK E_4$ theory. (a)~ The $(2I_5, 2I_1)$ massless configuration, with modular group $\Gamma^1(5)$ discussed in \cite{Closset:2021lhd}. (b)~The $(I_5,2III, I_1)$ with modular group $\Gamma^0(5)$ shown in gray. The colored regions are `absorbed' into the AD points.}
\end{figure}
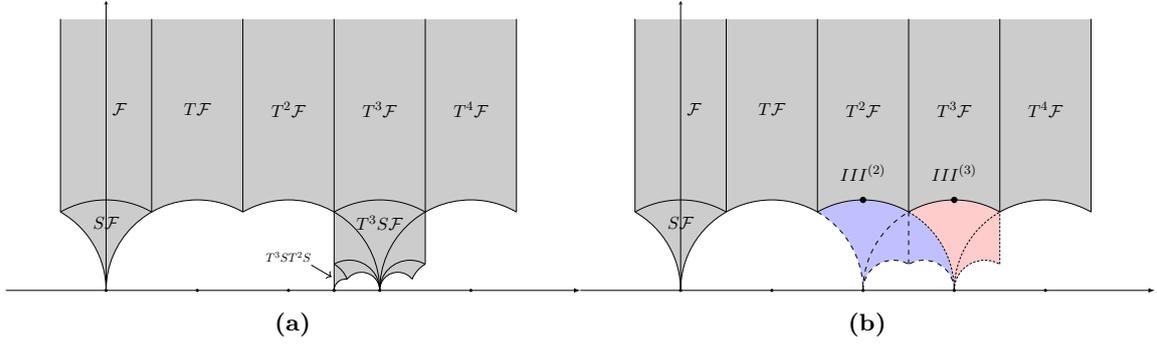%
The monodromy group is $\Gamma^0(5)$, with the fundamental domain shown in figure~\ref{fig: E4 Gamma05}. Let us point out that this configuration can be obtained for other values of the characters as well. In terms of the gauge theory parameters, a solution consistent with the above values is given by:
\be
    \left(\lambda, M_1, M_2, M_3 \right) = \left(4 + 4i, 1, 1, -i \right)~.
\ee
Setting $M_1 = M_2 = M$, one generically has the $(I_5, I_2, 5I_1)$ configuration, further enhanced to the $(I_5, 2I_2, 3I_1)$ configuration by setting $M = 1$. Then, for the above values of the parameters one recovers the $(I_5,2III, I_1)$ configuration. 
The light BPS states can be understood starting from the massless configuration as follows. Performing the mutation sequence $\hat{\gamma}_{3,4}-\hat{\gamma}_2$ on \eqref{quiver E4} we find:
\bea\label{quiver E4 I5 2III I1}
 \begin{tikzpicture}[baseline=1mm]
\node[] (1) []{$\CE_{\gamma_1=(1,0)}$};
\node[] (2) [right = of 1]{$\CE_{\gamma_{2}=(0,1)}$};
\node[] (7) [below = of 1]{$\CE_{\gamma_{5,6,7}=(1,-3)}$};
\node[] (3) [below = of 2]{$\CE_{\gamma_{3,4}=(-1,2)}$};
\draw[->-=0.5] (1) to   (2);
\draw[->-=0.5] (2) to   (3);
\draw[->-=0.5] (3) to   (7);
\draw[->>>-=0.5] (7) to   (1);
\draw[->-=0.3] (7) to   (2);
\draw[->>-=0.3] (1) to   (3);
\end{tikzpicture}
\eea
The type $III$ singularities are then formed from:
\bea
    III^{(2)}~:~ & \qquad 2(-1,2) \oplus (1,-3)~, \\
    III^{(3)}~:~ & \qquad (1,-3) \oplus (0,1) \oplus (1,-3)~, 
\eea 
conform \eqref{III Decomposition} and \eqref{III Decomposition v2}, which are indeed of type $I_2 \oplus I_1$.

\medskip

\noindent
\paragraph{$\boldsymbol{(I_6, I_2, 2II)}$ configuration for $\KK E_3$.} As a last configuration with ${\rm rk}(\Phi) = 1$, let us discuss the $(I_6, I_2, 2II)$ configuration with the $I_6$ fibre fixed at infinity. As showed in \cite{Closset:2021lhd}, this configuration can be found on the CB of the $\KK E_3$ theory, for:
\be
    \lambda = 1~, \qquad M_1 = -M_2 = {1 \ov 2\sqrt{2}}~.
\ee
In this case the discriminant and $j$-invariant become:
\bea
    \Delta(U) = -{1\ov 32} U^2(27-2U^2)^2~, \qquad j(U) = {(27 - 2U^2)(-3+2U^2)^3 \ov 2U^2}~.
\eea
A solution to the equation $J(U) = J(\tau)$ is then given by:
\be
    U(\tau) = {i\ov\sqrt{2}} \left( {\eta \left( {\tau - 1 \ov 3} \right) \ov \eta(\tau - 1)}  \right)^6~,
\ee
where the shift in $\tau$ is chosen for consistency with the massless configuration - that is we ensure that $U(\tau)$ satisfies \eqref{U choice}. This turns out to be a modular function for the congruence subgroup $6C^0$, in the notation of \cite{Cummins2003}. In this case, the $I_2$ singularity lies at $\tau = 1$, while the type $II$ singularities correspond to $\tau = 3 + e^{2\pi i \ov 3}$ and $\tau = 6 + e^{2\pi i \ov 3}$, respectively. 
We can further check that this is in agreement with the massless configuration by considering the BPS quiver of the theory; it was found in \cite{Closset:2021lhd} that the relevant quiver is given by:
\bea\label{quiver E3a}
 \begin{tikzpicture}[baseline=1mm]
\node[] (1) []{$\CE_{\gamma_1=(1,0)}$};
\node[] (2) [right = of 1]{$\CE_{\gamma_{2,3,4}=(-1,2)}$};
\node[] (3) [below = of 1]{$\CE_{\gamma_{5,6}=(1,-3)}$};
\draw[->>-=0.5] (1) to   (2);
\draw[->-=0.5] (2) to   (3);
\draw[->>>-=0.5] (3) to   (1);
\end{tikzpicture}
\eea
Performing the mutation sequence $\hat{\gamma}_5 - \gamma_{2,3} - \gamma_1$, we find the new quiver:
\bea\label{quiver E3 I6 I2 2II}
 \begin{tikzpicture}[baseline=1mm]
\node[] (4) []{$\CE_{\gamma_{4}=(0,-1)}$};
\node[] (7) [left = of 4] {};
\node[] (8) [right = of 4] {};
\node[] (1) [below = of 7]{$\CE_{\gamma_{1}=(-1,2)}$};
\node[] (6) [below = of 8]{$\CE_{\gamma_{6}=(1,-5)}$};
\node[] (5) [below = of 1]{$\CE_{\gamma_{3}=(0,1)}$};
\node[] (2) [below = of 6]{$\CE_{\gamma_{2, 3}=(1,-1)}$};
\draw[->>>-=0.5] (1) to   (6);
\draw[->>>>-=0.5] (6) to   (2);
\draw[->-=0.5] (2) to   (5);
\draw[->-=0.5] (5) to   (1);
\draw[->-=0.5] (1) to   (4);
\draw[->-=0.5] (4) to   (6);
\draw[->-=0.7] (6) to   (5);
\draw[->-=0.3] (2) to   (1);
\draw[->-=0.7] (4) to   (2);
\end{tikzpicture}
\eea
We can explain the mutation sequence as follows. To obtain the type $II$ singularities one needs to split the $I_2$ of the massless configuration, which is achieved here with the mutation on node 5. Then, we realise that the $I_2$ cusp of the new configuration comes from the decomposition of $I_3 \rightarrow I_2 \oplus I_1 $, which is achieved through the mutation $\gamma_{2,3}$. The states of the above BPS quiver will combine as follows:
\be
    II^{(3)}~:~ \qquad (0,1) \oplus (-1,2) ~, \qquad II^{(6)}~:~ \qquad (0,1) \oplus (1, -5) ~,
\ee
according to \eqref{II Decomposition}.


\subsubsection{${\rm rk}(\Phi) = 3$ Configurations}

Finally, the modular configurations of highest Mordell-Weil group rank are listed in table~\ref{tab:Congruence Rank3}. We will analyze again the configurations for `unusual' monodromy groups.%
\renewcommand{\arraystretch}{1}
\begin{table}[t]
\small
\centering
\begin{tabular}{ |c||c|c|c|c|} 
 \hline
$\{F_v\}$ & $\Phi_{\rm tor}$& \textit{Field theory} & $\Fg_F$ & $\Gamma$ 
\\
\hline 
\hline
\multirow{4}{*}{$I_0^*, III, II, I_1$} & \multirow{4}{*}{$-$} & \text{4d} $SU(2)\, N_f = 4$ & $A_1 \oplus 3\frak{u}(1)$ &  \multirow{4}{*}{$\Gamma(1)$}  \\ \cline{3-4}
&  & \text{MN} $E_7$ & $D_4 \oplus 3\frak{u}(1)$  &  \\ \cline{3-4}
&  & \text{MN} $E_8$ & $D_4 \oplus A_1 \oplus 3\frak{u}(1)$  &  \\ \cline{3-4}
&  & $\KK E_8$ & $D_4 \oplus A_1 \oplus 3\frak{u}(1)$  &  \\ \hline
 \hline
\multirow{3}{*}{$I_0^*, I_2, 2II$} & \multirow{3}{*}{$-$} & \text{4d} $SU(2)\, N_f = 4$ & $A_1 \oplus 3\frak{u}(1)$ &  \multirow{3}{*}{$\Gamma^2$}  \\ \cline{3-4}
&  & \text{MN} $E_8$ & $D_4 \oplus A_1 \oplus 3\frak{u}(1)$  & \\ \cline{3-4}
&  & $\KK E_7$ & $D_4 \oplus  3\frak{u}(1)$  &  \\ \hline
\hline
\multirow{2}{*}{$I_3, 3III$} & \multirow{2}{*}{$-$} & $\KK E_6$ & $3A_1 \oplus 3\frak{u}(1)$ &  \multirow{2}{*}{$\Gamma^3$}  \\ \cline{3-4}
&  & \text{MN} $E_7$ & $A_2\oplus 2A_1 \oplus 3\frak{u}(1)$  &  \\ \hline
 \hline
\multirow{3}{*}{$I_4, 2III, II$} & \multirow{3}{*}{$-$} & $\KK E_5$ & $2A_1 \oplus 3\frak{u}(1)$ &  \multirow{3}{*}{$4A^0$}   \\ \cline{3-4}
&  & \text{MN} $E_8$ & $A_3 \oplus 2A_1 \oplus 3\frak{u}(1)$  & \\ \cline{3-4}
&  & \text{MN} $E_7$ & $A_3 \oplus A_1 \oplus 3\frak{u}(1)$  & \\ \hline
\hline
\multirow{3}{*}{$I_5, III, 2II$} & \multirow{3}{*}{$-$} & $\KK E_4$ & $A_1 \oplus 3\frak{u}(1)$ &  \multirow{3}{*}{$5A^0$} \\ \cline{3-4}
&  & \text{MN} $E_8$ & $A_4 \oplus A_1 \oplus 3\frak{u}(1)$  & \\ \cline{3-4}
&  & \text{MN} $E_7$ & $A_4 \oplus 3\frak{u}(1)$  & \\ \hline
\hline
\multirow{2}{*}{$I_6, 3II$} & \multirow{2}{*}{$-$} & $\KK E_3$ & $3\frak{u}(1)$ &  \multirow{2}{*}{$6A^0$} \\ \cline{3-4}
&  & \text{MN} $E_8$ & $A_5 \oplus 3\frak{u}(1)$  & \\ \hline
\end{tabular}
    \caption{Modular RES with ${\rm rk}(\Phi)=3$ and corresponding field theories.}
    \label{tab:Congruence Rank3}
\end{table}

\medskip

\begin{figure}[t]
    \centering
    \centering
    \scalebox{0.7}{


\begin{tikzpicture}[scale=2]

    \pgfdeclarelayer{background}
    \pgfsetlayers{background,main}

    \pgfmathsetmacro{\myxlow}{-1}
    \pgfmathsetmacro{\myxhigh}{4}
    \pgfmathsetmacro{\myiterations}{1}
    
    \draw[-latex](\myxlow-0.1,0) -- (\myxhigh+0.2,0);
    \pgfmathsetmacro{\succofmyxlow}{\myxlow+0.5}
    
    \draw[-stealth, very thin] (0,0)--(0,3.2) node[above] {}; 
    \begin{scope}
            \clip
                (-0.5,0.866) 
                { -- (-0.5,0.866) arc(60:0:1)
             -- (0,0) arc(180:60:1/3)
             -- (0.5,0.29) arc(120:0:1/3)
             -- (1,0) arc(0:120:1)
             }
                -- (-0.5, 0.866) -- cycle;
            \fill[blue,opacity=0.25] (-1,0) rectangle (1,1);
    \end{scope}

    \begin{scope}
            \clip
                (0.5,0.866) 
                { -- (0.5,0.866) arc(60:0:1)
             -- (1,0) arc(180:60:1/3)
             -- (1.5,0.29) arc(120:0:1/3)
             -- (2,0) arc(0:120:1)
             }
                -- (0.5, 0.866) -- cycle;
            \fill[green,opacity=0.15] (0,0) rectangle (2,1);
    \end{scope}

    \begin{scope}
            \clip
                (1.5,0.866) 
                { -- (1.5,0.866) arc(60:0:1)
             -- (2,0) arc(180:120:1)
             -- (2.5,0.866) arc(60:0:1)
             -- (3,0) arc(180:120:1)
             -- (3.5, 0.866) arc(60:120:1)
             -- (2.5, 0.866) arc(60:120:1)
             }
                -- (1.5, 0.866) -- cycle
            ;
            \fill[red,opacity=0.2] (0,0) rectangle (4,1);
    \end{scope}
    
    \begin{scope}
    
    \foreach \i in {-0.5,0.5,1.5,2.5}
            {\draw[very thin, black] (\i,0.866) arc(120:60:1);
            \draw[very thin, black] (\i,0.866) -- (\i,3);
            }

            \draw[very thin, black] (3.5,0.866) -- (3.5,3);
            
    \end{scope}
    
    \draw[very thin, black, dash pattern=on 1.5pt off 1pt] (-0.5,0.866) arc(60:0:1);
        \draw[very thin, black, dash pattern=on 1.5pt off 1pt] (0,0) arc(180:120:1);
        \draw[very thin, black, dash pattern=on 1.5pt off 1pt] (0,0) arc(180:60:1/3);
        \draw[very thin, black, dash pattern=on 1.5pt off 1pt] (1,0) arc(0:120:1/3);
        \draw[very thin, black, dash pattern=on 1.5pt off 1pt] (0.5, 0.29) -- (0.5,0.866);

        \draw[very thin, black, dashed] (1,0) arc(180:60:1/3);
        \draw[very thin, black, dashed] (0.5,0.866) arc(60:0:1);
        \draw[very thin, black, dashed] (1,0) arc(180:120:1);
        \draw[very thin, black, dashed] (1.5, 0.29) -- (1.5,0.866);
        \draw[very thin, black, dashed] (2,0) arc(0:120:1/3);
        
        \draw[very thin, black, dash pattern=on 1.5pt off 1pt] (2,0) arc(180:120:1);
        \draw[very thin, black, dash pattern=on 1.5pt off 1pt] (1.5,0.866) arc(60:0:1);

        \draw[very thin, black, dash pattern=on 1.5pt off 1pt] (3,0) arc(180:120:1);
        \draw[very thin, black, dash pattern=on 1.5pt off 1pt] (2.5,0.866) arc(60:0:1);
        
    \begin{scope}
        \begin{pgfonlayer}{background}
            \clip
                (-0.5,3) 
                { -- (-0.5,0.866) arc(120:60:1)
             -- (0.5,0.866) arc(120:60:1)
             -- (1.5,0.866) arc(120:60:1)
             -- (2.5,0.866) arc(120:60:1)
             }
                -- (3.5, 3) -- cycle;
            \fill[gray,opacity=0.4] (-1,-1) rectangle (4.5,3);
        \end{pgfonlayer}
    \end{scope}

    \begin{scope}
        \node at (0.15,2) {$\mathcal{F}$};
        \node at (1,2) {$T\mathcal{F}$};
        \node at (2,2) {$T^2\mathcal{F}$};
        \node at (3,2) {$T^3\mathcal{F}$};
        
        \node at (-0.25,1.25) {$III^{(0)}$};
        \node at (1,1.25) {$III^{(1)}$};
        \node at (2.5,0.6) {$II^{(3)}$};

        \fill (0,0.99)  circle[radius=1pt];
        \fill (1,0.99)  circle[radius=1pt];
        \fill (2.5,0.866)  circle[radius=1pt];
        
        \foreach \i in {0,1,2,3}
        {
        \fill (\i,0)  circle[radius=0.5pt];
        }
        
    \end{scope}
    
\end{tikzpicture}

}
    \caption{Fundamental domain for the $(I_4,2III,II)$ configuration on the CB of $\KK E_5$. The fundamental domain of the monodromy group $4A^0$ is drawn in gray. The coloured regions are absorbed in the elliptic points.
    }
    \label{fig: E5 Configuration colour1}
\end{figure}
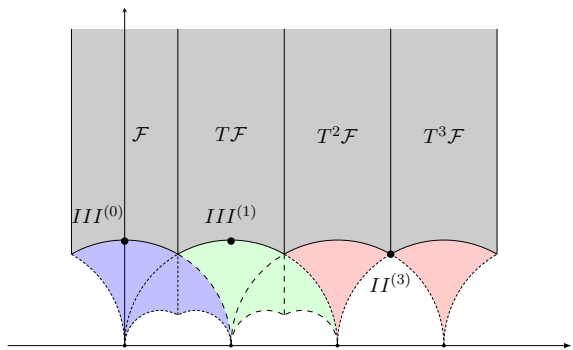

\noindent
\paragraph{$\boldsymbol{(I_4,2III, II)}$ configuration for $\KK E_5$.} Another configuration appearing on the CB of the $\KK E_5$ theory is the $(I_4,2III, II)$ configuration. This can be found for:
\be \label{E5 characters I4 2III II}
    \boldsymbol{\chi}^{E_5} = \left( {13i \ov 2}, -23, 12(1-i), -10 i, -4(1+i) \right)~,
\ee
in which case the two $III$ singularities reside at $U_{\pm} = 3(1+i) \pm 4e^{3\pi i \ov 4}$, while the $II$ singularity is at $U_{II} = -7(1+i)$. We then find that one root of $J = J(\tau)$ is given by:
\be
    U(\tau) = q^{- {1\ov 4}} - (1+i) + 12i q^{1\ov 4} + 32(1+i) q^{1\ov 2} + 66q^{3\ov 4} + \cO(q)~.
\ee
In order to find a closed-form expression for this, let us first define:
\be
   f(\tau) = \left({\eta(2\tau) \ov \eta(\tau)}\right)^{12}~,
\ee
which changes sign under a $T$ transformation. This function is essentially the square root of the Hauptmodul for $\Gamma^0(2)$. We then note that:
\be
    U(\tau) = -(1+i) + 32(1-i) f(\tau) + i\left( {1\ov f(\tau)} - 24i - 2048i f(\tau)^2 \right)^{1\ov 2} ~.
\ee
Moreover, $U(\tau)$ is invariant under $S$ and $T^4$ transformations. As $f(\tau)$ is invariant under $T^2$ transformations one might be inclined to believe that $U(\tau)$ also has this property, but, due to the square root in the above expression, this turns out to be false. The holomorphicity of the above function is `restored' by first using the identities:
\bea
    \vartheta_2(\tau) = {2\eta(2\tau)^2 \ov \eta(\tau)}~, \qquad \quad \vartheta_2(\tau) \vartheta_3(\tau) \vartheta_4(\tau) = 2\eta(\tau)^3~,
\eea
for which we refer the reader to appendix \ref{Appendix Modular forms}. This leads to the expression:
\be
    f(\tau) = {\vartheta_2(\tau)^4 \ov 16 \vartheta_3(\tau)^2  \vartheta_4(\tau)^2 } =  {\vartheta_3(\tau)^4 - \vartheta_4(\tau)^4  \ov 16 \vartheta_3(\tau)^2 \vartheta_4(\tau)^2 }~.
\ee
We then finally find:
\be \label{4A0}
    U(\tau) = -(1+i) + 2(1-i) {\vartheta_2(\tau)^6 + i\vartheta_3(\tau)^6 + \vartheta_4(\tau)^6 \ov \vartheta_2(\tau)^2 \vartheta_3(\tau)^2 \vartheta_4(\tau)^2}~.
\ee
One can then check that the $III$ singularities correspond to $\tau = i$ and $\tau = 3+i$, while the type $II$ singularity resides at $\tau = 2 + e^{2\pi i \ov 3}$, for the above solution. The fundamental domain of the monodromy group is depicted in figure~\ref{fig: E5 Configuration colour1}. This is in agreement with \cite{STROMBERG2019436}, in whose notation the group is:
\be
    4A^0 \cong \Gamma(4; 0,1,2,1;4)~.
\ee
Understanding the origin of the AD points in terms of BPS states is again a non-trivial task for this configuration. We can still understand this fundamental domain starting from the BPS quiver of the $\KK E_5$ theory in \eqref{quiver E5 orbi} and performing the mutation sequence $\hat{\gamma}_7 - \hat{\gamma}_3$. %
This corresponds to splitting the $I_4$ and one of the $I_2$ singularities, which can be achieved by turning on mass deformations. The resulting quiver is shown below:%
\bea\label{quiver E5 I4 2III II}
 \begin{tikzpicture}[baseline=1mm]
\node[] (8) []{$\CE_{\gamma_8=(1,-2)}$};
\node[] (11) [below = of 8]{};
\node[] (1) [left = of 11]{$\CE_{\gamma_{1,2}=(1,0)}$};
\node[] (4) [right = of 11]{$\CE_{\gamma_{4,5,6}=(0,-1)}$};
\node[] (3) [below = of 1]{$\CE_{\gamma_{3}=(0,1)}$};
\node[] (7) [below = of 4]{$\CE_{\gamma_{7}=(-1,1)}$};
\draw[->-=0.5] (1) to   (3);
\draw[->-=0.5] (3) to   (7);
\draw[->-=0.5] (7) to   (4);
\draw[->-=0.5] (4) to   (1);
\draw[->-=0.3] (1) to   (7);
\draw[->-=0.3] (7) to   (8);
\draw[->-=0.3] (4) to   (8);
\draw[->>-=0.3] (8) to   (1);
\draw[->-=0.3] (8) to   (3);
\end{tikzpicture}
\eea%
One can obtain the singularities corresponding to elliptic points as follows:
\be
    \mathbb{M}_{III^{(0)}} = \mathbb{M}_{(1,0)}^2\mathbb{M}_{(0,-1)}~, \qquad \mathbb{M}_{III^{(3)}} =  \mathbb{M}_{(1,-2)}\mathbb{M}_{(0,-1)}^2~, \qquad \mathbb{M}_{II^{(2)}} = \mathbb{M}_{(0,-1)} \mathbb{M}_{(-1,1)}~,
\ee
which is again based on the decompositions \eqref{II Decomposition} and \eqref{III Decomposition}.


\medskip

\noindent
\paragraph{$\boldsymbol{(I_5,III, 2II)}$ configuration for $\KK E_4$.} The other modular configuration of $\KK E_4$ different from the massless configuration with the monodromy group being a congruence subgroup is: $(I_5, III, 2II)$. This can be obtained for:
\be
    \boldsymbol{\chi}^{E_4} = \left( -{47 \ov 16}, {3\ov 2}, 6, -{5\ov2}\right)~,
\ee
in which case the discriminant and $j$-invariant read:
\be
    \Delta(U) = (2U - 7)^3(4U^2 + 16U + 151)^2~, \qquad j(U) = {1\ov 32}(2U - 1)^3(4U^2 + 16U + 151)~.
\ee
Hence, the $U$-plane singularities will be: $U_{III} = {7\ov 2}$ and $U_{II_{\pm}} = -2 \pm {3i\sqrt{15} \ov 2}$. One root of the equation $J = J(\tau)$ is given by:
\be
    U(\tau) = q^{- {1\ov 5}} - {1\ov 2} - 6q^{1\ov 5} + 20 q^{2\ov 5} + 15 q^{3\ov 5} + 36 q^{4\ov 5} + \cO(q)~.
\ee
This is the McKay-Thompson series of class $5a$, which can be reproduced in closed-form as follows. Introducing:
\be
    f_{5B} = \left( {\eta \left({\tau \ov 5}\right) \ov \eta(\tau)}\right)^6~, \qquad f_{25a} = {\eta \left({\tau \ov 5}\right) \ov \eta(5\tau)}~,
\ee
we express $U(\tau)$ as:\footnote{We are thankful to \texttt{oeis.org} for this expression.}
\be \label{5A0}
    U(\tau) = -{1\ov 2} + \left( 1 + {5\ov f_{25a}(\tau)}\right)\left(1+f_{5B}(\tau)\right) + 5\left( f_{25a}(\tau) - {5\ov f_{25a}(\tau)} \right) {f_{5B}(\tau) \ov f_{25a}(\tau)^3}~.
\ee
One then finds that the $III$ singularity corresponds to $\tau = i$, while the two type $II$ singularities will be at $\tau_- = 2 + e^{2\pi i\ov 3}$ and $\tau_+ = 4 + e^{2\pi i \ov 3}$, respectively. The mutually non-local light BPS states becoming massless at these loci can be found form the quiver \eqref{quiver E4} by following the mutation sequence $\hat{\gamma}_3 - \gamma_2$. Of course, this corresponds to turning on mass deformations and involves decompositions using $(0,1)$ states, as in \eqref{III Decomposition v2}.

\medskip

\noindent
\paragraph{$\boldsymbol{(I_6,3II)}$ configuration for $\KK E_3$.} The last configuration that we discuss is the $(I_6, 3II)$ configuration, which appears on the CB of the $\KK E_3$ theory. This can be found for:
\be
    \lambda = e^{2\pi i \ov 3}~, \qquad M_1 = \left(2-\sqrt{3}\right) e^{5\pi i \ov 6}~, \qquad M_2 = \left(2+\sqrt{3}\right) e^{11 \pi i \ov 6}~.
\ee
In this case both the discriminant and the $j$-invariant are quadratic polynomials in $u^3$. For instance, the latter reads:
\be
    j(U) = U^3\left(U^3 - 48i\sqrt{3}\right)~,
\ee
which can be easily solved for $U(\tau)$:
\be
    U(\tau) = \Big(24\sqrt{3}\left( i \pm \sqrt{J(\tau)-1}\right) \Big)^{1 \ov 3}~.
\ee
The apparent branch cuts of this solution resolve once we express it in terms of theta functions. This is done by using the modular $\lambda(\tau)$ function\footnote{This should not be confused with the gauge theory parameter $\lambda$.} defined in appendix~\ref{Appendix Modular forms}, such that:
\be
    U(\tau)^3 = \mp 16{ \left( 1 + e^{ \pm{2\pi i \ov 3}} \lambda(\tau) \right)^3 \ov \lambda(\tau)(1-\lambda(\tau))}~.
\ee
Expressing the denominator in terms of Dedekind-eta functions, this also becomes a cubic expression, with:
\be
     U(\tau)^3 = \mp 16{ \left( \vartheta_3(\tau)^4 + e^{ \pm{2\pi i \ov 3}} \vartheta_2(\tau)^4 \right)^3 \ov \vartheta_2(\tau)^4 \vartheta_3(\tau)^4 \vartheta_4(\tau)^4} = \mp \left({  \vartheta_3(\tau)^4 + e^{ \pm{2\pi i \ov 3}} \vartheta_2(\tau)^4 \ov \eta(\tau)^{4}}  \right)^3~.
\ee
One can check using \eqref{U choice} and \eqref{alpha E1, E3} that the asymptotics consistent with the massless configuration are $q^{-1/6}$, with no additional prefactor. Thus, the correct solution reads:
\be
    U(\tau) = {  \vartheta_3(\tau)^4 + e^{ {2\pi i \ov 3}} \vartheta_2(\tau)^4 \ov \eta(\tau)^4}~.
\ee
Note that this is again chosen such that it agrees with \eqref{U choice}, where the prefactor $\alpha$ in \eqref{alpha E1, E3} is trivial. For this solution the type $II$ singularities lie at $\tau = 2k + e^{2\pi i \ov 3}$, for $k = 0, 1$ and 2, respectively. Working with $w = U^3$, it becomes clear that the monodromy group on the $w$-plane is the unique index 2 subgroup of ${\rm PSL}(2,\mathbb{Z})$, which we denoted by $\Gamma^2$. The monodromy group on the $U$-plane is then an index $3$ subgroup of $\Gamma^2$. Note that this is different from the `commutator' subgroup $\Gamma' = \Gamma^2 \cap \Gamma^3$ of ${\rm PSL}(2,\mathbb{Z})$, which is also an index $6$ congruence subgroup, but it is torsion-free \cite{10.1215/ijm/1255632506}.



\subsection{Non-Congruence subgroups}

We will now consider the modular rational elliptic surfaces for which the relevant subgroup of ${\rm PSL}(2,\mathbb{Z})$ is non-congruence. It turns out that all such rational elliptic surfaces have either ${\rm rk}(\Phi) = 1$ or $2$. 

A common feature of many of the modular functions of congruence subgroups studied above was the integrality of the coefficients appearing in the $q$-expansion. In particular, it is known that for modular forms of congruence subgroups the Fourier expansions have bounded denominators. This property is lost at the level of non-congruence subgroups, which is usually referred to as the `unbounded denominator conjecture'. The systematic study of non-congruence modular forms was initiated in \cite{atkin1971modular}, where the authors observed that the Fourier coefficients of non-congruence modular forms satisfy a set of relations that are usually referred to as ASD congruences. Certain non-congruence subgroups of index $7$ and $9$ have been considered in this context in \cite{atkin1971modular, Scholl1988TheLR, Scholl2010OnLR} and more recently discussed in \cite{2020arXiv200701336F}. For a nice review of the existing literature see \cite{long2007arithmetic}.

In some cases, non-congruence subgroups are subgroups of certain congruence subgroups. In this way, modular forms for non-congruence subgroups of index $36$ have been discussed in \cite{2008arXiv0805.2144F}, in particular for index $3$ subgroups of certain index $12$ congruence subgroups of ${\rm PSL}(2,\mathbb{Z})$. There, the modular forms of the non-congruence subgroups could be deduced from those of the aforementioned index $12$ congruence subgroups, based on \eqref{Subgroups inclusions}. However, in our case, we cannot apply this logic to the index $7$ non-congruence subgroups of ${\rm PSL}(2,\mathbb{Z})$, but we can try to view the index $9$ non-congruence subgroups as index $3$ subgroups of $\Gamma^0(2)$, $\Gamma^3$ or their conjugates. We in fact find that:
\bea
    (I_6, III, I_2, I_1)~: & \qquad \Gamma(9; 0,3,1,0; 6) \subset \Gamma_0(2)~,\\
    (I_4, I_3, III, I_2)~: & \qquad \Gamma(9; 0,3,1,0; 12) \subset \Gamma_0(2) ~,
\eea
where by $\Gamma_0(2)$ we refer to one of the groups in this conjugacy class. The generators of the non-congruence subgroups are found from the fundamental domains obtained in \cite{STROMBERG2019436}. We will show these inclusions in the next subsections.

One can attempt a similar inclusion for the index $10$ subgroups into the unique index $5$ congruence subgroup of genus zero $5A^0$, or into the congruence subgroup $\Gamma^2$. However, we will see that this is not possible. Let us also mention that modular forms for non-congruence subgroups have been also found in \cite{2009arXiv0910.0739K} using computational methods. 


\subsubsection{${\rm rk}(\Phi) = 1$ Configurations.}

Consider first the surfaces with ${\rm rk}(\Phi) = 1$. These will be relevant for understanding the CB of the $\KK \t E_1$ and $\KK E_2$ theories, and are listed in table~\ref{tab:Non-Congruence 1}.%
\renewcommand{\arraystretch}{1}
\begin{table}[h]
\small
\centering
\begin{tabular}{ |c||c|c|c|c|} 
 \hline
$\{F_v\}$ & $\Phi_{\rm tor}$& \textit{Field theory} & $\Fg_F$ & $\Gamma$ 
\\
\hline 
 \hline
\multirow{3}{*}{$I_7, III, 2I_1$} & \multirow{3}{*}{$-$} & $\KK E_8$ & $A_6 \oplus A_1 \oplus \frak{u}(1)$ & \multirow{3}{*}{$\Gamma(9; 0, 3, 1, 0;7)$}    \\ \cline{3-4}
&  &$\KK E_2$ & $A_1 \oplus \frak{u}(1)$  &   \\ \cline{3-4}
&  & \text{MN} $E_7$ & $A_6 \oplus \frak{u}(1)$  &  \\ \hline 
\hline
\multirow{4}{*}{$I_6, III, I_2, I_1$} & \multirow{4}{*}{$\mathbb{Z}_2$} & $\KK E_8$ & $A_5 \oplus 2A_1 \oplus \frak{u}(1)$ &  \multirow{4}{*}{$\Gamma(9; 0, 3, 1, 0;6)$}   \\ \cline{3-4}
&  &$\KK E_7$ & $A_5 \oplus A_1 \oplus \frak{u}(1)$  &   \\ \cline{3-4}
&  &$\KK E_3$ & $2A_1 \oplus \frak{u}(1)$  &   \\ \cline{3-4}
&  &\text{MN} $E_7$ & $A_5 \oplus A_1 \oplus \frak{u}(1)$  &   \\ \hline 
\hline
\multirow{4}{*}{$I_5, I_3, III, I_1$} & \multirow{4}{*}{$-$} & $\KK E_8$ & $A_4 \oplus A_2 \oplus A_1 \oplus \frak{u}(1)$ &  \multirow{4}{*}{$\Gamma(9; 0, 3, 1, 0;15)$}   \\ \cline{3-4}
&  &$\KK E_6$ & $A_4 \oplus A_1 \oplus \frak{u}(1)$  &    \\ \cline{3-4}
&  &$\KK E_4$ & $A_2 \oplus A_1 \oplus \frak{u}(1)$  &   \\ \cline{3-4}
&  &\text{MN} $E_7$ & $A_4 \oplus A_2 \oplus \frak{u}(1)$  &   \\ \hline 
\hline
\multirow{4}{*}{$I_4, I_3, III, I_2$} & \multirow{4}{*}{$\mathbb{Z}_2$} & $\KK E_7$ & $A_3 \oplus A_2 \oplus A_1 \oplus \frak{u}(1)$ &  \multirow{4}{*}{$\Gamma(9; 0, 3, 1, 0; 12)$}  \\ \cline{3-4}
&  &$\KK E_6$ & $A_3 \oplus 2A_1 \oplus \frak{u}(1)$  &   \\ \cline{3-4}
&  &$\KK E_5$ & $A_2 \oplus 2A_1 \oplus \frak{u}(1)$   & \\ \cline{3-4}
&  &\text{MN} $E_7$ & $A_3 \oplus A_2 \oplus A_1 \oplus \frak{u}(1)$  &   \\ \hline 
\hline
\multirow{3}{*}{$I_8, II, 2I_1$} & \multirow{3}{*}{$-$} & $\KK E_8$ & $A_7 \oplus \frak{u}(1)$ &  \multirow{3}{*}{$\Gamma(10; 0, 3, 0, 1;8)$}    \\ \cline{3-4}
&  &$\KK \t E_1$ & $\frak{u}(1)$  &    \\ \cline{3-4}
&  & \text{MN} $E_8$ & $A_7 \oplus \frak{u}(1)$  &  \\ \hline 
\hline
\multirow{4}{*}{$I_7, I_2, II, I_1$} & \multirow{4}{*}{$-$} & $\KK E_8$ & $A_6 \oplus A_1 \oplus \frak{u}(1)$ &  \multirow{4}{*}{$\Gamma(10; 0, 3, 0, 1;14)$}   \\ \cline{3-4}
&  &$\KK E_7$ & $A_6 \oplus \frak{u}(1)$  &   \\ \cline{3-4}
&  &$\KK E_2$ & $A_1 \oplus \frak{u}(1)$  &  \\ \cline{3-4}
&  &\text{MN} $E_8$ & $A_6 \oplus A_1 \oplus \frak{u}(1)$  & \\ \hline 
\hline
\multirow{4}{*}{$I_5, I_4, II, I_1$} & \multirow{4}{*}{$-$} & $\KK E_8$ & $A_4 \oplus A_3 \oplus \frak{u}(1)$ &  \multirow{4}{*}{$\Gamma(10; 0, 3, 0, 1;20)$}   \\ \cline{3-4}
&  &$\KK E_4$ & $A_3 \oplus \frak{u}(1)$  &   \\ \cline{3-4}
&  &$\KK E_5$ & $A_4 \oplus \frak{u}(1)$  &  \\ \cline{3-4}
&  &\text{MN} $E_8$ & $A_4 \oplus A_3 \oplus \frak{u}(1)$  &   \\ \hline 
\hline
\multirow{4}{*}{$I_5, I_3, I_2, II$} & \multirow{4}{*}{$-$} & $\KK E_7$ & $A_4 \oplus A_2 \oplus \frak{u}(1)$ &  \multirow{4}{*}{$\Gamma(10; 0, 3, 0, 1;30)$}   \\ \cline{3-4}
&  &$\KK E_6$ & $A_4 \oplus A_1 \oplus \frak{u}(1)$  & \\ \cline{3-4}
&  &$\KK E_4$ & $A_2 \oplus A_1 \oplus \frak{u}(1)$  & \\ \cline{3-4}
&  &\text{MN} $E_8$ & $A_4 \oplus A_2 \oplus A_1 \oplus \frak{u}(1)$  & \\ \hline 
\end{tabular}
    \caption{Modular rational elliptic surfaces for non-congruence subgroups with ${\rm rk}(\Phi) = 1$. }
    \label{tab:Non-Congruence 1}
\end{table}%
We will again follow the notation of \cite{STROMBERG2019436}, which is also described in appendix~\ref{Appendix Permutations} and at the beginning of the section.

Among these configurations, we notice the $(I_8,II,2I_1)$ configuration, which is the only modular configuration appearing on the CB of the $\KK \t E_1$ theory. Additionally, the configurations $(I_7, III, 2I_1)$ and $(I_7, I_2, II, I_1)$ are two configurations found on the CB of the $\KK E_2$ theory.
\medskip

\noindent
\paragraph{$\boldsymbol{(I_6, III, I_2, I_1)}$ configuration for $\KK E_3$.} We consider the $\KK E_3$ theory, where this configuration was found in \cite{Closset:2021lhd}, for the parameters:
\be
    \lambda = 1~, \qquad M_1 = M_2 = {1\ov 2}~.
\ee
For these values the curve becomes:
\bea
    g_2(U) & = {1\ov 12}\left( U^4 - 18U^2 - 24U + 9\right)~, \\
    g_3(U) & = -{1\ov 216}(U+3)^2\left( U^4 - 6U^3 + 18U + 27\right)~, 
\eea
with the discriminant and $J$-invariant:
\bea
    \Delta(U)  = {1\ov 4}(U-5)(U+1)^2(U+3)^2~, \qquad J(U) = {\left(U^3  - 3U^2 - 9U +3\right)^3 \ov 432(U-5)(U+1)^2}~.
\eea
One can then compute $U(\tau)$ to arbitrary orders in $q = e^{2\pi i \tau}$ and check the unbounded denominators conjecture explicitly. At this stage, however, it is not clear how to draw the fundamental domain for this configuration, due to the lack of a closed form for $U(\tau)$. For the sake of the argument, let us pick a representative of this conjugacy class from \cite{STROMBERG2019436}, with the monodromies given by:
\bea
    \mathbb{M}_{I_1} = STS^{-1}~, \quad \mathbb{M}_{III} = T^3 S^{-1}T^{-3}~, \qquad \mathbb{M}_{I_2} = (T^4S)T^2(T^4S)^{-1}~, \qquad \mathbb{M}_{\infty} = T^6~.
\eea
It is then straightforward to check that these generators of the $\Gamma(9;0,3,1,0;6)$ non-congruence subgroup can be expressed in terms of the generators of $\Gamma^0(2)$, namely:
\be
    STS^{-1}~, \qquad TS^{-1}T~, \qquad T^2~,
\ee
and thus this non-congruence subgroup is an index $3$ subgroup of $\Gamma^0(2)$. As a result, every non-congruence subgroup in this conjugacy class will be an index $3$ subgroup of an element of the conjugacy class of $\Gamma^0(2)$.\footnote{There are four distinct index $3$ subgroups of ${\rm PSL}(2,\mathbb{Z})$, three of them being in the same conjugacy class: $\Gamma_0(2)$, $\Gamma^0(2)$ and $\Gamma_{\theta}$} Thus, the polynomial \eqref{Polynomial J(U)=J(t)} has an intermediate field and we can try to find closed-form expressions for the modular forms of $\Gamma(9; 0,3,1,0;6)$ from those of $\Gamma^0(2)$. We first solve $J = J(\tau)$, with one root given by:
\be
    U(\tau) = 2^{-{1 \ov 3}}q^{-{1\ov 6}} + 1 + 2^{7\ov 3}q^{1\ov 6} + {4\ov 3} 2^{2\ov 3} q^{1\ov 3} - {32 \ov 3} 2^{1\ov 3} q^{2\ov 3} - {2\ov 9}2^{2\ov 3}q^{5\ov 6} + \cO(q)~.
\ee
We then notice that:
\be \label{intermmediate field factorisation}
    -5 - 9U(\tau) - 3U^2(\tau) + U^3(\tau)~ =~ {1\ov 2} \left( {\eta\left({\tau \ov 2}\right)\ov \eta(\tau)}\right)^{24} ~=~ {1\ov 2} F_{\Gamma^0(2)}(\tau)~,
\ee
which can be proved explicitly from the $J$-invariant of the $\Gamma^0(2)$ curve. This then gives a closed-form expression for $U(\tau)$ in terms of $F_{\Gamma^0(2)}(\tau)$, which can be further simplified to:
\be
    U(\tau) = 1 + 2 \left( {\vartheta_3(\tau) \ov \vartheta_2(\tau)} \right)^{4\ov 3} +  2 \left( {\vartheta_2(\tau) \ov \vartheta_3(\tau)} \right)^{4\ov 3}~.
\ee
We remark, however, that despite this relatively simple form, the above expression still involves fractional powers of the Jacobi theta functions. As pointed out in \cite{Schultz}, the modular functions of non-congruence subgroups are not invariant under the action of any of the principal congruence subgroups $\Gamma(N)$, for $N\in \mathbb{N}^*$; as a result, these Hauptmoduln cannot be expressed in terms of the usual $q$-products. Thus, in some sense, the above expression is the `best' one can achieve using the usual modular functions. Such expressions were also found in \cite{2009arXiv0910.0739K} using computational methods.

Note that since the configuration is modular, one does not expect any branch points or branch cuts on the fundamental domain. This is indeed the case, as a consequence of the theta functions being holomorphic and non-zero functions on the upper half-plane. As a result, the radicands cannot vanish and thus $U(\tau)$ is also holomorphic on $\mathbb{H}$.

\medskip

\noindent
\paragraph{$\boldsymbol{(I_4, I_3, III, I_2)}$ configuration for $\KK E_5$.} Another index $9$ subgroup of ${\rm PSL}(2,\mathbb{Z})$ corresponds to the configuration $(I_4, I_3, III, I_2)$. This can be obtained, for instance, on the CB of the $\KK E_5$ theory, by setting the characters to:
\be
    \boldsymbol{\chi}^{E_5} = \left(-6,~ {7\ov 3},~ {16 i \ov 3\sqrt{3}},~ {872\ov 27},~ {16i\ov 3\sqrt{3}} \right)~.
\ee
In this case, the $j$-invariant becomes:
\be
    j(U_1) = {\left( U_1^3 + 12U_1^2 - 384 \right)^3 \ov (U_1+4)^3(3U_1 + 28)^2}~, \qquad U_1 = i\sqrt{3} U~.
\ee
One can show that this non-congruence subgroup is an index $3$ subgroup of $\Gamma^0(2)$ and thus the polynomial \eqref{Polynomial J(U)=J(t)} has an intermediate field. In fact, from the $J$-invariants we find:
\be
    {\left(U_1(\tau)+4 \right)^3 \ov 3U_1(\tau)+28} =  \left({\eta\left({\tau \ov 2}\right)\ov \eta(\tau)}\right)^{24} =  F_{\Gamma^0(2)}(\tau)~.
\ee
This leads to the following expression for $U(\tau)$:
\be
    U(\tau) = -{4\ov i\sqrt{3}} + 2\left(1+{i\ov \sqrt{3}}\right) {\vartheta_4^8 \ov \vartheta_2^2 \vartheta_3^2}  \left(\vartheta_4^8 (\vartheta_2^2 + \vartheta_3^2)^2 \right)^{ -{1\ov 3}} + 2\left(1-{i\ov \sqrt{3}}\right) { \left(\vartheta_4^8 (\vartheta_2^2 + \vartheta_3^2)^2 \right)^{ {1\ov 3}} \ov \vartheta_2^2 \vartheta_3^2}~.
\ee
As before, the radicands are non-vanishing on $\mathbb{H}$ and $U(\tau)$ is holomorphic. Fundamental domains for this conjugacy class were derived in \cite{STROMBERG2019436}, for instance. These can then be used to show explicitly that $\Gamma(9; 0,3,1,0;12)$ is indeed a subgroup of $\Gamma^0(2)$, or other groups in this conjugacy class.

\medskip

\noindent
\paragraph{$\KK E_2$ configurations.} We would now like to discuss the modular configurations appearing on the CB of the $\KK E_2$ theory. From the list of modular configurations of \cite{Doran:1998hm}, it is clear that there are three modular configurations for the $\KK E_2$ theory, which we list again below:
\be \label{E2 Configurations}
\renewcommand{\arraystretch}{1.3}
\small
\begin{tabular}{ |c||c|c|c|c|c|c|} 
\hline
 $\{F_v\}$ &  $\lambda$ & $M_1$  & $\mathfrak{g}_F$ & ${\rm rk}(\Phi)$ & $\Phi_{\rm tor}$ & $\Gamma \in {\rm PSL}(2,\mathbb{Z})$ \\
\hline \hline
$I_7, 2II, I_1$ & $e^{i\theta_0}$ & $2^{{7\ov4}} e^{i\theta_1}$ & $2 \frak{u}(1)$ &  $2$  & $-$ & $\Gamma^0(7)$\\ \hline
$I_7, I_2, II, I_1$ & $1$ & $\pm \frac{2}{3\sqrt{3}}$ & $A_1 \oplus \frak{u}(1)$ &  $1$  & $-$ & $\Gamma_{721}$\\ \hline
$I_7, III, 2I_1$ & $1$ & $2i$ & $A_1 \oplus \frak{u}(1)$ &  $1$  & $-$ & $\Gamma_{711}$\\ \hline
\end{tabular} 
\ee \renewcommand{\arraystretch}{1}%
in the conventions of \cite{Closset:2021lhd}. One of them corresponds to a congruence subgroup and was also discussed in \cite{Closset:2021lhd}. In particular, the modular function of this congruence subgroup is given by:
\be
    F_{\Gamma^0(7)} = \left( {\eta\left({\tau \ov 7}\right) \ov \eta(\tau)} \right)^4~.
\ee
However, since this configuration already involves two elliptic points, it is difficult to find the BPS spectrum from the modular properties. The reason for this is that the decompositions \eqref{II Decomposition} are not unique and thus involve further physical insights. Alternatively, one can use the methods developed in \cite{Aspman:2021evt} by introducing branch cuts on the upper half-plane, but we leave this discussion for future work.

For the conjugacy classes of the non-congruence subgroups of the other two configurations, fundamental domains were be obtained in \cite{STROMBERG2019436}, for instance. As mentioned before, closed-forms for the modular forms of these non-congruence subgroups are not known and, as a result, their modular transformations are difficult to obtain. Thus, even though we can obtain the series expansion for $U(\tau)$ and its modular transformations from the Seiberg-Witten geometry, matching these with their transformations under the coset representatives is a non-trivial task. We can choose the modular function for $E_2$ with the asymptotics:
\be
    U(\tau) = \left(M \lambda^2\right)^{1\ov 7} q^{- {1\ov 7}} + \ldots~.
\ee
For the $(I_7, I_2, II, I_1)$ configuration, the solution to the equation $J(U) = J(\tau)$ is then given by:
\be
    U(\tau) =  \frak{q}^{- {1\ov 7}} + {5\ov 6\sqrt{3}} + {11\ov 3} \frak{q}^{1\ov 7} + {23\ov 27\sqrt{3}} \frak{q}^{2\ov 7} + {1115 \ov 648} \frak{q}^{3\ov 7} - {5561 \ov 972\sqrt{3}} \frak{q}^{4\ov 7} + \ldots~,
\ee
where we introduced:
\be
    \frak{q} = {2\ov 3\sqrt{3}}q~,
\ee
to simplify the expressions. The non-congruence subgroup associated to the final modular configuration of $\KK E_2$, namely $(I_7, III, 2I_1)$, was studied extensively in \cite{atkin1971modular, Scholl1985, Scholl1988TheLR}, where it was denoted by $\Gamma_{711}$. This group has a unique cusp form of weight $4$ which can be determined from a certain weight $4$ newform on $\Gamma_0(14)$. This is an index $24$ subgroup of ${\rm PSL}(2,\mathbb{Z})$, with four cusps of widths $(14,1,7,2)$ and no elliptic points. The relation between the two subgroups is rather surprising, as $\Gamma_0(14)$ is not a subgroup of $\Gamma_{711}$. The modular function of this subgroup has the following asymptotics:
\be
    U(\tau) = \frak{q}^{-{1\ov 7}} + {i\ov 2} + 5 \frak{q}^{1\ov 7} + 7i \frak{q}^{2\ov 7} + {35\ov 8} \frak{q}^{3\ov 7} - {77i\ov 4} \frak{q}^{4\ov 7} - {231\ov 16}\frak{q}^{5\ov 7} + {129i\ov 112}\frak{q}^{6\ov 7} + \cO(\frak{q})~,
\ee
where we now use:
\be
    \frak{q} = 2i\, q~.
\ee
Let us also note that there is no congruence subgroup that includes $\Gamma_{711}$. This follows from the fact that the only possibilities for such constructions would be index $2$ or $5$ subgroups of ${\rm PSL}(2,\mathbb{Z})$, but ramifications of neither possibility can lead to a width-seven cusp. 

\medskip

\noindent
\paragraph{$\boldsymbol{(I_8, II, 2I_1)}$ configuration for $\KK \t E_1$.} As discussed in \cite{Closset:2021lhd}, this configuration appears on the CB of the $\KK \t E_1$ theory, for:
\be
    \lambda = \pm {16 i \ov 3\sqrt{3}}~.
\ee
In this case, one of the roots of $J = J(\tau)$ has the series expansion:
\be
    U(\tau) = {2 e^{\pi i \ov 8} \ov 3^{3\ov 8} q^{1\ov 8}} + {2i \ov 3 \sqrt{3}} - {4 e^{7\pi i \ov 8} \ov 3^{5\ov 8}} q^{1\ov 8} + {160 e^{\pi i \ov 4} \ov 9~3^{3\ov 4}} q^{1\ov 4} - {64 e^{5\pi i \ov 8} \ov 9~3^{7\ov 8}} q^{3\ov 8} - {512 \ov 81} \sqrt{q} + \ldots ~.
\ee
which, as in some of the previous examples, does not have integer coefficients anymore. The relevant monodromy group was denoted by $\Gamma(10; 0,3,0,1;14)$ in \cite{STROMBERG2019436}. Note that this is not a subgroup of $5A^0$, the unique index $5$ congruence subgroup of ${\rm PSL}(2,\mathbb{Z})$, as the generator $T^8 \not\in 5A^0$. As for the previous configurations, we leave a more detailed study of the $U$-plane of the $\KK \t E_1$ theory for future work.

\medskip

\noindent 
\paragraph{Modular configurations for $\KK E_4$.} We will discuss the remaining three configurations in the context of the $\KK E_4$ theory. From table~\ref{tab:Non-Congruence 1}, these are $(I_5, I_3, III, I_1)$, $(I_5, I_4, II, I_1)$ and $(I_5, I_3, I_2, II)$, with the monodromy groups that we will denote by $\Gamma_{531}$, $\Gamma_{541}$ and $\Gamma_{532}$, respectively. These configurations can be found for:
\bea
    \Gamma_{531}~:~ & \boldsymbol{\chi}^{E_4} = \left( {3\times 3^{3\ov 5} \ov 2^{1\ov 5}}~, {43\times 3^{1\ov 5} \ov 4\times 2^{2\ov 5}}~, {17 \ov 2^{4\ov 5}\times 3^{3\ov 5}}~, {115 \ov 6\times 2^{3\ov 5} \times 3^{1\ov 5}} \right)~,\\
    \Gamma_{541}~:~ & \boldsymbol{\chi}^{E_4} = \left( {155 \ov 16\times 3^{3\ov 5}}~, {25 \ov 2\times 3^{1\ov 5}}~, {70 \ov 9\times 3^{2\ov 5}}~, {145 \ov 6\times 3^{4\ov 5}} \right)~,\\
    \Gamma_{532}~:~ & \boldsymbol{\chi}^{E_4} = \left( {5 \ov 2^{1\ov 5}}~,  {25 \ov 4\times 2^{1\ov 5}}~, -{5 \ov 2^{4\ov 5}}~, -{5 \ov 2\times 2^{3\ov 5}} \right)~.
\eea
As usual, these can be found from the branching of the characters of $E_4$ into representations of the unbroken flavour symmetry. The first few coefficients of the Fourier expansion are given by:
\bea    \label{E4 generic expansion}
    U(\tau) =&  q^{-{1\ov 5}} + {\chi_4 \ov 5} + {1\ov 25}(60\chi_1 + 2\chi_2\chi_3 - 3\chi_4^2)q^{1\ov 5} + \\
    & + {1\ov 125}(25\chi_1\chi_2^2 + 25\chi_3^2 - 120\chi_1\chi_4 + 14\chi_4^3 + 900\chi_2 - 35\chi_2\chi_3\chi_4) q^{2\ov 5} + \ldots
\eea
While we do not have closed-form expressions for the above modular functions, we can still pick the correct fundamental domains from the respective conjugacy classes. This is done by using the massless configuration, as well as the BPS quiver, as follows. Recall that the massless configuration had monodromy group $\Gamma^1(5)$, with the bulk cusps $(I_1, I_1, I_5)$ positioned at $\tau = \left(1, {5\ov 2}, 3\right)$ and the associated BPS states:
\be
    I_1: (1,0)~, \qquad I_1:~ (-2,5)~, \qquad I_5:~ 5(1,-3)~.
\ee
For the $\Gamma_{531}$ configuration, merging the cusp at $\tau = {5\ov 2}$ with two states of the $I_5$ cusp at $\tau = 3$ leads to:
\bea
    \Gamma_{531}~:~ \quad (I_1, I_3, III)~:~ \qquad \tau = \left( 0, 3, {5 + i \ov 2} \right)~,
\eea
where the position of the $III$ elliptic point is given by the action of $T^3ST$ on $\tau_{III^{(0)}} = i$. Then, the $III$ monodromy reads:
\bea
    \Gamma_{531}~:~ & \qquad \mathbb{M}_{III} = \left(T^3ST\right) S^{-1} \left(T^3ST\right)^{-1} = \mathbb{M}_{(-2,5)} \left(\mathbb{M}_{(1,-3)}\right)^2~.
\eea
The resulting fundamental domain is indeed part of the conjugacy class of $\Gamma_{531}$ in \cite{STROMBERG2019436}. Similarly, for the $\Gamma_{541}$ group, `merging' the $(-2,5)$ state with one of the $(1,-3)$ states leads to:
\bea
    \Gamma_{541}~:~ \quad (I_1, I_4, II)~:~ \qquad \tau = \left( 0, 3, {5 + 3 \tau_{II^{(0)}} \ov 2 + \tau_{II^{(0)}} } \right)~,
\eea
where the position of the $II$ elliptic point is given by the action of $T^3ST^2$ on $\tau_{II^{(0)}} = e^{2\pi i \ov 3}$. The monodromy of the $II$ elliptic point becomes:
\bea
    \Gamma_{541}~:~ & \qquad \mathbb{M}_{II} = \left(T^3ST^2\right) \left(ST\right)^{-1} \left(T^3ST^2\right)^{-1} = \mathbb{M}_{(-2,5)} \mathbb{M}_{(1,-3)}~.
\eea
As before, there exists such a fundamental domain in the conjugacy class of $\Gamma_{541}$ in the results of \cite{STROMBERG2019436}. Finally, a possible choice of fundamental domain of $\Gamma_{532}$ can be inferred from the BPS quiver \eqref{quiver E4 I5 2III I1}, leading to:
\bea
    \Gamma_{532}~:~ \quad (II, I_2, I_3)~:~ \qquad \tau = \left( e^{2\pi i \ov 3}, 2, 3 \right)~,
\eea
which is again in agreement with \cite{STROMBERG2019436}. Note that the $II$ singularity at $\tau_{II^{(0)}} = e^{2\pi i \ov 3}$ is constructed from:
\bea
    \Gamma_{532}~:~ & \qquad \mathbb{M}_{II} =  \left(ST\right)^{-1}  = \mathbb{M}_{(1,0)} \mathbb{M}_{(0,1)}~.
\eea


\subsubsection{${\rm rk}(\Phi) = 2$ Configurations.}

The remaining modular surfaces are listed in table~\ref{tab:Non-Congruence 2}.%
\renewcommand{\arraystretch}{1}
\begin{table}[h]
\small
\centering
\begin{tabular}{ |c||c|c|c|c|} 
 \hline
$\{F_v\}$ & $\Phi_{\rm tor}$& \textit{Field theory} & $\Fg_F$ & $\Gamma$ 
\\
\hline 
 \hline
\multirow{4}{*}{$I_6, III, II, I_1$} & \multirow{4}{*}{$-$} & $\KK E_8$ & $A_5 \oplus A_1 \oplus 2\frak{u}(1)$ &  \multirow{4}{*}{$\Gamma(7; 0, 2, 1, 1; 6)$}   \\ \cline{3-4}
&  &$\KK E_3$ & $A_1 \oplus 2\frak{u}(1)$  &    \\ \cline{3-4}
&  &\text{MN} $E_8$ & $A_5 \oplus A_1 \oplus 2\frak{u}(1)$  &   \\ \cline{3-4}
&  &\text{MN} $E_7$ & $A_5\oplus 2\frak{u}(1)$  &   \\ \hline 
\hline
\multirow{4}{*}{$I_5, III, I_2, II$} & \multirow{4}{*}{$-$} & $\KK E_7$ & $A_4 \oplus 2A_1 \oplus 2\frak{u}(1)$ &  \multirow{4}{*}{$\Gamma(7; 0, 2, 1, 1; 10)$}   \\ \cline{3-4}
&  &$\KK E_4$ & $2A_1 \oplus 2\frak{u}(1)$  &    \\ \cline{3-4}
&  &\text{MN} $E_8$ & $A_4 \oplus 2A_1 \oplus 2\frak{u}(1)$  &   \\ \cline{3-4}
&  &\text{MN} $E_7$ & $A_4 \oplus A_1 \oplus 2\frak{u}(1)$  &  \\ \hline 
\hline
\multirow{4}{*}{$I_4, I_3, III, II$} & \multirow{4}{*}{$-$} & $\KK E_6$ & $A_3 \oplus A_1 \oplus 2\frak{u}(1)$ &  \multirow{4}{*}{$\Gamma(7; 0, 2, 1, 1; 12)$}   \\ \cline{3-4}
&  &$\KK E_5$ & $A_2 \oplus A_1 \oplus 2\frak{u}(1)$  &    \\ \cline{3-4}
&  &\text{MN} $E_8$ & $A_3 \oplus A_2 \oplus A_1 \oplus 2\frak{u}(1)$  &  \\ \cline{3-4}
&  &\text{MN} $E_7$ & $A_3 \oplus A_2 \oplus 2\frak{u}(1)$  &  \\ \hline 
\end{tabular}
    \caption{Modular rational elliptic surfaces for non-congruence subgroups with ${\rm rk}(\Phi) = 2$. }
    \label{tab:Non-Congruence 2}
\end{table}
These are all index $7$ subgroups of ${\rm PSL}(2,\mathbb{Z})$ and cannot be subgroups of other congruence groups. These are in fact the simplest examples of non-congruence subgroups. The subgroups associated with the configurations $(I_5, III, I_2, II)$ and $(I_4, I_3, III, II)$ were considered in \cite{Scholl2010OnLR}, where they were denoted by $\Gamma_{52}$ and $\Gamma_{43}$, respectively. It was shown there that the spaces of cusp forms of weight $4$ for these groups are generated by certain cusp forms of $\Gamma_0(35)$ and $\Gamma_0(28)$, respectively.

We will not spend too much time describing these groups, but rather refer the reader to some recent work from \cite{2020arXiv200701336F}, where all index $7$ non-congruence subgroups have been described. As before, they can be found on the CB of the KK theories from the branching of the $E_n$ characters.

Since the previous section ended with configurations of $\KK E_4$, let us start with the $\boldsymbol{(I_5, III, I_2, II)}$ configuration, which can be easily found by working with the gauge theory parameters. Setting $M_1 = M_2$ and $\lambda = 1$ we first find the $(I_5,2I_2, 3I_1)$ configuration and by tuning the remaining parameters to one of the possibilities:
\be
    M_3 = -{1\ov 4M_1^3} + {1\ov M_1}~, \qquad M_1 = \pm {1\ov 4}\sqrt{7\ov 2}~,
\ee
we recover the $(I_5, III, I_2, II)$ configuration. In terms of the characters, one has, for example:
\bea
    \Gamma_{52}~:~  \boldsymbol{\chi}^{E_4} = \left( {433 \ov 16\times 7^{4\ov 5}}~, {323\ov 14 \times 7^{3\ov 5}}~,-{26\ov 7\times 7^{1\ov5}}~, -{115\ov 14 \times 7^{2\ov 5}}  \right)~.
\eea
with the asymptotics of $U(\tau)$ given in \eqref{E4 generic expansion}. Once again, we try to determine the fundamental domain from the previously studied configurations and check agreement with allowed domains in \cite{STROMBERG2019436}. As such, a possibility is given by:
\bea
    \Gamma_{52}~:~ \quad (II^{(0)}, III^{(2)}, I_2)~:~ \qquad \tau = \left( e^{2\pi i \ov 3}, 2+i, 3 \right)~,
\eea
which is again described by the BPS quiver \eqref{quiver E4 I5 2III I1}, with the elliptic points having monodromies:
\bea
    \Gamma_{52}~:~ & \qquad \mathbb{M}_{II^{(0)}} =  \left(ST\right)^{-1}  = \mathbb{M}_{(1,0)} \mathbb{M}_{(0,1)}~, \\
    & \qquad \mathbb{M}_{III^{(2)}} =  T^2S^{-1}T^{-2}  = \left(\mathbb{M}_{(1,-2)}\right)^2 \mathbb{M}_{(1,-3)}~.
\eea
The next configuration, $\boldsymbol{(I_6, III, II, I_1)}$, can be found on the CB of $\KK E_3$, for the parameter values \cite{Closset:2021lhd}:
\bea
   \Gamma_{61}~:~ \left( \lambda, M_1, M_2 \right) = (1, \delta - i, \delta + i)~, \qquad \delta = {3^{3\ov 4} \ov 2}e^{5\pi i \ov 12}~,
\eea
with the modular function:
\be
    U(\tau) = {\left(-1+3i\sqrt{3}\right)^{1\ov 6} \ov \sqrt{2}} q^{-{1\ov 6}} + {(-3)^{3\ov 4} \ov 4 + 3(-1)^{5\ov 6}\sqrt{3}} - {48i\sqrt{2}(-2i+\sqrt{3}) \ov (-1+3i\sqrt{3})^{13\ov 6}} q^{1\ov 6} + \ldots ~.
\ee
Let us recall that the massless $\KK E_3$ configuration had bulk cusps $(I_1, I_3, I_2)$ at $\tau = (0,2,3)$, with light BPS states $(1,0)$, $3(-1,2)$ and $2(1,-3)$, and the quiver depicted in \eqref{quiver E3a}. Mutating on $\hat{\gamma}_5$ leads to the slightly modified quiver:
\bea
 \begin{tikzpicture}[baseline=1mm]
\node[] (1) []{$\CE_{\gamma_1=(1,0)}$};
\node[] (2) [right = of 1]{$\CE_{\gamma_{2,3,4}=(0,-1)}$};
\node[] (5) [below = of 1]{$\CE_{\gamma_{5}=(-1,3)}$};
\node[] (6) [below = of 2]{$\CE_{\gamma_{6}=(1,-3)}$};
\draw[->>>-=0.5] (1) to   (5);
\draw[->-=0.3] (5) to   (2);
\draw[->-=0.5] (2) to   (6);
\draw[->-=0.5] (2) to   (1);
\draw[->>>-=0.3] (6) to   (1);
\end{tikzpicture}
\eea
This quiver then implies that a possible fundamental domain is given by:
\bea
    \Gamma_{61}~:~ \quad (I_1, III^{(2)}, II^{(4)})~:~ \qquad \tau = \left( 0, 2+i, 4+e^{2\pi i \ov 3} \right)~,
\eea
with the monodromies:
\bea
    \Gamma_{61}~:~ & \qquad \mathbb{M}_{III^{(2)}} =  T^2S^{-1}T^{-2} = \mathbb{M}_{(-1,3)} \left(\mathbb{M}_{(0,-1)}\right)^2~, \\
    & \qquad \mathbb{M}_{II^{(4)}} =   \left(ST\right)^{-1}  = \mathbb{M}_{(0,-1)} \mathbb{M}_{(1,-3)}~.
\eea
Finally, let us also mention the $\boldsymbol{(I_4, I_3, III, II)}$ configuration, which can be found on the CB of $\KK E_5$, for the characters:
\bea
    \Gamma_{43}~:~  \boldsymbol{\chi}^{E_5} = \left( {178i \ov 7\sqrt{7}}~, -{10805\ov 343}~, -{912(-1)^{3\ov 4} \ov 49\times 7^{1\ov 4}}~, -{27176i \ov 343\sqrt{7}}~, -{176(-1)^{1\ov 4} \ov 7\times 7^{3\ov 4}}  \right)~.
\eea
Its modular function has the asymptotics:
\be
    U(\tau) = q^{-{1\ov 4}} - {44(-1)^{1\ov 4} \ov 7\times 7^{3\ov 4}} + {2076i \ov 49\sqrt{7}} q^{1\ov 4} - {88064(-1)^{3\ov 4} \ov 2401\times 7^{1\ov 4}} q^{1\ov 2} + {206562 \ov 117649} q^{3\ov 4} + \cO(q)~,
\ee
with the coefficients appearing to satisfy the unbounded denominator conjecture.


\section{Rank-one SCFTs} \label{Rank-one SCFTs}

In this section, we apply similar arguments to some of the 4d $\CN=2$ rank-one SCFTs. The language of rational elliptic surfaces should include the theories whose Coulomb branches involve `undeformable' singularities, but it is not clear at the moment what is the interpretation of the Mordell-Weil group in this context. Additionally, the flavour algebra cannot be directly obtained using the standard F-theory arguments, as depicted in table~\ref{tab:kodaira}, but it was still determined in \cite{Caorsi:2018ahl} from the so-called `flavour root system'. 

Let us first note that the language of rational elliptic surfaces can still be used to find the allowed configurations of singular fibers as follows. To a given (maximal) deformation pattern we associate in the usual way a naive fibre contribution $T_{def}$ to the trivial lattice \cite{schuttshioda}. Then, we claim that the allowed configurations of singular fibres are those whose naive flavour lattice contain this $T_{def}$ lattice as a sublattice. Obviously, this condition is equivalent to having frozen singularities on the Coulomb branch. We refer to section \ref{section: intro} for more details.

Here, we will discuss the Argyres-Douglas theories \cite{Argyres:1995jj, Argyres:1995xn}, as well as the $\cN=2^*$ theory. Our analysis can be extended to the so-called $I_4$ series of \cite{Argyres:2015ffa, Argyres:2015gha, Argyres:2016xmc, Argyres:2016xua}, which involves the Argyres-Wittig SCFTs \cite{Argyres:2007tq}. We will see that we can bypass the issues outlined above and study the modular properties and BPS quivers of these theories.


\subsection{Argyres-Douglas Theories}

The theories with trivial maximally deformed Coulomb branch discussed in \cite{Closset:2021lhd} serve as a tool in deriving the BPS quivers of the theories with undeformable singularities. Conformal theories are, in particular, different compared to the $\KK E_n$ theories considered so far, having an additive fiber at infinity rather than a multiplicative one, which is usually equivalent to an elliptic point, rather than a cusp. We start first by applying our logic to the Argyres-Douglas theories, with the aim of clarifying the decompositions \eqref{II Decomposition}, \eqref{III Decomposition} and \eqref{III Decomposition v2}.  

\medskip

\noindent
\paragraph{$H_0$ theory.} The $H_0$ theory is the first SCFT of this type, being found originally on the CB of the pure $SU(3)$ theory, as well as on the CB of the $SU(2)$ $N_f = 1$ theory \cite{Argyres:1995jj, Argyres:1995xn}. Its Seiberg-Witten geometry is given by:
\be \label{SW curve AD H0}
    y^2 = x^3 + x M + u~.
\ee
The theory has no flavour symmetry, being described by the fixed fiber at infinity $F_{\infty} = II^*$. There are thus only two allowed configurations, namely $(II^*,II)$ and $(II^*, 2I_1)$, which correspond to the massless configuration and generic values for the mass parameter. Neither of these is modular, but the BPS quiver can be deduced by embedding the theory on the CB of SQCD theories, for instance. 

In \cite{Closset:2021lhd, Aspman:2021vhs} for example, by embedding the AD theory on the CB of $SU(2)\, N_f = 1$, it was deduced from the mutually non-local light BPS states that the BPS quiver becomes:
\bea\label{quiver AD H0}
 \begin{tikzpicture}[baseline=1mm]
\node[] (1) []{$\CE_{\gamma_{1}}$};
\node[] (2) [right = of 1] {$\CE_{\gamma_{2}}$};
\draw[->-=0.5] (1) to   (2);
\end{tikzpicture}
\eea
in agreement with older results \cite{Alim:2011kw}. This is also in agreement with the more general `decomposition' of type $II$ singularities proposed in \eqref{II Decomposition}. That is, for an elliptic point situated at $\tau = n+e^{2\pi i \ov 3}$, for some integer $n$, we had the following possible scenarios based on the monodromies:
\be 
    II^{(n)}~:~ (1, 1-n) \oplus (1,-n)~, \qquad (1,-n) \oplus (0,1) ~, \qquad (0,1) \oplus (1,1-n)~,
\ee
which all reproduce the \eqref{quiver AD H0} quiver.

\medskip

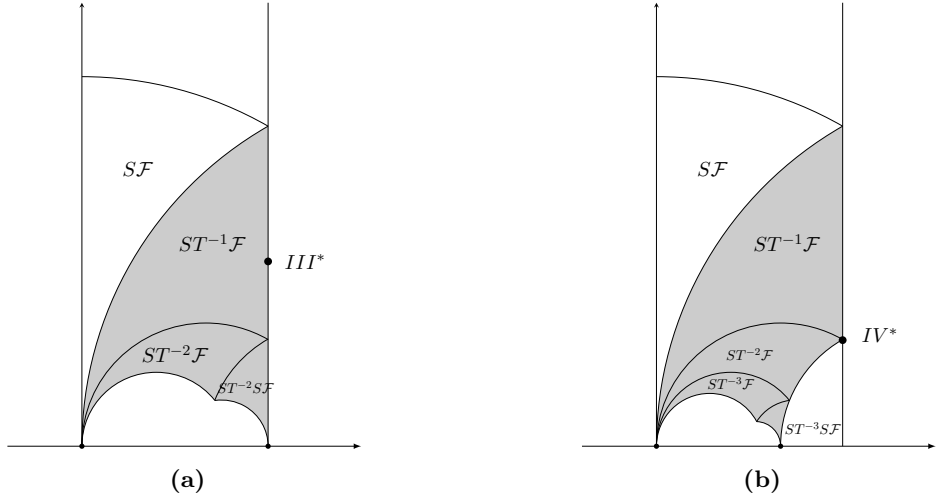
\begin{figure}[t]
    \centering
    \begin{subfigure}{0.5\textwidth}
    \centering
     \scalebox{0.7}{


\begin{tikzpicture}[scale=7]

    \pgfdeclarelayer{background}
    \pgfsetlayers{background,main}

    \pgfmathsetmacro{\myxlow}{-0.1}
    \pgfmathsetmacro{\myxhigh}{0.5+0.1}
    \pgfmathsetmacro{\myiterations}{1}

    \draw[-latex](\myxlow-0.1,0) -- (\myxhigh+0.15,0);
    \pgfmathsetmacro{\succofmyxlow}{\myxlow+0.5}
    
    \draw[-stealth, very thin] (0,0)--(0,1.2) node[above] {}; 
    \begin{scope}
        
        \draw[very thin, black] (0,0) arc(180:120:1);
        \draw[very thin, black] (0,0) arc(180:60:1/3);
        \draw[very thin, black] (0,0) arc(180:38.3:1/5);
        \draw[very thin, black] (0.5,0.866) arc(60:90:1);
        
        \draw[very thin, black] (1/2, 0) arc(0:98.3:1/8);
        \draw[very thin, black] (0.5, 0.29) arc(120:120+38.4:1/3); 
    
        \draw[very thin, black] (1/2,0) -- (1/2, 1.2);

    \end{scope}

    \begin{scope}
        \begin{pgfonlayer}{background}
            \clip
                (0,0) 
                {-- (0,0) arc(180:120:1)
             -- (1/2,0) arc(0:98.3:1/8)
             -- (+0.357, 0.124) arc(38.3:180:1/5)
             }
                -- (0, 0) -- cycle
            ;
            \fill[gray,opacity=0.4] (-0.1,-0.1) rectangle (0.6,1.2);
        \end{pgfonlayer}
    \end{scope}

    \begin{scope}
        \node at (0.15,0.75) {$S\mathcal{F}$};
        \node at (0.35,0.55) {$ST^{-1}\mathcal{F}$};
        \node at (0.25,0.25) {$ST^{-2}\mathcal{F}$};
        \node at (0.44,0.16) [scale=0.7] {$ST^{-2}S\mathcal{F}$};
        
        \fill (0,0)  circle[radius=0.2pt];
        \fill (1/2,0)  circle[radius=0.2pt];
        \fill (1/2,1/2)  circle[radius=0.3pt];
        
        \node at (0.6, 0.5) {$III^*$};
        
    \end{scope}
    
\end{tikzpicture}

}
    \caption{ }
    \label{fig:H1-IIIs-I2-I1 Configuration}
    \end{subfigure}%
    \begin{subfigure}{0.5\textwidth}
    \centering
    \scalebox{0.7}{


\begin{tikzpicture}[scale=7]

    \pgfdeclarelayer{background}
    \pgfsetlayers{background,main}

    \pgfmathsetmacro{\myxlow}{-0.1}
    \pgfmathsetmacro{\myxhigh}{0.5+0.1}
    \pgfmathsetmacro{\myiterations}{1}

    \draw[-latex](\myxlow-0.1,0) -- (\myxhigh+0.15,0);
    \pgfmathsetmacro{\succofmyxlow}{\myxlow+0.5}
    
    \draw[-stealth, very thin] (0,0)--(0,1.2) node[above] {}; 
    \begin{scope}
        
        \draw[very thin, black] (0,0) arc(180:120:1);
        \draw[very thin, black] (0,0) arc(180:60:1/3);
        \draw[very thin, black] (0,0) arc(180:38.3:1/5);
        \draw[very thin, black] (0,0) arc(180:27.7:1/7);
        \draw[very thin, black] (0.5,0.866) arc(60:90:1);
        
        \draw[very thin, black] (+0.357, 0.124) arc(98.3:148:1/8);
        \draw[very thin, black] (1/3, 0) arc(180:120:1/3); 
        \draw[very thin, black] (1/3, 0) arc(0:88:1/15); 
        
        \draw[very thin, black] (1/2,0) -- (1/2, 1.2);
        
    \end{scope}

    \begin{scope}
        \begin{pgfonlayer}{background}
            \clip
                (0,0) 
                {-- (0,0) arc(180:27.7:1/7)
             -- (0.269,0.0666) arc(88:0:1/15)
             -- (1/3,0) arc(180:120:1/3)
             -- (1/2, 0.866) arc(120:180:1)
             }
                -- (0, 0) -- cycle
            ;
            \fill[gray,opacity=0.4] (-0.1,-0.1) rectangle (0.6,1.2);
        \end{pgfonlayer}
    \end{scope}

    \begin{scope}
        \node at (0.15,0.75) {$S\mathcal{F}$};
        \node at (0.35,0.55) {$ST^{-1}\mathcal{F}$};
        \node at (0.25,0.25) [scale=0.7] {$ST^{-2}\mathcal{F}$};
        \node at (0.2,0.17) [scale=0.7] {$ST^{-3}\mathcal{F}$};
        \node at (0.42,0.05) [scale=0.7] {$ST^{-3}S\mathcal{F}$};
        
        \fill (0,0)  circle[radius=0.2pt];
        \fill (1/3,0)  circle[radius=0.2pt];
        \fill (1/2,0.287)  circle[radius=0.3pt];
        
        \node at (0.6, 0.3) {$IV^*$};
        
    \end{scope}
    
\end{tikzpicture}

}
    \caption{ }
    \label{fig:H2-IVS-I3-I2 Configuration}
    \end{subfigure}
    \caption{Fundamental domains for  $(a)~, (III^*, I_2, I_1)$ configuration on the CB of the $H_1$ AD theory, $(b)~, (IV^*, I_3, I_1)$ configuration on the CB of the $H_2$ AD theory. Monodromy groups are conjugate to $\Gamma_0(2)$ and $\Gamma_0(3)$, respectively.}
\end{figure}

\noindent
\paragraph{$H_1$ theory.} The next AD theory is usually referred to as the $H_1$ theory, having $A_1$ flavour algebra and $SU(2)/\mathbb{Z}_2$ flavour symmetry \cite{Closset:2021lhd, Buican:2021xhs}, with the SW geometry given by:
\be \label{SW curve AD H1}
    y^2 = x^3 + ux + u M_{2/3} - M_2~.
\ee
The theory is described by $F_{\infty} = III^*$, leading to four allowed configurations, depending on the values of the two mass parameters. Out of these four, two configurations are modular, namely $(III^*, I_2, I_1)$ and $(III^*, II, I_1)$. %
\be \label{AD H1 Configurations}
\renewcommand{\arraystretch}{1.3}
\small
\begin{tabular}{ |c||c|c|c|c|c|c|} 
\hline
 $\{F_v\}$ &  $M_{2/3}$ & $M_2$  & $\mathfrak{g}_F$ & ${\rm rk}(\Phi)$ & $\Phi_{\rm tor}$ & $\Gamma \in {\rm PSL}(2,\mathbb{Z})$ \\
\hline \hline
$III^*, III$ & $0$ & $0$ & $A_1$ &  $0$  & $\mathbb{Z}_2$ & $-$\\ \hline
$III^*, I_2, I_1$ & $M$ & $-M^3$ & $A_1$ &  $0$  & $\mathbb{Z}_2$ & $\Gamma_0(2)$\\ \hline
$III^*, II, I_1$ & $M$ & $0$ & $\frak{u}(1)$ &  $1$  & $-$ & $\Gamma(1)$\\ \hline
$III^*, 3I_1$ & $M_{2/3}$ & $M_2$ & $\frak{u}(1)$ &  $1$  & $-$ & $-$\\ \hline
\end{tabular}
\ee 
\renewcommand{\arraystretch}{1}%
As pointed out in \eqref{AD H1 Configurations}, the $(III^*, I_2, I_1)$ configuration appears for $M_2 = -M_{2/3}^3$, so let us set $M_{2/3} = M$. Then, the $J$-invariant reads:
\be
    J(u) = { 4u^3 \ov (u+3M^2)^2 (4u+3M^2)}~. 
\ee
It is not directly obvious that $u(\tau)$ will be a modular function for a congruence subgroup in the conjugacy class of $\Gamma_0(2)$, but this can be proved as follows. First, the Hauptmodul of $\Gamma_0(2)$ satisfies:\footnote{This can be shown by finding the same configuration on the CB of $\KK E_8$. A possibility is $\boldsymbol{\chi}^{E_8} = \{ 3, -2496, -494, -12320, -17290, -1216, 140, 24 \}$.}
\be
    F_{\Gamma_0(2)}(\tau) = \left( {\eta\left({\tau}\right) \ov \eta(2\tau)} \right)^{24}~, \qquad j(\tau) = {\left(256 + F_{\Gamma_0(2)}\right)^3 \ov \left(F_{\Gamma_0(2)}\right)^2}~.
\ee
Thus, one can immediately see that a possible solution for $u(\tau)$ is given by:
\be
    u_0(\tau) = -{3M^2 \ov 4} \, {  256 + F_{\Gamma_0(2)} \ov 64 + F_{\Gamma_0(2)}}~.
\ee
Some care needs to be taken when identifying $u(\tau)$. As the monodromy group is conjugate to $\Gamma_0(2)$, let us fix the $I_2$ cusp at $\tau = 0$, for the sake of the argument. For the above expression, we have:
\be
   u_0(\tau) + {3M^2 \ov 4}  = -192\,{3M^2 \ov 4}  \left( q - 40 q^2 + 1324 q^3 + \cO(q^4) \right)~,
\ee
which corresponds to the expansion around the cusp of width-one at $u = -{3M^2 \ov 4}$. As a result, we take:
\be \label{H1 Solution u}
    u(\tau) = -{3M_2 \ov 4} \, {256 + F_1 \ov 64 + F_1}~, \qquad \quad F_1 = \left( \sqrt{2} \, {\eta\left({\tau }\right) \ov \eta\left({\tau \ov 2}\right)} \right)^{24}~,
\ee
where $F_1$ is the $S$-transformation of $F_{\Gamma_0(2)}$. It also follows that the type $III^*$ elliptic point lies in the $T$-transform of the region corresponding to \eqref{H1 Solution u}, with the fundamental domain shown in figure~\ref{fig:H1-IIIs-I2-I1 Configuration}. %
This domain can be obtained from that of $\Gamma_0(2)$ by realizing that the  $T^0 \cF$ and $S \cF$ regions are identified with $ST^{-2}S\cF$ and $ST^{-2}\CF$, respectively, since:
\be
    ST^{-2}S = \left( \begin{matrix} -1 & 0 \\ -2 & -1 \end{matrix} \right) \in \Gamma_0(2)~.
\ee
From this domain we can then find the associated BPS quiver:
\bea\label{quiver AD H1 v1}
 \begin{tikzpicture}[baseline=1mm]
\node[] (1) []{$\CE_{\gamma_{1,2} = (1,0)}$};
\node[] (3) [right = of 1] {$\CE_{\gamma_{3} = (-2,1)}$};
\draw[->-=0.5] (1) to   (3);
\end{tikzpicture}
\eea
which, after the mutation sequence $\gamma_1 - \gamma_3$ leads to the usual quiver of the $H_1$ AD theory \cite{Alim:2011kw}, namely:
\bea
 \begin{tikzpicture}[baseline=1mm]
\node[] (1) []{$\CE_{\gamma_{1}} $};
\node[] (3) [right = of 1] {$\CE_{\gamma_{3}}$};
\node[] (2) [below= of 1] {$\CE_{\gamma_{3}}$};
\draw[->-=0.5] (1) to   (3);
\draw[->-=0.5] (3) to   (2);
\draw[->-=0.5] (2) to   (1);
\end{tikzpicture}
\eea
As for the $H_0$ theory, these quivers can be also obtained by embedding the theory on the Coulomb branches of SQCDs. At this stage let us remark that:
\be
    \mathbb{M}_{(1,0)}^2 \mathbb{M}_{(-2,1)} = (ST^{-1}) S^{-1} (ST^{-1})^{-1}~,
\ee
which, similar to \eqref{III Decomposition}, \eqref{III Decomposition v2}, corresponds to a decomposition of the $III$ singular point into the states $2(1,0) \oplus (-2,1)$. Thus, we see that these decompositions can be argued from the fundamental domains of the Argyres-Douglas theories. The other possibilities correspond to different choices of fundamental domains and, in certain cases, to different congruence subgroups in the same conjugacy class. For instance, the decomposition into $2(1,0) \oplus (1,-1)$ can be seen from a fundamental domain of $\Gamma_{\theta} \cong \Gamma_0(2)$, which is the index $3$ congruence subgroup generated by $T^2$ and $S$, with the coset representatives: $(S, ST^{-1}, TST)$.

\medskip
\noindent
\paragraph{$H_2$ theory.} The final AD theory is referred to as the $H_2$ theory and has $SU(3)/\mathbb{Z}_3$ flavour symmetry \cite{Closset:2021lhd}, with the SW geometry given by:
\be \label{SW curve AD H2}
    y^2 = x^3 + x( u M_{1/2} +  M_{2}) + (u^2 + M_3)~.
\ee
The theory is described by $F_{\infty} = IV^*$, with $8$ allowed configurations in Persson's classification. Out of these, only $3$ are modular: $(IV^*, I_3, I_1)$, $(IV^*, III, I_1)$ and $(IV^*, I_2, II)$. We will focus on the former, which is extremal and was analysed in \cite{Closset:2021lhd} in the context of the $\KK E_6$ theory. This configuration can be found for instance for:
\be
    M_{1/2} = M~, \qquad M_{2} = {1\ov 48}M^4~, \qquad M_3 = -{1\ov 1728}M^6~,
\ee
in which case the discriminant and $J$-invariant become:
\be
    \Delta(u) = (24u + M)^3 (216u + 5M^3)~, \qquad \quad J(u) = {4M^3(48u + M^3)^3 \ov (24u+M^3)^3 (216 u +5M^3)}~.
\ee
We proceed as before, by finding first the same configuration on the CB of $\KK E_8$, such that:\footnote{This can be found for $\boldsymbol{\chi}^{E_8} = \{ -13, 135, -8, 411, 390, 71, 5, 5 \}$.}
\be
    F_1(\tau) = \left( \sqrt{3} \, { \eta(\tau) \ov \eta\left( {\tau \ov 3}\right) }\right)^{12}~, \qquad \quad j(\tau) = { (F_1 + 27)(F_1 + 243)^3 \ov F_1^3}~.
\ee
Thus, we find that:
\be
    u(\tau) = - {M_1^3 \ov 216}\, {243 + 5F_1 \ov 27 + F_1}~.
\ee
The fundamental domain for this configuration is shown in figure~\ref{fig:H2-IVS-I3-I2 Configuration}. The BPS quiver for the configuration $(I_3, I_1)$ becomes: 
\bea\label{quiver AD H2 v2}
 \begin{tikzpicture}[baseline=1mm]
\node[] (1) []{$\CE_{\gamma_{1,2,3} = (1,0)}$};
\node[] (4) [right = of 1] {$\CE_{\gamma_{4} = (-3,1)}$};
\draw[->-=0.5] (1) to   (4);
\end{tikzpicture}
\eea
which, after the mutation sequence $\gamma_1 - \gamma_4 - \gamma_2$ leads to:
\bea
 \begin{tikzpicture}[baseline=1mm]
\node[] (1) []{$\CE_{\gamma_1} $};
\node[] (2) [right = of 1] {$\CE_{\gamma_2}$};
\node[] (4) [below = of 1] {$\CE_{\gamma_4}$};
\node[] (3) [below = of 2] {$\CE_{\gamma_3}$};
\draw[->-=0.5] (1) to   (2);
\draw[->-=0.3] (2) to   (4);
\draw[->-=0.5] (4) to   (3);
\draw[->-=0.3] (3) to   (1);
\end{tikzpicture}
\eea
This is a known quiver for $H_2$ -- see \textit{e.g.} \cite{Cecotti:2013sza}, where this quiver is denoted by $Q(1,1)$.


\subsection{$I_0^*$ SCFTs}

In this subsection, we finally discuss the $\cN=2^*$ theory, which is the first example of a theory with undeformable singularities on the Coulomb branch. The Coulomb branch of this theory shows some similarities to the $SU(2)$ $N_f = 4$ theory, as recently discussed in \cite{Aspman:2021evt}, both being described by the $(2I_0^*)$ rational elliptic surface at the conformal point. There are in fact three such SW geometries \cite{Argyres:2015ffa, Argyres:2015gha, Argyres:2016xmc, Argyres:2016xua}, as the $\cN = 4$ theory, with flavour algebra $sp(2)$, has two allowed deformation patterns, namely $(2I_1, I_4)$ and $(3I_2)$, respectively. The two geometries are related by a change of normalization of the electric and magnetic charges, first noted in \cite{Seiberg:1994aj}, with the rational elliptic surfaces related by a two-isogeny \cite{Moore:1997pc}. 


\medskip

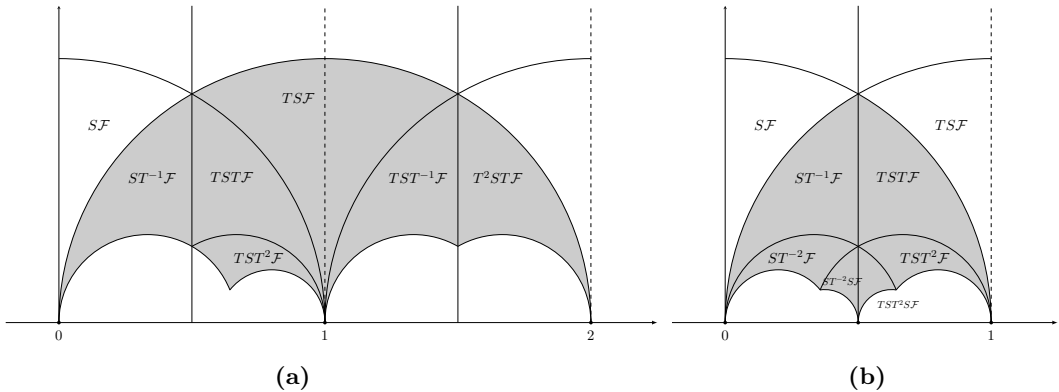
\begin{figure}[t]
    \centering
    \begin{subfigure}{0.5\textwidth}
    \centering
     \scalebox{0.5}{


\begin{tikzpicture}[scale=7]

    \pgfdeclarelayer{background}
    \pgfsetlayers{background,main}

    \pgfmathsetmacro{\myxlow}{-0.1}
    \pgfmathsetmacro{\myxhigh}{2+0.1}
    \pgfmathsetmacro{\myiterations}{1}

    \draw[-latex](\myxlow-0.1,0) -- (\myxhigh+0.15,0);
    \pgfmathsetmacro{\succofmyxlow}{\myxlow+0.5}
    
    \draw[-stealth, very thin] (0,0)--(0,1.2) node[above] {}; 
    \begin{scope}
        
        \draw[very thin, black] (0,0) arc(180:0:1);
        \draw[very thin, black] (0,0) arc(180:60-38.4:1/3);


        \draw[very thin, black] (1/2,0) -- (1/2, 1.2);
        \draw[very thin, black, dashed] (1,0) -- (1, 1.2);
        \draw[very thin, black] (3/2,0) -- (3/2, 1.2);
        \draw[very thin, black, dashed] (2,0) -- (2, 1.2);
        
        \draw[very thin, black] (1,0) arc(0:90:1);
        \draw[very thin, black] (1,0) arc(0:120:1/3);
        \draw[very thin, black] (1,0) arc(0:180-38.3:1/5);
        
        \draw[very thin, black] (1,0) arc(180:60:1/3);
        \draw[very thin, black] (1,0) arc(180:90:1);
        
        \draw[very thin, black] (2,0) arc(0:120:1/3);

    \end{scope}

    \begin{scope}
        \begin{pgfonlayer}{background}
            \clip
                (0,0) 
                {-- (0,0) arc(180:60-38.4:1/3)
                -- (1-0.357, 0.124) arc(180-38.3:0:1/5)
             -- (1,0) arc(180:60:1/3)
             -- (1.5, 0.288675) arc(120:0:1/3)
             -- (2,0) arc(0:180:1)
             }
                -- (0, 0) -- cycle
            ;
            \fill[gray,opacity=0.4] (-0.1,-0.1) rectangle (2.2,1.2);
        \end{pgfonlayer}
    \end{scope}

    \begin{scope}
        \node at (0.15,0.75) {$S\mathcal{F}$};
        \node at (0.35,0.55) {$ST^{-1}\mathcal{F}$};
        
        \node at (1-0.1,0.85) {$TS\mathcal{F}$};
        \node at (1-0.35,0.55) {$TST\mathcal{F}$};
        \node at (1+0.35,0.55) {$TST^{-1}\mathcal{F}$};
        \node at (1-0.25,0.25) {$TST^2\mathcal{F}$};
        \node at (2-0.35,0.55) {$T^2ST\mathcal{F}$};
        
        \fill (0,0)  circle[radius=0.2pt];
        \fill (2,0)  circle[radius=0.2pt];
        \fill (1,0)  circle[radius=0.2pt];
        
        \node at (0, -0.05) {0};
        \node at (1, -0.05) {1};
        \node at (2, -0.05) {2};
    \end{scope}
    
\end{tikzpicture}

}
    \caption{ }
    \label{fig:D4-I0s-I4-2I1 Configuration}
    \end{subfigure}%
    \begin{subfigure}{0.5\textwidth}
    \centering
    \scalebox{0.5}{


\begin{tikzpicture}[scale=7]

    \pgfdeclarelayer{background}
    \pgfsetlayers{background,main}

    \pgfmathsetmacro{\myxlow}{-0.1}
    \pgfmathsetmacro{\myxhigh}{1+0.1}
    \pgfmathsetmacro{\myiterations}{1}

    \draw[-latex](\myxlow-0.1,0) -- (\myxhigh+0.15,0);
    \pgfmathsetmacro{\succofmyxlow}{\myxlow+0.5}
    
    \draw[-stealth, very thin] (0,0)--(0,1.2) node[above] {}; 
    \begin{scope}
        
        \draw[very thin, black] (0,0) arc(180:90:1);
        \draw[very thin, black] (0,0) arc(180:60-38.4:1/3);
        \draw[very thin, black] (0,0) arc(180:38.3:1/5);

        \draw[very thin, black] (1/2, 0) arc(0:98.3:1/8);

        \draw[very thin, black] (1/2,0) -- (1/2, 1.2);
        \draw[very thin, black, dashed] (1,0) -- (1, 1.2);

        \draw[very thin, black] (1,0) arc(0:90:1);
        \draw[very thin, black] (1,0) arc(0:120+38.4:1/3);
        \draw[very thin, black] (1,0) arc(0:180-38.3:1/5);
        \draw[very thin, black] (1/2, 0) arc(180:180-98.3:1/8);

    \end{scope}

    \begin{scope}
        \begin{pgfonlayer}{background}
            \clip
                (0,0) 
                {-- (0,0) arc(180:120:1)
                -- (1/2, 0.866) arc(60:0:1)
                -- (1,0) arc(0:180-38.3:1/5)
                -- (1-0.357, 0.124) arc(180-98.3:180:1/8)
             -- (1/2,0) arc(0:98.3:1/8)
             -- (+0.357, 0.124) arc(38.3:180:1/5)
             }
                -- (0, 0) -- cycle
            ;
            \fill[gray,opacity=0.4] (-0.1,-0.1) rectangle (1.2,1.2);
        \end{pgfonlayer}
    \end{scope}

    \begin{scope}
        \node at (0.15,0.75) {$S\mathcal{F}$};
        \node at (0.35,0.55) {$ST^{-1}\mathcal{F}$};
        \node at (0.25,0.25) {$ST^{-2}\mathcal{F}$};
        \node at (0.44,0.16) [scale=0.7] {$ST^{-2}S\mathcal{F}$};
        
        \node at (1-0.15,0.75) {$TS\mathcal{F}$};
        \node at (1-0.35,0.55) {$TST\mathcal{F}$};
        \node at (1-0.25,0.25) {$TST^2\mathcal{F}$};
        \node at (1-0.35,0.07) [scale=0.7] {$TST^{2}S\mathcal{F}$};
        
        \fill (0,0)  circle[radius=0.2pt];
        \fill (1/2,0)  circle[radius=0.2pt];
        \fill (1,0)  circle[radius=0.2pt];
        
        \node at (0, -0.05) {0};
        \node at (1, -0.05) {1};
        
    \end{scope}
    
\end{tikzpicture}

}
    \caption{ }
    \label{fig:D4-I0s-3I2 Configuration}
    \end{subfigure}
    \caption{Fundamental domains for (a) $\Gamma^0(4)$ and (b) $\Gamma(2)$ modular configurations of the CB of 4d $SU(2)$ $N_f = 4$, based on \cite{Aspman:2021vhs}.}
\end{figure}

\noindent
\paragraph{$SU(2)\, N_f = 4$ theory.} For reasons that will become clear momentarily, let us first review the $SU(2)~N_f = 4$ theory, with the SW curve in the conventions of \cite{Manschot:2019pog} given by:
\be
    y^2 = \left( {T(x) \ov 1-\mathfrak{q}}\right)^2 - {4\mathfrak{q} \ov (1-\mathfrak{q})^2} \prod_{i=1}^{4} (x+m_i)~,
\ee
where we introduced:
\be
    T(x) = (1+\mathfrak{q}) x^2 +\mathfrak{q} \sum_{i} m_i x - (1-\mathfrak{q})u~,
\ee
and $\mathfrak{q} = e^{2\pi i \tau_{uv}}$ is the instanton counting parameter for the `weak-coupling' description. Out of the 19 possible configurations for this SCFT, 6 of them turn out to be modular. The $(I_0^*, I_4, 2I_1)$ and $(I_0^*, 3I_2)$ configurations will be of particular interest to us, as they are also allowed configurations for the $\CN = 2^*$ theories. Their monodromy groups are $\Gamma_0(4)$ and $\Gamma(2)$, respectively, with the modular properties discussed in \cite{Aspman:2021evt}. They arise for mass parameters $(m,m,m,m)$ and $(m,m,0,0)$, for instance. Fundamental domains for the two configurations are drawn in figures~\ref{fig:D4-I0s-I4-2I1 Configuration} and~\ref{fig:D4-I0s-3I2 Configuration}. %
The BPS quivers of the $N_f = 4$ theory can be formed from the basis of BPS states read from these fundamental domains. The first quiver corresponds to the $(I_4, 2I_1)$ configuration, with the $I_1$ cusps at $\tau = 0$ and $\tau = 2$, and the $I_4$ cusp at $\tau = 1$, namely:
\bea\label{quiver Nf=4 v1}
 \begin{tikzpicture}[baseline=1mm]
\node[] (1) []{$\CE_{\gamma_{1,2,3,4}=(1,-1)} $};
\node[] (5) [right = of 1] {$\CE_{\gamma_{5}=(-1,0)}$};
\node[] (6) [below = of 1] {$\CE_{\gamma_6 = (-1,2)}$};
\draw[->-=0.5] (1) to   (6);
\draw[->>-=0.5] (6) to   (5);
\draw[->-=0.5] (5) to   (1);
\end{tikzpicture}
\eea
This quiver was discussed in some detail in \cite{Alim:2011kw}. For the $(3I_2)$ bulk configuration, we instead have cusps at $\tau = 0$, ${1\ov 2}$ and $1$, with the quiver:
\bea\label{quiver Nf=4 v2}
 \begin{tikzpicture}[baseline=1mm]
\node[] (1) []{$\CE_{\gamma_{1,2}=(1,0)} $};
\node[] (3) [right = of 1] {$\CE_{\gamma_{3,4}=(-2,1)}$};
\node[] (5) [below = of 1] {$\CE_{\gamma_{5,6} = (1,-1)}$};
\draw[->-=0.5] (1) to   (3);
\draw[->-=0.5] (3) to   (5);
\draw[->-=0.5] (5) to   (1);
\end{tikzpicture}
\eea
The two quivers are related by a mutation sequence, such as $\gamma_{3, 4}- \gamma_5 - \gamma_{1,2}$, for example, with some additional mutations needed to recover the correct BPS states.


\medskip

\noindent
\paragraph{$\CN= 2^*$ curves.} We now pass on to the $\cN=2^*$ curves. As stated before, there are two different mass deformations of the $\CN= 4$ theory, namely $(3I_2)$ and $(I_4, 2I_1)$. Both of these have Mordell-Weil rank-one, being the only allowed configurations for each deformation pattern, other than the $(2I_0^*)$ configuration. Additionally, they are both modular, being the first examples of modular rational elliptic surfaces with `frozen' singularities. 

Thus, we will make use of the refined correspondence between BPS states and cusps on the upper half-plane discussed in section \ref{sec: BPS states - Singularities identification}. Recall that for a frozen $I_n$ singularity, we have a different charge normalization of the BPS states, namely $Q = \sqrt{n}$ \cite{Argyres:2015ffa}. Thus, we have the generalized correspondence between cusp-positions on the upper half-plane $\mathbb{H}$ and BPS states:
\be
    \tau = {p\ov m}~ \Longleftrightarrow Q(m,-p)~.
\ee
First, for the $(I_4, 2I_1)$ deformation, from the fundamental domain in figure~\ref{fig:D4-I0s-I4-2I1 Configuration}, we have the basis of BPS states:
\be
    \sqrt{4}(1,-1)~, \qquad (-1,0)~, \qquad (-1,2)~,
\ee
with the associated BPS quiver reading:
\bea\label{quiver N=2* I4-2I1}
 \begin{tikzpicture}[baseline=1mm]
\node[] (1) []{$\CE_{\gamma_{1}=(2,-2)} $};
\node[] (2) [right = of 1] {$\CE_{\gamma_{2}=(-1,0)}$};
\node[] (3) [below = of 1] {$\CE_{\gamma_{3} = (-1,2)}$};
\draw[->>-=0.5] (1) to   (3);
\draw[->>-=0.5] (3) to   (2);
\draw[->>-=0.5] (2) to   (1);
\end{tikzpicture}
\eea
This is also a known quiver for the $\CN = 2^*$ theory, discussed for instance in \cite{Alim:2011kw}. Finally, for the BPS quiver of the $(3I_2)$ deformation, from the fundamental domain in figure~\ref{fig:D4-I0s-3I2 Configuration}, we have a basis of BPS states given by:
\be
    \sqrt{2}(1,0)~, \qquad \sqrt{2}(-2,1)~, \qquad \sqrt{2}(1,-1)~,
\ee
with the BPS quiver:
\bea\label{quiver N=2* 3I2}
 \begin{tikzpicture}[baseline=1mm]
\node[] (1) []{$\CE_{\gamma_{1}=\sqrt{2}(1,0)} $};
\node[] (2) [right = of 1] {$\CE_{\gamma_{2}=\sqrt{2}(-2,1)}$};
\node[] (3) [below = of 1] {$\CE_{\gamma_{3}=\sqrt{2}(1,-1)}$};
\draw[->>-=0.5] (1) to   (2);
\draw[->>-=0.5] (2) to   (3);
\draw[->>-=0.5] (3) to   (1);
\end{tikzpicture}
\eea
Note that this is the same quiver as \eqref{quiver N=2* I4-2I1}, up to the renormalization of the BPS states. The charge normalization is rather peculiar in this case, an issue that we hope to clarify in future work. Let us finally mention that this analysis can be done for Argyres-Wittig SCFTs \cite{Argyres:2007tq}, reproducing known BPS quivers of these theories \cite{Caorsi:2019vex}, based on quivers of the Minahan-Nemeschanski theories \cite{Cecotti:2013sza}.





\subsection*{Acknowledgements}
I am very grateful to Johannes Aspman, Elias Furrer, Jan Manschot and especially Cyril Closset for interesting discussions, feedback and correspondences on related topics. The author would also like to thank Carmen Jorge-Diaz, Joseph McGovern and Palash Singh for helpful discussions on modular forms. The work of HM is supported by a Royal Society Research Grant for Research Fellows.


\appendix

\section{Subgroups of ${\rm PSL}(2,\mathbb{Z})$}

\subsection{Permutations} \label{Appendix Permutations}

In \cite{10.1112/jlms/s2-1.1.351} Millington showed that there is a one-to-one correspondence between conjugacy classes of finite index $\mu \in \mathbb{N}$ subgroups of ${\rm PSL}(2,\mathbb{Z})$ and `pairs' of permutations in $S_{\mu}$. To make this statement more precise, we follow the presentation of \cite{STROMBERG2019436} and represent such subgroups $\Gamma$ by a set of integers:
\be
    \Gamma \cong \Gamma(\mu; g, c, e_2, e_3; N)~,
\ee
where $c$ is the number of cusps, $e_{2,3}$ are the number of elliptic points of orders $2$ and $3$, respectively, and $g$ is the genus of the associated Riemann surface, given by \eqref{genus}. Here $N$ is the \emph{generalized level} of $\Gamma$, defined as the least common multiple of the widths of the cusps of $\Gamma$. When $\Gamma$ is a congruence subgroup, \textit{i.e.} $\Gamma(N) \subset \Gamma$, then $N$ is the same as the usual notion of level \cite{10.1215/ijm/1256059574}. 

The information about a subgroup $\Gamma$ of the modular group can be encoded in a pair $(\sigma_S, \sigma_R)$ of permutations, with $\sigma_R,~\sigma_S \in S_{\mu}$. Such a pair is `legitimate' if:
\be
    \sigma^2_S = \sigma^3_R = Id~,
\ee
and the subgroup generated by the pair of permutations is transitive on $\{1, 2, \ldots, \mu \}$. Legitimate pairs form an equivalence class, with the equivalence relation given by:
\be
    (\sigma_S, \sigma_R) \sim (\sigma_S', \sigma_R') \quad \Longrightarrow \quad \exists\, \pi \in S_{\mu}~: \quad (\sigma_S', \sigma_R') = (\pi^{-1}\sigma_S\pi,  \pi^{-1}\sigma_R\pi)~.
\ee
Millington proved that these equivalence classes are in one-to-one correspondence with the subgroups $\Gamma$ of index $\mu$ defined above. The details of $\Gamma$ can be read as follows. The number of elliptic elements $e_{2,3}$ are the elements fixed by the permutations $\sigma_S$ and $\sigma_R$, respectively, while the cusps of widths $(\mu_1, \mu_2, \ldots, \mu_c)$ are the $c$ disjoint cycles of lengths $\mu_1, \mu_2, \ldots, \mu_c$ of $\sigma_T = \sigma_S \sigma_R$.

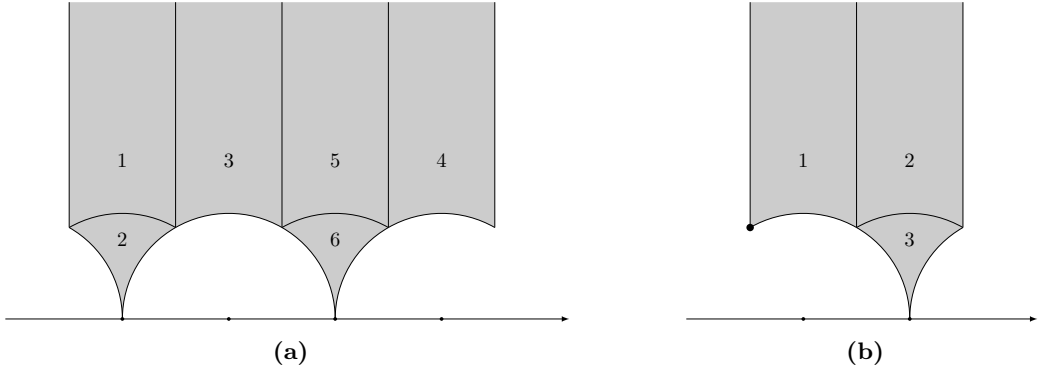
\begin{figure}[t]
    \centering
    \begin{subfigure}{0.5\textwidth}
    \centering
    \scalebox{0.7}{


\begin{tikzpicture}[scale=2]

    \pgfdeclarelayer{background}
    \pgfsetlayers{background,main}

    \pgfmathsetmacro{\myxlow}{-1}
    \pgfmathsetmacro{\myxhigh}{4}
    \pgfmathsetmacro{\myiterations}{1}
    
    \draw[-latex](\myxlow-0.1,0) -- (\myxhigh+0.2,0);
    \pgfmathsetmacro{\succofmyxlow}{\myxlow+0.5}

    \begin{scope}
    
    \foreach \i in {-0.5,0.5,1.5,2.5}
            {\draw[very thin, black] (\i,0.866) arc(120:60:1);
            \draw[very thin, black] (\i,0.866) -- (\i,3);
            }

            \draw[very thin, black] (3.5,0.866) -- (3.5,3);
            
            \draw[very thin, black] (-0.5,0.866) arc(60:0:1);
            \draw[very thin, black] (0,0) arc(180:120:1);
        
            \draw[very thin, black] (2-0.5,0.866) arc(60:0:1);
            \draw[very thin, black] (2,0) arc(180:120:1);
            
    \end{scope}

    \begin{scope}
        \begin{pgfonlayer}{background}
            \pgfmathsetmacro{\myradius}{pow(1/3,1)}
            \clip
                (-0.5,3) 
                { -- (-0.5,0.866) arc(60:0:1)
             -- (0, 0) arc(180:0:1)
             -- (2,0) arc(180:60:1)
             }
                -- (3.5, 3) -- cycle
            ;
            \fill[gray,opacity=0.4] (-1,-1) rectangle (3.5,3);
        \end{pgfonlayer}
    \end{scope}

    \begin{scope}

        \node at (0,1.5) {$1$};
        \node at (1,1.5) {$3$};
        \node at (2,1.5) {$5$};
        \node at (3,1.5) {$4$};
        
        \node at (0,0.75) {$2$};
        \node at (2,0.75) {$6$};

        \fill (0,0)  circle[radius=0.5pt];
        \fill (1,0)  circle[radius=0.5pt];
        \fill (2,0)  circle[radius=0.5pt];
        \fill (3,0)  circle[radius=0.5pt];

    \end{scope}
    
\end{tikzpicture}

}
    \caption{ }
    \label{fig: Permutations Gamma04}
    \end{subfigure}%
    \begin{subfigure}{0.5\textwidth}
    \centering
    \scalebox{0.7}{


\begin{tikzpicture}[scale=2]

    \pgfdeclarelayer{background}
    \pgfsetlayers{background,main}

    \pgfmathsetmacro{\myxlow}{-1}
    \pgfmathsetmacro{\myxhigh}{2}
    \pgfmathsetmacro{\myiterations}{1}
    
    \draw[-latex](\myxlow-0.1,0) -- (\myxhigh+0.2,0);
    \pgfmathsetmacro{\succofmyxlow}{\myxlow+0.5}

    \begin{scope}
    
        \foreach \i in {-0.5,0.5}
            {\draw[very thin, black] (\i,0.866) arc(120:60:1);
            \draw[very thin, black] (\i,0.866) -- (\i,3);
            }

            \draw[very thin, black] (1.5,0.866) -- (1.5,3);
            \draw[very thin, black] (1-0.5,0.866) arc(60:0:1);
            \draw[very thin, black] (1,0) arc(180:120:1);
            
    \end{scope}

    \begin{scope}
        \begin{pgfonlayer}{background}
            \clip
                (-0.5,3) 
                { -- (-0.5,0.866) arc(120:0:1)
                -- (1, 0) arc(180:120:1)
                }
                -- (1.5, 3) -- cycle;
            \fill[gray,opacity=0.4] (-1,-1) rectangle (3.5,3);
        \end{pgfonlayer}
    \end{scope}
    
    \begin{scope}

        \node at (0,1.5) {$1$};
        \node at (1,1.5) {$2$};

        \node at (1,0.75) {$3$};

        \fill (0,0)  circle[radius=0.5pt];
        \fill (1,0)  circle[radius=0.5pt];
        
        \fill (-0.5,0.866)  circle[radius=1pt];
        
    \end{scope}
    
\end{tikzpicture}

}
    \caption{ }
    \label{fig: Permutations Gamma02}
    \end{subfigure}
    \caption{Possible fundamental domains for congruence subgroups in the conjugacy classes $\Gamma(6; 0, 3, 0, 0; 4)$ and $\Gamma(3; 0, 2,1, 0; 2)$, respectively.}
\end{figure}

As a simple example, consider the conjugacy class that contains $\Gamma^0(4)$, for which the signature is given by:
\be
    \Gamma^0(4) \cong \Gamma(6; 0, 3, 0, 0; 4)~.
\ee
Following \cite{STROMBERG2019436}, this class is described by:
\be
    \sigma_S = (1~2)(3~4)(5~6)~, \qquad \sigma_R = (1~2~3)(4~5~6)~,
\ee
 with $\sigma_T = \sigma_S \sigma_R$ given by:
 \be
    \sigma_T = (1~3~5~4)(2)(6)~.
 \ee
One quickly notices that $\sigma_S$ and $\sigma_R$ do not have any length one cycles and, thus, $\Gamma^0(4)$ has no elliptic points. Additionally, $\sigma_T$ has three cycles of widths $4, 1$ and $1$, which are the widths of the cusps of the congruence subgroup. The fundamental domain is encoded in $\sigma_T$ and is shown in figure~\ref{fig: Permutations Gamma04}. 

Let us also consider a subgroup containing elliptic points. Take, for instance, $\Gamma^0(2)$ with signature:
\be
    \Gamma^0(2) \cong \Gamma(3; 0, 2,1, 0; 2)~,
\ee
described by:
\be
    \sigma_S = (1~2~3)~, \qquad \sigma_R = (1)(2~3)~, \qquad \sigma_T = (1~2)(3)~.
\ee
In this case, there are two cusps of widths $2$ and $1$, as well as an elliptic point of order 3, due to the one-cycle in $\sigma_R$. Recall that these are the points in the orbit of $\tau = e^{2\pi i \ov 3}$. The fundamental domain for this conjugacy class is shown in figure~\ref{fig: Permutations Gamma02}. Note that the domain depicted here is reminiscent to $\Gamma_{\theta}$, which is, of course, in the same equivalence class with $\Gamma^0(2)$.

\subsection{Modular Forms} \label{Appendix Modular forms}

The ring of modular forms for the ${\rm SL}(2,\mathbb{Z})$ group is generated by the holomorphic Eisenstein series of weights 4 and 6, defined as:
\bea
    E_4(\tau)  = 1 + 240 \sum_{n=1}^{\infty} {n^3 q^n \ov 1-q^n}~, \qquad \quad E_6(\tau)  = 1 - 504 \sum_{n=1}^{\infty} {n^5 q^n \ov 1-q^n}~,
\eea
for $q = e^{2\pi i \tau}$. Their $T$ and $S$ transformations are:
\be
    E_k(\tau + 1) = E_k (\tau)~, \qquad E_k\left( -{1\ov \tau}\right) = \tau^k E_k(\tau)~.
\ee
while their zeroes lie at $e^{2\pi i \ov 3}$ (or $e^{i \pi \ov 3}$) and $i$, respectively. The modular $j$-function in ${\rm PSL}(2,\mathbb{Z})$ is a holomorphic map away from the cusp of ${\rm PSL}(2,\mathbb{Z})$, defined as:
\be
    j(\tau) = 1728{E_4(\tau)^3 \ov E_4(\tau)^3 - E_6(\tau)^2}~,
\ee
being a bijection between $\mathbb{H}/\Gamma(1)$ and $\mathbb{C}$ which parameterizes isomorphism classes of elliptic curves. From the zeroes of the Eisenstein series $E_{4,6}$, we find that:
\be
    J(i) = {1 \ov 1728} j(i) = 1~, \qquad J\left(e^{2\pi i \ov 3}\right) = {1 \ov 1728} j\left(e^{2\pi i\ov3}\right) = 0~.
\ee
An alternative definition for the $j$-invariant is given by:
\be
    j(\tau) = 1728 {g_2(\tau)^3 \ov \Delta(\tau)}~, \qquad \Delta(\tau) = g_2(\tau)^3 - 27g_3(\tau)^2~,
\ee
where $g_2(\tau)$ and $g_3(\tau)$ are constant multiples of the Eisenstein series, making the relation to elliptic curves explicit. As a result, the function $\Delta(\tau)$ is referred to as the modular discriminant, being a nowhere vanishing modular form of weight $12$. The rings of modular forms $M_{\star}\left( \Gamma\right)$ for congruence subgroups usually have more generators. In order to describe those, we first introduce the Dedekind $\eta$-function:
\be
    \eta(\tau) = q^{1\ov24}\prod_{j=1}^{\infty}(1-q^j)~,
\ee
which is a holomorphic function on the upper half-plane, satisfying:
\be
    \eta(\tau+1)=e^{i\pi\ov12}\eta(\tau)~, \qquad \eta\left(-{1\ov\tau}\right)=\sqrt{-i\tau}\, \eta(\tau)~.
\ee
Additionally, we have \cite{Harnad:1998hh}:
\bea
    \eta\left(\tau + {1\ov2}\right) & = e^{{i\pi \ov 24}} {\eta(2\tau)^3 \ov \eta(\tau) \eta(4\tau)}~, \\
    \eta\left(\tau + {1\ov3}\right)^3 & = e^{{i\pi \ov 12}} \eta(\tau)^3 + 3\sqrt{3}\,e^{-{i\pi \ov 12}}\eta(9\tau)^3~.
\eea
Let us note that this is related to the modular discriminant as follows:
\be
    \Delta(\tau) = (2\pi)^{12}\eta(\tau)^{24}~,
\ee
which shows that the Dedekind eta function is also a non-vanishing function on the upper half-plane. A useful theorem used throughout the text states that the $\eta$-quotient $f(\tau) = \prod_{\delta|N} \eta(\delta \tau)^{r_{\delta}}$ satisfying:
\be
    \sum_{\delta|N} \delta r_{\delta} = 0~{\rm mod}~24~, \qquad  \sum_{\delta|N} {N\ov \delta}r_{\delta} = 0~{\rm mod}~24~,
\ee
with $k = {1\ov 2}\sum_{\delta|N} r_{\delta} \in \mathbb{Z}$ is a weakly holomorphic weight $k$ modular form for $\Gamma_0(N)$, \textit{i.e.} that is:
\be
    f\left({a\tau + b \ov c\tau+d}\right) = \chi(d)(c\tau+d)^k f(\tau)~,
\ee
with the Dirichlet character $\chi(d) = \left( {(-1)^k s \ov d}\right)$, where $s = \prod_{\delta|N} \delta^{r_{\delta}}$. When the associated elliptic curve $X_{\Gamma}$ has genus zero, there is only one modular form of weight $0$, called the Hauptmodul of $\Gamma$. We also introduce the Jacobi theta forms:
\be
    \vartheta_2(\tau) = \sum_{n \in \mathbb{Z}+{1\ov 2}} q^{n^2 \ov 2}~, \qquad \vartheta_3(\tau) = \sum_{n\in\mathbb{Z}} q^{n^2 \ov 2}~, \qquad \vartheta_4(\tau) = \sum_{n\in \mathbb{Z}}(-1)^n q^{n^2 \ov 2}~,
\ee
with the usual expression for $q = e^{2\pi i \tau}$. Under the $T$ and $S$ transformations of ${\rm SL}(2,\mathbb{Z})$, these behave as follows:
\bea
    T: &\qquad  \vartheta_2 \rightarrow e^{i \pi \ov 4}\vartheta_2~, && \qquad \vartheta_3 \rightarrow \vartheta_4~, && \qquad \vartheta_4 \rightarrow \vartheta_3~, \\
    S: &\qquad  \vartheta_2 \rightarrow \sqrt{-i\tau}\, \vartheta_4~, && \qquad \vartheta_3 \rightarrow \sqrt{-i\tau}\,\vartheta_3~, && \qquad \vartheta_4 \rightarrow \sqrt{-i\tau}\,\vartheta_4~. 
\eea
Note that the zeros of these functions lie along the real axis. Moreover, they satisfy the following identity:
\be
    \vartheta_2(\tau)^4 + \vartheta_4(\tau)^4 = \vartheta_3(\tau)^4~.
\ee
It is sometimes useful to rewrite them in terms of Dedekind-eta functions, which can be done as follows:
\bea
    \vartheta(\tau) = {2\eta(2\tau)^2 \ov \eta(\tau)}~, \qquad \vartheta_3(\tau) = {\eta(\tau)^5 \ov \eta\left({\tau \ov 2}\right)^2 \eta(2\tau)^2}~, \qquad \vartheta_4(\tau) = {\eta\left({\tau\ov 2}\right)^2 \ov \eta(\tau)}~,
\eea
from which one finds:
\be
    \vartheta_2(\tau) \vartheta_3(\tau) \vartheta_4(\tau) = 2\eta(\tau)^3~.
\ee
We also introduce the modular $\lambda$-function:
\be
    \lambda(\tau) = {\vartheta_2(\tau)^4 \ov \vartheta_3(\tau)^4}~,
\ee
which is related to the $j$-invariant by:
\be
    j(\tau) = {256\left(1-\lambda(\tau) + \lambda(\tau)^2\right)^3 \ov \lambda(\tau)^2 \left(1-\lambda(\tau)\right)^2}~.
\ee
Finally, let us point out that the Eisenstein series can be also expressed in terms of the theta functions, with:
\bea
    E_4 & = {1\ov 2} \left(\vartheta_2^8 + \vartheta_3^8 + \vartheta_4^8 \right)~, \\
    E_6 & = {1\ov 2}\left(\vartheta_3^4 - 2\vartheta_4^4 \right) \left(\vartheta_4^4 - 2\vartheta_3^4 \right) \left(\vartheta_3^4 + \vartheta_4^4 \right)~.
\eea


\bibliographystyle{JHEP} 
\bibliography{bib5d}

\providecommand{\href}[2]{#2}\begingroup\raggedright\begin{thebibliography}{100}

\bibitem{Seiberg:1994aj}
N.~Seiberg and E.~Witten, \emph{{Monopoles, duality and chiral symmetry
  breaking in N=2 supersymmetric QCD}},
  \href{http://dx.doi.org/10.1016/0550-3213(94)90214-3}{\emph{Nucl. Phys. B}
  {\bf 431} (1994) 484--550}, [\href{https://arxiv.org/abs/hep-th/9408099}{{\tt
  hep-th/9408099}}].

\bibitem{Seiberg:1994rs}
N.~Seiberg and E.~Witten, \emph{{Electric - magnetic duality, monopole
  condensation, and confinement in N=2 supersymmetric Yang-Mills theory}},
  \href{http://dx.doi.org/10.1016/0550-3213(94)90124-4}{\emph{Nucl. Phys. B}
  {\bf 426} (1994) 19--52}, [\href{https://arxiv.org/abs/hep-th/9407087}{{\tt
  hep-th/9407087}}].

\bibitem{Persson:1990}
U.~Persson, \emph{Configurations of kodaira fibers on rational elliptic
  surfaces}, {\emph{Mathematische Zeitschrift} {\bf 205} (1990) 1--47}.

\bibitem{Miranda:1990}
R.~Miranda, \emph{Persson's list of singular fibers for a rational elliptic
  surface}, {\emph{Mathematische Zeitschrift} {\bf 205} (1990) 191--211}.

\bibitem{Caorsi:2018ahl}
M.~Caorsi and S.~Cecotti, \emph{{Special Arithmetic of Flavor}},
  \href{http://dx.doi.org/10.1007/JHEP08(2018)057}{\emph{JHEP} {\bf 08} (2018)
  057}, [\href{https://arxiv.org/abs/1803.00531}{{\tt 1803.00531}}].

\bibitem{Closset:2021lhd}
C.~Closset and H.~Magureanu, \emph{{The $U$-plane of rank-one 4d
  $\mathcal{N}=2$ KK theories}},
  \href{http://dx.doi.org/10.21468/SciPostPhys.12.2.065}{\emph{SciPost Phys.}
  {\bf 12} (2022) 065}, [\href{https://arxiv.org/abs/2107.03509}{{\tt
  2107.03509}}].

\bibitem{Argyres:1995jj}
P.~C. Argyres and M.~R. Douglas, \emph{{New phenomena in SU(3) supersymmetric
  gauge theory}},
  \href{http://dx.doi.org/10.1016/0550-3213(95)00281-V}{\emph{Nucl. Phys. B}
  {\bf 448} (1995) 93--126}, [\href{https://arxiv.org/abs/hep-th/9505062}{{\tt
  hep-th/9505062}}].

\bibitem{Argyres:1995xn}
P.~C. Argyres, M.~R. Plesser, N.~Seiberg and E.~Witten, \emph{{New N=2
  superconformal field theories in four-dimensions}},
  \href{http://dx.doi.org/10.1016/0550-3213(95)00671-0}{\emph{Nucl. Phys. B}
  {\bf 461} (1996) 71--84}, [\href{https://arxiv.org/abs/hep-th/9511154}{{\tt
  hep-th/9511154}}].

\bibitem{Minahan:1996cj}
J.~A. Minahan and D.~Nemeschansky, \emph{{Superconformal fixed points with E(n)
  global symmetry}},
  \href{http://dx.doi.org/10.1016/S0550-3213(97)00039-4}{\emph{Nucl. Phys. B}
  {\bf 489} (1997) 24--46}, [\href{https://arxiv.org/abs/hep-th/9610076}{{\tt
  hep-th/9610076}}].

\bibitem{Minahan:1996fg}
J.~A. Minahan and D.~Nemeschansky, \emph{{An N=2 superconformal fixed point
  with E(6) global symmetry}},
  \href{http://dx.doi.org/10.1016/S0550-3213(96)00552-4}{\emph{Nucl. Phys. B}
  {\bf 482} (1996) 142--152}, [\href{https://arxiv.org/abs/hep-th/9608047}{{\tt
  hep-th/9608047}}].

\bibitem{Banks:1996nj}
T.~Banks, M.~R. Douglas and N.~Seiberg, \emph{{Probing F theory with branes}},
  \href{http://dx.doi.org/10.1016/0370-2693(96)00808-8}{\emph{Phys. Lett. B}
  {\bf 387} (1996) 278--281}, [\href{https://arxiv.org/abs/hep-th/9605199}{{\tt
  hep-th/9605199}}].

\bibitem{Mikhailov:1998bx}
A.~Mikhailov, N.~Nekrasov and S.~Sethi, \emph{{Geometric realizations of BPS
  states in N=2 theories}},
  \href{http://dx.doi.org/10.1016/S0550-3213(98)80001-1}{\emph{Nucl. Phys. B}
  {\bf 531} (1998) 345--362}, [\href{https://arxiv.org/abs/hep-th/9803142}{{\tt
  hep-th/9803142}}].

\bibitem{Aspinwall:1998xj}
P.~S. Aspinwall and D.~R. Morrison, \emph{{Nonsimply connected gauge groups and
  rational points on elliptic curves}},
  \href{http://dx.doi.org/10.1088/1126-6708/1998/07/012}{\emph{JHEP} {\bf 07}
  (1998) 012}, [\href{https://arxiv.org/abs/hep-th/9805206}{{\tt
  hep-th/9805206}}].

\bibitem{Mayrhofer:2014opa}
C.~Mayrhofer, D.~R. Morrison, O.~Till and T.~Weigand, \emph{{Mordell-Weil
  Torsion and the Global Structure of Gauge Groups in F-theory}},
  \href{http://dx.doi.org/10.1007/JHEP10(2014)016}{\emph{JHEP} {\bf 10} (2014)
  016}, [\href{https://arxiv.org/abs/1405.3656}{{\tt 1405.3656}}].

\bibitem{Cvetic:2017epq}
M.~Cvetic and L.~Lin, \emph{{The Global Gauge Group Structure of F-theory
  Compactification with U(1)s}},
  \href{http://dx.doi.org/10.1007/JHEP01(2018)157}{\emph{JHEP} {\bf 01} (2018)
  157}, [\href{https://arxiv.org/abs/1706.08521}{{\tt 1706.08521}}].

\bibitem{Gaiotto:2014kfa}
D.~Gaiotto, A.~Kapustin, N.~Seiberg and B.~Willett, \emph{{Generalized Global
  Symmetries}}, \href{http://dx.doi.org/10.1007/JHEP02(2015)172}{\emph{JHEP}
  {\bf 02} (2015) 172}, [\href{https://arxiv.org/abs/1412.5148}{{\tt
  1412.5148}}].

\bibitem{Morrison:2020ool}
D.~R. Morrison, S.~Schafer-Nameki and B.~Willett, \emph{{Higher-Form Symmetries
  in 5d}}, \href{http://dx.doi.org/10.1007/JHEP09(2020)024}{\emph{JHEP} {\bf
  09} (2020) 024}, [\href{https://arxiv.org/abs/2005.12296}{{\tt 2005.12296}}].

\bibitem{Albertini:2020mdx}
F.~Albertini, M.~Del~Zotto, I.~Garc\'\i{}a~Etxebarria and S.~S. Hosseini,
  \emph{{Higher Form Symmetries and M-theory}},
  \href{http://dx.doi.org/10.1007/JHEP12(2020)203}{\emph{JHEP} {\bf 12} (2020)
  203}, [\href{https://arxiv.org/abs/2005.12831}{{\tt 2005.12831}}].

\bibitem{Apruzzi:2021vcu}
F.~Apruzzi, L.~Bhardwaj, J.~Oh and S.~Schafer-Nameki, \emph{{The Global Form of
  Flavor Symmetries and 2-Group Symmetries in 5d SCFTs}},
  \href{https://arxiv.org/abs/2105.08724}{{\tt 2105.08724}}.

\bibitem{Bhardwaj:2021ojs}
L.~Bhardwaj, \emph{{Global Form of Flavor Symmetry Groups in 4d N=2 Theories of
  Class S}},  \href{https://arxiv.org/abs/2105.08730}{{\tt 2105.08730}}.

\bibitem{Closset:2020scj}
C.~Closset, S.~Schafer-Nameki and Y.-N. Wang, \emph{{{Coulomb and Higgs
  Branches from Canonical Singularities: Part 0}}},
  \href{http://dx.doi.org/10.1007/JHEP02(2021)003}{\emph{JHEP} {\bf 02} (2021)
  003}, [\href{https://arxiv.org/abs/2007.15600}{{\tt 2007.15600}}].

\bibitem{DelZotto:2020esg}
M.~Del~Zotto, I.~Garc\'\i{}a~Etxebarria and S.~S. Hosseini, \emph{{Higher form
  symmetries of Argyres-Douglas theories}},
  \href{http://dx.doi.org/10.1007/JHEP10(2020)056}{\emph{JHEP} {\bf 10} (2020)
  056}, [\href{https://arxiv.org/abs/2007.15603}{{\tt 2007.15603}}].

\bibitem{Buican:2021xhs}
M.~Buican and H.~Jiang, \emph{{1-Form Symmetry, Isolated N=2 SCFTs, and
  Calabi-Yau Threefolds}},  \href{https://arxiv.org/abs/2106.09807}{{\tt
  2106.09807}}.

\bibitem{Bhardwaj:2020phs}
L.~Bhardwaj and S.~Sch\"afer-Nameki, \emph{{Higher-form symmetries of 6d and 5d
  theories}}, \href{http://dx.doi.org/10.1007/JHEP02(2021)159}{\emph{JHEP} {\bf
  02} (2021) 159}, [\href{https://arxiv.org/abs/2008.09600}{{\tt 2008.09600}}].

\bibitem{Bhardwaj:2021pfz}
L.~Bhardwaj, M.~Hubner and S.~Schafer-Nameki, \emph{{1-form Symmetries of 4d
  N=2 Class S Theories}},
  \href{http://dx.doi.org/10.21468/SciPostPhys.11.5.096}{\emph{SciPost Phys.}
  {\bf 11} (2021) 096}, [\href{https://arxiv.org/abs/2102.01693}{{\tt
  2102.01693}}].

\bibitem{Bhardwaj:2021mzl}
L.~Bhardwaj, S.~Giacomelli, M.~H\"ubner and S.~Sch\"afer-Nameki,
  \emph{{Relative Defects in Relative Theories: Trapped Higher-Form Symmetries
  and Irregular Punctures in Class S}},
  \href{https://arxiv.org/abs/2201.00018}{{\tt 2201.00018}}.

\bibitem{Shioda:1972}
T.~Shioda, \emph{{On elliptic modular surfaces}},
  \href{http://dx.doi.org/10.2969/jmsj/02410020}{\emph{Journal of the
  Mathematical Society of Japan} {\bf 24} (1972) 20 -- 59}.

\bibitem{Moore:1997pc}
G.~W. Moore and E.~Witten, \emph{{Integration over the u plane in Donaldson
  theory}}, \href{http://dx.doi.org/10.4310/ATMP.1997.v1.n2.a7}{\emph{Adv.
  Theor. Math. Phys.} {\bf 1} (1997) 298--387},
  [\href{https://arxiv.org/abs/hep-th/9709193}{{\tt hep-th/9709193}}].

\bibitem{Labastida:2005zz}
J.~Labastida and M.~Marino, \emph{{Topological quantum field theory and four
  manifolds}}, vol.~25.
\newblock Springer, Dordrecht, 2005,
  \href{http://dx.doi.org/10.1007/1-4020-3177-7}{10.1007/1-4020-3177-7}.

\bibitem{Malmendier:2008db}
A.~Malmendier and K.~Ono, \emph{{{SO(3)-Donaldson invariants of CP**2 and Mock
  Theta Functions}}},
  \href{http://dx.doi.org/10.2140/gt.2012.16.1767}{\emph{Geom. Topol.} {\bf 16}
  (2012) 1767--1833}, [\href{https://arxiv.org/abs/0808.1442}{{\tt
  0808.1442}}].

\bibitem{Manschot:2021qqe}
J.~Manschot and G.~W. Moore, \emph{{Topological correlators of $SU(2)$,
  $\mathcal{N}=2^*$ SYM on four-manifolds}},
  \href{https://arxiv.org/abs/2104.06492}{{\tt 2104.06492}}.

\bibitem{Korpas:2019cwg}
G.~Korpas, J.~Manschot, G.~W. Moore and I.~Nidaiev, \emph{{{Mocking the
  $u$-plane integral}}},  \href{https://arxiv.org/abs/1910.13410}{{\tt
  1910.13410}}.

\bibitem{Aspman:2021kfp}
J.~Aspman, E.~Furrer, G.~Korpas, Z.-C. Ong and M.-C. Tan, \emph{{{The $u$-plane
  integral as a bridge between enumerative geometry and number theory}}},
  \href{https://arxiv.org/abs/2109.04302}{{\tt 2109.04302}}.

\bibitem{Aspman:2020lmf}
J.~Aspman, E.~Furrer and J.~Manschot, \emph{{Elliptic Loci of SU(3) Vacua}},
  \href{http://dx.doi.org/10.1007/s00023-021-01040-5}{\emph{Annales Henri
  Poincare} {\bf 22} (2021) 2775--2830},
  [\href{https://arxiv.org/abs/2010.06598}{{\tt 2010.06598}}].

\bibitem{Aspman:2021evt}
J.~Aspman, E.~Furrer and J.~Manschot, \emph{{Four flavors, triality, and
  bimodular forms}},
  \href{http://dx.doi.org/10.1103/PhysRevD.105.025017}{\emph{Phys. Rev. D} {\bf
  105} (2022) 025017}, [\href{https://arxiv.org/abs/2110.11969}{{\tt
  2110.11969}}].

\bibitem{Aspman:2021vhs}
J.~Aspman, E.~Furrer and J.~Manschot, \emph{{{Cutting and gluing with running
  couplings in N=2 QCD}}},
  \href{http://dx.doi.org/10.1103/PhysRevD.105.025021}{\emph{Phys. Rev. D} {\bf
  105} (2022) 025021}, [\href{https://arxiv.org/abs/2107.04600}{{\tt
  2107.04600}}].

\bibitem{Matone:1995jr}
M.~Matone, \emph{{Koebe 1/4 theorem and inequalities in N=2 supersymmetric
  QCD}}, \href{http://dx.doi.org/10.1103/PhysRevD.53.7354}{\emph{Phys. Rev. D}
  {\bf 53} (1996) 7354--7358},
  [\href{https://arxiv.org/abs/hep-th/9506181}{{\tt hep-th/9506181}}].

\bibitem{Matone:1995rx}
M.~Matone, \emph{{Instantons and recursion relations in N=2 SUSY gauge
  theory}}, \href{http://dx.doi.org/10.1016/0370-2693(95)00920-G}{\emph{Phys.
  Lett. B} {\bf 357} (1995) 342--348},
  [\href{https://arxiv.org/abs/hep-th/9506102}{{\tt hep-th/9506102}}].

\bibitem{Fiol:2000wx}
B.~Fiol and M.~Marino, \emph{{BPS states and algebras from quivers}},
  \href{http://dx.doi.org/10.1088/1126-6708/2000/07/031}{\emph{JHEP} {\bf 07}
  (2000) 031}, [\href{https://arxiv.org/abs/hep-th/0006189}{{\tt
  hep-th/0006189}}].

\bibitem{Denef:2002ru}
F.~Denef, \emph{{Quantum quivers and Hall / hole halos}},
  \href{http://dx.doi.org/10.1088/1126-6708/2002/10/023}{\emph{JHEP} {\bf 10}
  (2002) 023}, [\href{https://arxiv.org/abs/hep-th/0206072}{{\tt
  hep-th/0206072}}].

\bibitem{Alim:2011kw}
M.~Alim, S.~Cecotti, C.~Cordova, S.~Espahbodi, A.~Rastogi and C.~Vafa,
  \emph{{$\mathcal{N} = 2$ quantum field theories and their BPS quivers}},
  \href{http://dx.doi.org/10.4310/ATMP.2014.v18.n1.a2}{\emph{Adv. Theor. Math.
  Phys.} {\bf 18} (2014) 27--127}, [\href{https://arxiv.org/abs/1112.3984}{{\tt
  1112.3984}}].

\bibitem{Alim:2011ae}
M.~Alim, S.~Cecotti, C.~Cordova, S.~Espahbodi, A.~Rastogi and C.~Vafa,
  \emph{{BPS Quivers and Spectra of Complete N=2 Quantum Field Theories}},
  \href{http://dx.doi.org/10.1007/s00220-013-1789-8}{\emph{Commun. Math. Phys.}
  {\bf 323} (2013) 1185--1227}, [\href{https://arxiv.org/abs/1109.4941}{{\tt
  1109.4941}}].

\bibitem{Closset:2019juk}
C.~Closset and M.~Del~Zotto, \emph{{On 5d SCFTs and their BPS quivers. Part I:
  B-branes and brane tilings}},  \href{https://arxiv.org/abs/1912.13502}{{\tt
  1912.13502}}.

\bibitem{Herzog:2003zc}
C.~P. Herzog, \emph{{Exceptional collections and del Pezzo gauge theories}},
  \href{http://dx.doi.org/10.1088/1126-6708/2004/04/069}{\emph{JHEP} {\bf 04}
  (2004) 069}, [\href{https://arxiv.org/abs/hep-th/0310262}{{\tt
  hep-th/0310262}}].

\bibitem{Franco:2005rj}
S.~Franco, A.~Hanany, K.~D. Kennaway, D.~Vegh and B.~Wecht, \emph{{Brane dimers
  and quiver gauge theories}},
  \href{http://dx.doi.org/10.1088/1126-6708/2006/01/096}{\emph{JHEP} {\bf 01}
  (2006) 096}, [\href{https://arxiv.org/abs/hep-th/0504110}{{\tt
  hep-th/0504110}}].

\bibitem{Hanany:2005ss}
A.~Hanany and D.~Vegh, \emph{{Quivers, tilings, branes and rhombi}},
  \href{http://dx.doi.org/10.1088/1126-6708/2007/10/029}{\emph{JHEP} {\bf 10}
  (2007) 029}, [\href{https://arxiv.org/abs/hep-th/0511063}{{\tt
  hep-th/0511063}}].

\bibitem{Closset:2018bjz}
C.~Closset, M.~Del~Zotto and V.~Saxena, \emph{{Five-dimensional SCFTs and gauge
  theory phases: an M-theory/type IIA perspective}},
  \href{http://dx.doi.org/10.21468/SciPostPhys.6.5.052}{\emph{SciPost Phys.}
  {\bf 6} (2019) 052}, [\href{https://arxiv.org/abs/1812.10451}{{\tt
  1812.10451}}].

\bibitem{Banerjee:2018syt}
S.~Banerjee, P.~Longhi and M.~Romo, \emph{{Exploring 5d BPS Spectra with
  Exponential Networks}},
  \href{http://dx.doi.org/10.1007/s00023-019-00851-x}{\emph{Annales Henri
  Poincare} {\bf 20} (2019) 4055--4162},
  [\href{https://arxiv.org/abs/1811.02875}{{\tt 1811.02875}}].

\bibitem{Banerjee:2020moh}
S.~Banerjee, P.~Longhi and M.~Romo, \emph{{Exponential BPS graphs and D-brane
  counting on toric Calabi-Yau threefolds: Part II}},
  \href{https://arxiv.org/abs/2012.09769}{{\tt 2012.09769}}.

\bibitem{Bonelli:2020dcp}
G.~Bonelli, F.~Del~Monte and A.~Tanzini, \emph{{BPS quivers of five-dimensional
  SCFTs, Topological Strings and q-Painlev\'e equations}},
  \href{https://arxiv.org/abs/2007.11596}{{\tt 2007.11596}}.

\bibitem{Mozgovoy:2020has}
S.~Mozgovoy and B.~Pioline, \emph{{Attractor invariants, brane tilings and
  crystals}},  \href{https://arxiv.org/abs/2012.14358}{{\tt 2012.14358}}.

\bibitem{Beaujard:2020sgs}
G.~Beaujard, J.~Manschot and B.~Pioline, \emph{{{Vafa\textendash{}Witten
  Invariants from Exceptional Collections}}},
  \href{http://dx.doi.org/10.1007/s00220-021-04074-2}{\emph{Commun. Math.
  Phys.} {\bf 385} (2021) 101--226},
  [\href{https://arxiv.org/abs/2004.14466}{{\tt 2004.14466}}].

\bibitem{Longhi:2021qvz}
P.~Longhi, \emph{{Instanton Particles and Monopole Strings in 5D SU(2)
  Supersymmetric Yang-Mills Theory}},
  \href{http://dx.doi.org/10.1103/PhysRevLett.126.211601}{\emph{Phys. Rev.
  Lett.} {\bf 126} (2021) 211601},
  [\href{https://arxiv.org/abs/2101.01681}{{\tt 2101.01681}}].

\bibitem{DelMonte:2021ytz}
F.~Del~Monte and P.~Longhi, \emph{{Quiver symmetries and wall-crossing
  invariance}},  \href{https://arxiv.org/abs/2107.14255}{{\tt 2107.14255}}.

\bibitem{Seiberg:1996bd}
N.~Seiberg, \emph{{Five-dimensional SUSY field theories, nontrivial fixed
  points and string dynamics}},
  \href{http://dx.doi.org/10.1016/S0370-2693(96)01215-4}{\emph{Phys. Lett.}
  {\bf B388} (1996) 753--760},
  [\href{https://arxiv.org/abs/hep-th/9608111}{{\tt hep-th/9608111}}].

\bibitem{Morrison:1996xf}
D.~R. Morrison and N.~Seiberg, \emph{{Extremal transitions and five-dimensional
  supersymmetric field theories}},
  \href{http://dx.doi.org/10.1016/S0550-3213(96)00592-5}{\emph{Nucl. Phys. B}
  {\bf 483} (1997) 229--247}, [\href{https://arxiv.org/abs/hep-th/9609070}{{\tt
  hep-th/9609070}}].

\bibitem{Nekrasov:1996cz}
N.~Nekrasov, \emph{{Five dimensional gauge theories and relativistic integrable
  systems}}, \href{http://dx.doi.org/10.1016/S0550-3213(98)00436-2}{\emph{Nucl.
  Phys.} {\bf B531} (1998) 323--344},
  [\href{https://arxiv.org/abs/hep-th/9609219}{{\tt hep-th/9609219}}].

\bibitem{Jefferson:2017ahm}
P.~Jefferson, H.-C. Kim, C.~Vafa and G.~Zafrir, \emph{{Towards Classification
  of 5d SCFTs: Single Gauge Node}},
  \href{https://arxiv.org/abs/1705.05836}{{\tt 1705.05836}}.

\bibitem{Jefferson:2018irk}
P.~Jefferson, S.~Katz, H.-C. Kim and C.~Vafa, \emph{{On Geometric
  Classification of 5d SCFTs}},
  \href{http://dx.doi.org/10.1007/JHEP04(2018)103}{\emph{JHEP} {\bf 04} (2018)
  103}, [\href{https://arxiv.org/abs/1801.04036}{{\tt 1801.04036}}].

\bibitem{Apruzzi:2018nre}
F.~Apruzzi, L.~Lin and C.~Mayrhofer, \emph{{Phases of 5d SCFTs from M-/F-theory
  on Non-Flat Fibrations}},
  \href{http://dx.doi.org/10.1007/JHEP05(2019)187}{\emph{JHEP} {\bf 05} (2019)
  187}, [\href{https://arxiv.org/abs/1811.12400}{{\tt 1811.12400}}].

\bibitem{Apruzzi:2019enx}
F.~Apruzzi, C.~Lawrie, L.~Lin, S.~Schafer-Nameki and Y.-N. Wang, \emph{{Fibers
  add Flavor, Part II: 5d SCFTs, Gauge Theories, and Dualities}},
  \href{https://arxiv.org/abs/1909.09128}{{\tt 1909.09128}}.

\bibitem{Bhardwaj:2018yhy}
L.~Bhardwaj and P.~Jefferson, \emph{{Classifying 5d SCFTs via 6d SCFTs: Rank
  one}},  \href{https://arxiv.org/abs/1809.01650}{{\tt 1809.01650}}.

\bibitem{Bhardwaj:2018vuu}
L.~Bhardwaj and P.~Jefferson, \emph{{Classifying 5d SCFTs via 6d SCFTs:
  Arbitrary rank}},
  \href{http://dx.doi.org/10.1007/JHEP10(2019)282}{\emph{JHEP} {\bf 10} (2019)
  282}, [\href{https://arxiv.org/abs/1811.10616}{{\tt 1811.10616}}].

\bibitem{Apruzzi:2019opn}
F.~Apruzzi, C.~Lawrie, L.~Lin, S.~Schafer-Nameki and Y.-N. Wang, \emph{{Fibers
  add Flavor, Part I: Classification of 5d SCFTs, Flavor Symmetries and BPS
  States}}, \href{http://dx.doi.org/10.1007/JHEP11(2019)068}{\emph{JHEP} {\bf
  11} (2019) 068}, [\href{https://arxiv.org/abs/1907.05404}{{\tt 1907.05404}}].

\bibitem{Bhardwaj:2019jtr}
L.~Bhardwaj, \emph{{On the classification of $5d$ SCFTs}},
  \href{https://arxiv.org/abs/1909.09635}{{\tt 1909.09635}}.

\bibitem{Bhardwaj:2019ngx}
L.~Bhardwaj, \emph{{Dualities of 5d gauge theories from S-duality}},
  \href{https://arxiv.org/abs/1909.05250}{{\tt 1909.05250}}.

\bibitem{Closset:2020afy}
C.~Closset, S.~Giacomelli, S.~Schafer-Nameki and Y.-N. Wang, \emph{{{5d and 4d
  SCFTs: Canonical Singularities, Trinions and S-Dualities}}},
  \href{http://dx.doi.org/10.1007/JHEP05(2021)274}{\emph{JHEP} {\bf 05} (2021)
  274}, [\href{https://arxiv.org/abs/2012.12827}{{\tt 2012.12827}}].

\bibitem{Bhardwaj:2020gyu}
L.~Bhardwaj and G.~Zafrir, \emph{{Classification of 5d N=1 gauge theories}},
  \href{https://arxiv.org/abs/2003.04333}{{\tt 2003.04333}}.

\bibitem{Bhardwaj:2020kim}
L.~Bhardwaj, \emph{{More 5d KK theories}},
  \href{https://arxiv.org/abs/2005.01722}{{\tt 2005.01722}}.

\bibitem{vanBeest:2020kou}
M.~van Beest, A.~Bourget, J.~Eckhard and S.~Schafer-Nameki, \emph{{(Symplectic)
  Leaves and (5d Higgs) Branches in the Poly(go)nesian Tropical Rain Forest}},
  \href{https://arxiv.org/abs/2008.05577}{{\tt 2008.05577}}.

\bibitem{Hubner:2020uvb}
M.~Hubner, \emph{{5d SCFTs from $(E_n,E_m)$ Conformal Matter}},
  \href{https://arxiv.org/abs/2006.01694}{{\tt 2006.01694}}.

\bibitem{Apruzzi:2019kgb}
F.~Apruzzi, S.~Schafer-Nameki and Y.-N. Wang, \emph{{5d SCFTs from Decoupling
  and Gluing}},  \href{https://arxiv.org/abs/1912.04264}{{\tt 1912.04264}}.

\bibitem{Bhardwaj:2020ruf}
L.~Bhardwaj, \emph{{Flavor Symmetry of 5d SCFTs, Part 1: General Setup}},
  \href{https://arxiv.org/abs/2010.13230}{{\tt 2010.13230}}.

\bibitem{vanBeest:2020civ}
M.~van Beest, A.~Bourget, J.~Eckhard and S.~Schafer-Nameki, \emph{{(5d RG-flow)
  Trees in the Tropical Rain Forest}},
  \href{https://arxiv.org/abs/2011.07033}{{\tt 2011.07033}}.

\bibitem{vanBeest:2021xyt}
M.~van Beest and S.~Giacomelli, \emph{{Connecting 5d Higgs branches via
  Fayet-Iliopoulos deformations}},
  \href{http://dx.doi.org/10.1007/JHEP12(2021)202}{\emph{JHEP} {\bf 12} (2021)
  202}, [\href{https://arxiv.org/abs/2110.02872}{{\tt 2110.02872}}].

\bibitem{Closset:2021lwy}
C.~Closset, S.~Sch\"afer-Nameki and Y.-N. Wang, \emph{{Coulomb and Higgs
  Branches from Canonical Singularities, Part 1: Hypersurfaces with Smooth
  Calabi-Yau Resolutions}},  \href{https://arxiv.org/abs/2111.13564}{{\tt
  2111.13564}}.

\bibitem{Tian:2021cif}
J.~Tian and Y.-N. Wang, \emph{{5D and 6D SCFTs from $\mathbb{C}^3$ orbifolds}},
   \href{https://arxiv.org/abs/2110.15129}{{\tt 2110.15129}}.

\bibitem{Hori:2000ck}
K.~Hori, A.~Iqbal and C.~Vafa, \emph{{D-branes and mirror symmetry}},
  \href{https://arxiv.org/abs/hep-th/0005247}{{\tt hep-th/0005247}}.

\bibitem{Hori:2000kt}
K.~Hori and C.~Vafa, \emph{{Mirror symmetry}},
  \href{https://arxiv.org/abs/hep-th/0002222}{{\tt hep-th/0002222}}.

\bibitem{Ganor:1996pc}
O.~J. Ganor, D.~R. Morrison and N.~Seiberg, \emph{{Branes, Calabi-Yau spaces,
  and toroidal compactification of the N=1 six-dimensional E(8) theory}},
  \href{http://dx.doi.org/10.1016/S0550-3213(96)00690-6}{\emph{Nucl. Phys. B}
  {\bf 487} (1997) 93--127}, [\href{https://arxiv.org/abs/hep-th/9610251}{{\tt
  hep-th/9610251}}].

\bibitem{Eguchi:2002fc}
T.~Eguchi and K.~Sakai, \emph{{Seiberg-Witten curve for the E string theory}},
  \href{http://dx.doi.org/10.1088/1126-6708/2002/05/058}{\emph{JHEP} {\bf 05}
  (2002) 058}, [\href{https://arxiv.org/abs/hep-th/0203025}{{\tt
  hep-th/0203025}}].

\bibitem{Eguchi:2002nx}
T.~Eguchi and K.~Sakai, \emph{{Seiberg-Witten curve for E string theory
  revisited}}, \href{http://dx.doi.org/10.4310/ATMP.2003.v7.n3.a3}{\emph{Adv.
  Theor. Math. Phys.} {\bf 7} (2003) 419--455},
  [\href{https://arxiv.org/abs/hep-th/0211213}{{\tt hep-th/0211213}}].

\bibitem{Katz:1996fh}
S.~H. Katz, A.~Klemm and C.~Vafa, \emph{{Geometric engineering of quantum field
  theories}},
  \href{http://dx.doi.org/10.1016/S0550-3213(97)00282-4}{\emph{Nucl. Phys. B}
  {\bf 497} (1997) 173--195}, [\href{https://arxiv.org/abs/hep-th/9609239}{{\tt
  hep-th/9609239}}].

\bibitem{Klemm:1996bj}
A.~Klemm, W.~Lerche, P.~Mayr, C.~Vafa and N.~P. Warner, \emph{{Self-dual
  strings and N=2 supersymmetric field theory}},
  \href{http://dx.doi.org/10.1016/0550-3213(96)00353-7}{\emph{Nucl. Phys. B}
  {\bf 477} (1996) 746--766}, [\href{https://arxiv.org/abs/hep-th/9604034}{{\tt
  hep-th/9604034}}].

\bibitem{Douglas:1996xp}
M.~R. Douglas, S.~H. Katz and C.~Vafa, \emph{{Small instantons, Del Pezzo
  surfaces and type I-prime theory}},
  \href{http://dx.doi.org/10.1016/S0550-3213(97)00281-2}{\emph{Nucl. Phys.}
  {\bf B497} (1997) 155--172},
  [\href{https://arxiv.org/abs/hep-th/9609071}{{\tt hep-th/9609071}}].

\bibitem{Katz:1997eq}
S.~Katz, P.~Mayr and C.~Vafa, \emph{{Mirror symmetry and exact solution of 4-D
  N=2 gauge theories: 1.}},
  \href{http://dx.doi.org/10.4310/ATMP.1997.v1.n1.a2}{\emph{Adv. Theor. Math.
  Phys.} {\bf 1} (1998) 53--114},
  [\href{https://arxiv.org/abs/hep-th/9706110}{{\tt hep-th/9706110}}].

\bibitem{Doran:1998hm}
C.~F. Doran, \emph{{Picard-Fuchs uniformization: Modularity of the mirror map
  and mirror moonshine}},  \href{https://arxiv.org/abs/math/9812162}{{\tt
  math/9812162}}.

\bibitem{Sebbar:2001}
A.~Sebbar, \emph{{Torsion-free genus zero congruence subgroups of $PSL(2,R)$}},
  \href{http://dx.doi.org/10.1215/S0012-7094-01-11028-4}{\emph{Duke
  Mathematical Journal} {\bf 110} (2001) 377 -- 396}.

\bibitem{Cummins2003}
C.~J. Cummins and S.~Pauli, \emph{{Congruence Subgroups of $PSL(2,Z)$ of Genus
  Less than or Equal to 24}},
  \href{http://dx.doi.org/em/1067634734}{\emph{Experimental Mathematics} {\bf
  12} (2003) 243 -- 255}.

\bibitem{STROMBERG2019436}
F.~Strömberg, \emph{Noncongruence subgroups and maass waveforms},
  \href{http://dx.doi.org/https://doi.org/10.1016/j.jnt.2018.11.020}{\emph{Journal
  of Number Theory} {\bf 199} (2019) 436--493}.

\bibitem{Conway:1979qga}
J.~H. Conway and S.~P. Norton, \emph{{Monstrous Moonshine}},
  \href{http://dx.doi.org/10.1112/blms/11.3.308}{\emph{Bull. London Math. Soc.}
  {\bf 11} (1979) 308--339}.

\bibitem{Harnad:1998hh}
J.~Harnad and J.~McKay, \emph{{Modular solutions to equations of generalized
  Halphen type}}, \href{http://dx.doi.org/10.1098/rspa.2000.0517}{\emph{Proc.
  Roy. Soc. Lond. A} {\bf 456} (2000) 261--294},
  [\href{https://arxiv.org/abs/solv-int/9804006}{{\tt solv-int/9804006}}].

\bibitem{Maier2006}
R.~S. {Maier}, \emph{{On Rationally Parametrized Modular Equations}},
  {\emph{arXiv Mathematics e-prints} (Nov., 2006) math/0611041},
  [\href{https://arxiv.org/abs/math/0611041}{{\tt math/0611041}}].

\bibitem{2016arXiv160503988S}
A.~V. {Sutherland} and D.~{Zywina}, \emph{{Modular curves of prime-power level
  with infinitely many rational points}}, {\emph{arXiv e-prints} (May, 2016)
  arXiv:1605.03988}, [\href{https://arxiv.org/abs/1605.03988}{{\tt
  1605.03988}}].

\bibitem{10.1112/jlms/s2-1.1.351}
M.~H. Millington, \emph{{Subgroups of the Classical Modular Group}},
  \href{http://dx.doi.org/10.1112/jlms/s2-1.1.351}{\emph{Journal of the London
  Mathematical Society} {\bf s2-1} (01, 1969) 351--357},
  [\href{https://arxiv.org/abs/https://academic.oup.com/jlms/article-pdf/s2-1/1/351/2786190/s2-1-1-351.pdf}{{\tt
  https://academic.oup.com/jlms/article-pdf/s2-1/1/351/2786190/s2-1-1-351.pdf}}].

\bibitem{long2007arithmetic}
L.~Long, \emph{The arithmetic subgroups and their modular forms}, .

\bibitem{atkin1971modular}
A.~O.~L. Atkin and H.~P.~F. Swinnerton-Dyer, \emph{Modular forms on
  noncongruence subgroups}, {\emph{Combinatorics} {\bf 19} (1971) 1--25}.

\bibitem{2021arXiv210909040C}
F.~{Calegari}, V.~{Dimitrov} and Y.~{Tang}, \emph{{The Unbounded Denominators
  Conjecture}}, {\emph{arXiv e-prints} (Sept., 2021) arXiv:2109.09040},
  [\href{https://arxiv.org/abs/2109.09040}{{\tt 2109.09040}}].

\bibitem{Scholl1985}
A.~Scholl, \emph{Modular forms and de rham cohomology; atkin-swinnerton-dyer
  congruences.}, {\emph{Inventiones mathematicae} {\bf 79} (1985) 49--78}.

\bibitem{Scholl1988TheLR}
A.~Scholl, \emph{The l-adic representations attached to a certain noncongruence
  subgroup.}, {\emph{Journal f{\"u}r die reine und angewandte Mathematik
  (Crelles Journal)} {\bf 1988} (1988) 1 -- 15}.

\bibitem{Scholl2010OnLR}
A.~Scholl, \emph{On l-adic representations attached to non-congruence subgroups
  ii},  2010.

\bibitem{schuttshioda}
M.~Sch{\"u}tt and T.~Shioda, \emph{Mordell--Weil Lattices}.
\newblock Ergebnisse der Mathematik und ihrer Grenzgebiete. 3. Folge / A Series
  of Modern Surveys in Mathematics. Springer Singapore, 2019.

\bibitem{Argyres:2015ffa}
P.~Argyres, M.~Lotito, Y.~L\"u and M.~Martone, \emph{{Geometric constraints on
  the space of $ \mathcal{N} $ = 2 SCFTs. Part I: physical constraints on
  relevant deformations}},
  \href{http://dx.doi.org/10.1007/JHEP02(2018)001}{\emph{JHEP} {\bf 02} (2018)
  001}, [\href{https://arxiv.org/abs/1505.04814}{{\tt 1505.04814}}].

\bibitem{Argyres:2015gha}
P.~C. Argyres, M.~Lotito, Y.~L\"u and M.~Martone, \emph{{Geometric constraints
  on the space of $ \mathcal{N} $ = 2 SCFTs. Part II: construction of special
  K\"ahler geometries and RG flows}},
  \href{http://dx.doi.org/10.1007/JHEP02(2018)002}{\emph{JHEP} {\bf 02} (2018)
  002}, [\href{https://arxiv.org/abs/1601.00011}{{\tt 1601.00011}}].

\bibitem{Argyres:2016xmc}
P.~Argyres, M.~Lotito, Y.~L\"u and M.~Martone, \emph{{Geometric constraints on
  the space of $ \mathcal{N}$ = 2 SCFTs. Part III: enhanced Coulomb branches
  and central charges}},
  \href{http://dx.doi.org/10.1007/JHEP02(2018)003}{\emph{JHEP} {\bf 02} (2018)
  003}, [\href{https://arxiv.org/abs/1609.04404}{{\tt 1609.04404}}].

\bibitem{Argyres:2016xua}
P.~C. Argyres, M.~Lotito, Y.~L\"u and M.~Martone, \emph{{Expanding the
  landscape of $ \mathcal{N} $ = 2 rank 1 SCFTs}},
  \href{http://dx.doi.org/10.1007/JHEP05(2016)088}{\emph{JHEP} {\bf 05} (2016)
  088}, [\href{https://arxiv.org/abs/1602.02764}{{\tt 1602.02764}}].

\bibitem{Bonelli:2016idi}
G.~Bonelli, A.~Grassi and A.~Tanzini, \emph{{Seiberg\textendash{}Witten theory
  as a Fermi gas}},
  \href{http://dx.doi.org/10.1007/s11005-016-0893-z}{\emph{Lett. Math. Phys.}
  {\bf 107} (2017) 1--30}, [\href{https://arxiv.org/abs/1603.01174}{{\tt
  1603.01174}}].

\bibitem{Bonelli:2016qwg}
G.~Bonelli, O.~Lisovyy, K.~Maruyoshi, A.~Sciarappa and A.~Tanzini, \emph{{On
  Painlev\'e/gauge theory correspondence}},
  \href{https://arxiv.org/abs/1612.06235}{{\tt 1612.06235}}.

\bibitem{Bonelli:2017gdk}
G.~Bonelli, A.~Grassi and A.~Tanzini, \emph{{Quantum curves and $q$-deformed
  Painlev\'e equations}},
  \href{http://dx.doi.org/10.1007/s11005-019-01174-y}{\emph{Lett. Math. Phys.}
  {\bf 109} (2019) 1961--2001}, [\href{https://arxiv.org/abs/1710.11603}{{\tt
  1710.11603}}].

\bibitem{Argyres:2018taw}
P.~C. Argyres and M.~Lotito, \emph{{Flavor symmetries and the topology of
  special K\"ahler structures at rank 1}},
  \href{http://dx.doi.org/10.1007/JHEP02(2019)026}{\emph{JHEP} {\bf 02} (2019)
  026}, [\href{https://arxiv.org/abs/1811.00016}{{\tt 1811.00016}}].

\bibitem{Nahm:1996di}
W.~Nahm, \emph{{On the Seiberg-Witten approach to electric - magnetic
  duality}},  \href{https://arxiv.org/abs/hep-th/9608121}{{\tt
  hep-th/9608121}}.

\bibitem{Verrill2000}
H.~Verrill, ``Fundamental domain drawer.''
  \url{https://wstein.org/Tables/fundomain/}.

\bibitem{Douglas:2000ah}
M.~R. Douglas, B.~Fiol and C.~Romelsberger, \emph{{Stability and BPS branes}},
  \href{http://dx.doi.org/10.1088/1126-6708/2005/09/006}{\emph{JHEP} {\bf 09}
  (2005) 006}, [\href{https://arxiv.org/abs/hep-th/0002037}{{\tt
  hep-th/0002037}}].

\bibitem{Douglas:2000qw}
M.~R. Douglas, B.~Fiol and C.~Romelsberger, \emph{{The Spectrum of BPS branes
  on a noncompact Calabi-Yau}},
  \href{http://dx.doi.org/10.1088/1126-6708/2005/09/057}{\emph{JHEP} {\bf 09}
  (2005) 057}, [\href{https://arxiv.org/abs/hep-th/0003263}{{\tt
  hep-th/0003263}}].

\bibitem{Aspinwall:2004jr}
P.~S. Aspinwall, \emph{{D-branes on Calabi-Yau manifolds}},  in
  \emph{{Theoretical Advanced Study Institute in Elementary Particle Physics
  (TASI 2003): Recent Trends in String Theory}}, 3, 2004.
\newblock \href{https://arxiv.org/abs/hep-th/0403166}{{\tt hep-th/0403166}}.
\newblock \href{http://dx.doi.org/10.1142/9789812775108_0001}{DOI}.

\bibitem{Hanany:2012hi}
A.~Hanany and R.-K. Seong, \emph{{Brane Tilings and Reflexive Polygons}},
  \href{http://dx.doi.org/10.1002/prop.201200008}{\emph{Fortsch. Phys.} {\bf
  60} (2012) 695--803}, [\href{https://arxiv.org/abs/1201.2614}{{\tt
  1201.2614}}].

\bibitem{Caorsi:2019vex}
M.~Caorsi and S.~Cecotti, \emph{{Homological classification of 4d $ \mathcal{N}
  $ = 2 QFT. Rank-1 revisited}},
  \href{http://dx.doi.org/10.1007/JHEP10(2019)013}{\emph{JHEP} {\bf 10} (2019)
  013}, [\href{https://arxiv.org/abs/1906.03912}{{\tt 1906.03912}}].

\bibitem{Cecotti:2021ouq}
S.~Cecotti, M.~Del~Zotto, M.~Martone and R.~Moscrop, \emph{{{The Characteristic
  Dimension of Four-dimensional $\mathcal{N} = 2$ SCFTs}}},
  \href{https://arxiv.org/abs/2108.10884}{{\tt 2108.10884}}.

\bibitem{Bilal:1996sk}
A.~Bilal and F.~Ferrari, \emph{{Curves of marginal stability, and weak and
  strong coupling BPS spectra in N=2 supersymmetric QCD}},
  \href{http://dx.doi.org/10.1016/S0550-3213(96)00480-4}{\emph{Nucl. Phys. B}
  {\bf 480} (1996) 589--622}, [\href{https://arxiv.org/abs/hep-th/9605101}{{\tt
  hep-th/9605101}}].

\bibitem{Ferrari:1996sv}
F.~Ferrari and A.~Bilal, \emph{{The Strong coupling spectrum of the
  Seiberg-Witten theory}},
  \href{http://dx.doi.org/10.1016/0550-3213(96)00150-2}{\emph{Nucl. Phys. B}
  {\bf 469} (1996) 387--402}, [\href{https://arxiv.org/abs/hep-th/9602082}{{\tt
  hep-th/9602082}}].

\bibitem{Bilal:1997st}
A.~Bilal and F.~Ferrari, \emph{{The BPS spectra and superconformal points in
  massive N=2 supersymmetric QCD}},
  \href{http://dx.doi.org/10.1016/S0550-3213(98)00052-2}{\emph{Nucl. Phys. B}
  {\bf 516} (1998) 175--228}, [\href{https://arxiv.org/abs/hep-th/9706145}{{\tt
  hep-th/9706145}}].

\bibitem{Conrad}
K.~Conrad, ``{{$SL(2,\mathbb{Z})$}}.''
  \url{https://kconrad.math.uconn.edu/blurbs/}.

\bibitem{Klemm:1995wp}
A.~Klemm, W.~Lerche and S.~Theisen, \emph{{Nonperturbative effective actions of
  N=2 supersymmetric gauge theories}},
  \href{http://dx.doi.org/10.1142/S0217751X96001000}{\emph{Int. J. Mod. Phys.
  A} {\bf 11} (1996) 1929--1974},
  [\href{https://arxiv.org/abs/hep-th/9505150}{{\tt hep-th/9505150}}].

\bibitem{Huang:2013yta}
M.-X. Huang, A.~Klemm and M.~Poretschkin, \emph{{Refined stable pair invariants
  for E-, M- and $[p, q]$-strings}},
  \href{http://dx.doi.org/10.1007/JHEP11(2013)112}{\emph{JHEP} {\bf 11} (2013)
  112}, [\href{https://arxiv.org/abs/1308.0619}{{\tt 1308.0619}}].

\bibitem{Kim:2014nqa}
S.-S. Kim and F.~Yagi, \emph{{5d E$_{n}$ Seiberg-Witten curve via toric-like
  diagram}}, \href{http://dx.doi.org/10.1007/JHEP06(2015)082}{\emph{JHEP} {\bf
  06} (2015) 082}, [\href{https://arxiv.org/abs/1411.7903}{{\tt 1411.7903}}].

\bibitem{2006math.....12100K}
C.~{Kurth} and L.~{Long}, \emph{{On modular forms for some noncongruence
  arithmetic subgroups}}, {\emph{arXiv Mathematics e-prints} (Dec., 2006)
  math/0612100}, [\href{https://arxiv.org/abs/math/0612100}{{\tt
  math/0612100}}].

\bibitem{Cecotti:2015qha}
S.~Cecotti and M.~Del~Zotto, \emph{{{Galois covers of $\mathcal{N}=2$ BPS
  spectra and quantum monodromy}}},
  \href{http://dx.doi.org/10.4310/ATMP.2016.v20.n6.a1}{\emph{Adv. Theor. Math.
  Phys.} {\bf 20} (2016) 1227--1336},
  [\href{https://arxiv.org/abs/1503.07485}{{\tt 1503.07485}}].

\bibitem{Bourget:2017goy}
A.~Bourget and J.~Troost, \emph{{Permutations of Massive Vacua}},
  \href{http://dx.doi.org/10.1007/JHEP05(2017)042}{\emph{JHEP} {\bf 05} (2017)
  042}, [\href{https://arxiv.org/abs/1702.02102}{{\tt 1702.02102}}].

\bibitem{Ashok:2006br}
S.~K. Ashok, F.~Cachazo and E.~Dell'Aquila, \emph{{Children's drawings from
  Seiberg-Witten curves}},
  \href{http://dx.doi.org/10.4310/CNTP.2007.v1.n2.a1}{\emph{Commun. Num. Theor.
  Phys.} {\bf 1} (2007) 237--305},
  [\href{https://arxiv.org/abs/hep-th/0611082}{{\tt hep-th/0611082}}].

\bibitem{He:2020eva}
Y.-H. He, E.~Hirst and T.~Peterken, \emph{{Machine-learning dessins
  d\textquoteright{}enfants: explorations via modular and
  Seiberg\textendash{}Witten curves}},
  \href{http://dx.doi.org/10.1088/1751-8121/abbc4f}{\emph{J. Phys. A} {\bf 54}
  (2021) 075401}, [\href{https://arxiv.org/abs/2004.05218}{{\tt 2004.05218}}].

\bibitem{Bao:2021vxt}
J.~Bao, O.~Foda, Y.-H. He, E.~Hirst, J.~Read, Y.~Xiao et~al., \emph{{Dessins
  d\textquoteright{}enfants, Seiberg-Witten curves and conformal blocks}},
  \href{http://dx.doi.org/10.1007/JHEP05(2021)065}{\emph{JHEP} {\bf 05} (2021)
  065}, [\href{https://arxiv.org/abs/2101.08843}{{\tt 2101.08843}}].

\bibitem{Alim:2013eja}
M.~Alim, E.~Scheidegger, S.-T. Yau and J.~Zhou, \emph{{Special Polynomial
  Rings, Quasi Modular Forms and Duality of Topological Strings}},
  \href{http://dx.doi.org/10.4310/ATMP.2014.v18.n2.a4}{\emph{Adv. Theor. Math.
  Phys.} {\bf 18} (2014) 401--467},
  [\href{https://arxiv.org/abs/1306.0002}{{\tt 1306.0002}}].

\bibitem{10.1215/ijm/1255632506}
M.~Newman, \emph{{The structure of some subgroups of the modular group}},
  \href{http://dx.doi.org/10.1215/ijm/1255632506}{\emph{Illinois Journal of
  Mathematics} {\bf 6} (1962) 480 -- 487}.

\bibitem{2020arXiv200701336F}
A.~{Fiori} and C.~{Franc}, \emph{{{The unbounded denominator conjecture for the
  noncongruence subgroups of index $7$}}}, {\emph{arXiv e-prints} (July, 2020)
  arXiv:2007.01336}, [\href{https://arxiv.org/abs/2007.01336}{{\tt
  2007.01336}}].

\bibitem{2008arXiv0805.2144F}
L.~{Fang}, J.~W. {Hoffman}, B.~{Linowitz}, A.~{Rupinski} and H.~{Verrill},
  \emph{{Modular forms on noncongruence subgroups and Atkin-Swinnerton-Dyer
  relations}}, {\emph{arXiv e-prints} (May, 2008) arXiv:0805.2144},
  [\href{https://arxiv.org/abs/0805.2144}{{\tt 0805.2144}}].

\bibitem{2009arXiv0910.0739K}
L.~J.~P. {Kilford}, \emph{{Experimental finding of modular forms for
  noncongruence subgroups}}, {\emph{arXiv e-prints} (Oct., 2009)
  arXiv:0910.0739}, [\href{https://arxiv.org/abs/0910.0739}{{\tt 0910.0739}}].

\bibitem{Schultz}
D.~Schultz, ``{Notes on Modular Forms}.''
  \url{https://faculty.math.illinois.edu/~schult25/}.

\bibitem{Argyres:2007tq}
P.~C. Argyres and J.~R. Wittig, \emph{{Infinite coupling duals of N=2 gauge
  theories and new rank 1 superconformal field theories}},
  \href{http://dx.doi.org/10.1088/1126-6708/2008/01/074}{\emph{JHEP} {\bf 01}
  (2008) 074}, [\href{https://arxiv.org/abs/0712.2028}{{\tt 0712.2028}}].

\bibitem{Cecotti:2013sza}
S.~Cecotti and M.~Del~Zotto, \emph{{{The BPS spectrum of the 4d $\mathcal{N}$ =
  2 SCFT's $H_1, H_2, D_4, E_6, E_7, E_8$}}},
  \href{http://dx.doi.org/10.1007/JHEP06(2013)075}{\emph{JHEP} {\bf 06} (2013)
  075}, [\href{https://arxiv.org/abs/1304.0614}{{\tt 1304.0614}}].

\bibitem{Manschot:2019pog}
J.~Manschot, G.~W. Moore and X.~Zhang, \emph{{Effective gravitational couplings
  of four-dimensional $ \mathcal{N} $ = 2 supersymmetric gauge theories}},
  \href{http://dx.doi.org/10.1007/JHEP06(2020)150}{\emph{JHEP} {\bf 06} (2020)
  150}, [\href{https://arxiv.org/abs/1912.04091}{{\tt 1912.04091}}].

\bibitem{10.1215/ijm/1256059574}
K.~Wohlfahrt, \emph{{An extension of F. Klein's level concept}},
  \href{http://dx.doi.org/10.1215/ijm/1256059574}{\emph{Illinois Journal of
  Mathematics} {\bf 8} (1964) 529 -- 535}.

\end{thebibliography}\endgroup

\end{document}